\documentclass{elsart}

\usepackage{natbib}

\usepackage{graphics, graphicx}

\usepackage{amssymb}

\begin{document}

\begin{frontmatter}


\title{Lopsided spiral galaxies}
\author{Chanda J. Jog$^{*}$ \& Francoise Combes$^{**}$}
 
\address{$^*$ Department of Physics, Indian Institute of Science, Bangalore 560012, India}
\address{$^{**}$ Observatoire de Paris, LERMA, 61 Av. de l'Observatoire, Paris F-75014, France}
\footnote{Corresponding author: Chanda J. Jog (email: cjjog@physics.iisc.ernet.in)}








\address{}

\begin{abstract}
The light distribution in the disks of many galaxies is
non-axisymmetric or `lopsided' with a spatial extent much larger
along one half of a galaxy than the other, as in M101. Recent near-IR 
observations show that
lopsidedness is common. The stellar disks in nearly 30 \% of galaxies
have significant lopsidedness of $> 10 \% $ measured as the Fourier amplitude of the m=1 component normalized to the average value.
This asymmetry is traced
particularly well by the atomic hydrogen gas distribution lying in
the outer parts. 
The lopsidedness also occurs in the nuclear regions, where the nucleus is
offset with respect to the outer isophotes. The galaxies in a group 
environment show higher lopsidedness.
The origin of lopsidedness
could be due to the disk response to a tidally distorted halo, 
or via gas accretion. An m=1 perturbation in a
 disk leads to a shift in the center of mass in the disk, and this
then acts as an indirect force on the original center of the
disk. The disk is inherently supportive of an m=1
mode, which is a particular feature only of lopsided modes, and
which makes their dynamical study interesting
and challenging. 

The lopsidedness has a large impact on the dynamics
of the galaxy, its evolution, the star formation in it, and on the
growth of the central black hole and on the nuclear fueling, merging
of binary black holes etc.  The disk lopsidedness can be used as a diagnostic to study
the halo asymmetry.
This is an emerging area in galactic structure and dynamics. 
In this review, the observations
to measure the lopsided distribution, as well as the theoretical progress
made so far to understand its origin and properties, and the related 
open problems will be discussed. 
\end{abstract}

\begin{keyword}
galaxies : kinematics and dynamics - galaxies :
ISM  -  galaxies : spiral - galaxies : structure - galaxies :
individual: M101 

\end{keyword}

\end{frontmatter}

\newpage

\noindent   {\bf Contents:}

\noindent 1. Introduction

\noindent 2. Observations of lopsidedness in galactic disks

  2.1 Morphological lopsidedness

2.1.1 Morphological lopsidedness in HI gas

2.1.2 Morphological lopsidedness in old stars

2.2 Kinematical lopsidedness

2.3 Phase of the disk lopsidedness

2.4 Observations of off-centered nuclear disks

2.4.1 The case of the M31 nuclear disk

2.4.2 Other off-centered nuclei

3. Theoretical models for the origin of lopsidedness

3.1 Kinematical model for the origin of lopsidedness

3.2 Dynamical models for the origin of lopsidedness

3.2.1 Tidal encounters, and disk response to distorted 
halo

3.2.1.a  Orbits and isophotes in a perturbed disk

3.2.1.b  Kinematics in a perturbed disk

3.2.1.c Radius for the onset of lopsidedness

3.2.1.d   Comparison of A$_1$ vs. Tidal parameter

3.2.2 Gas Accretion, and other mechanisms

3.2.3 Lopsidedness as a global instability

3.2.4  Effect of inclusion of 
           rotation and a live halo

3.3 Comparison between origin of m=1 and m=2, stars and gas

3.4 A summary of the various mechanisms

4. Lopsidedness in the central region

4.1 Stability of central nuclear disks

4.1.1 Slow stable modes, damping slowly

4.2 Double nuclei due to infalling bodies

4.3 Core wandering

4.4 Other mechanisms

5.  Lopsidedness in galaxies in groups, clusters, and mergers

5.1 Lopsidedness in galaxies in groups

5.2 Lopsidedness in galaxies in clusters

5.3 Lopsidedness in centers of advanced mergers

6. Related topics

6.1 Relative strengths of lopsidedness (m=1) and spiral arms/bars (m=2)

6.1.1 Observed amplitudes of m=1 and m=2 components

6.2 Asymmetry in the dark matter halo

6.3 Comparison with warps

6.4 Implications for high redshift galaxies

7. Effect of lopsidedness on  galaxy evolution

8. Summary \& future directions

\newpage

\section{Introduction}

It has been known for a long time that the light and hence the mass
distribution in disks of spiral galaxies is not strictly
axisymmetric, as for example in M 101 or in NGC 1637 (Sandage 1961),
where the isophotes are elongated in one half of the galaxy.
Despite this, however, astronomers have largely tended to ignore
this fact and to assume the disks to be axisymmetric  because it is
much simpler to study the dynamics of axisymmetric disks.
 This phenomenon was first highlighted
in the pioneering paper by Baldwin, Lynden-Bell, \& Sancisi (1980),
where they detected an asymmetry in the spatial extent of the atomic
hydrogen gas in the outer regions in the two  halves of some
galaxies, and gave these the apt name of `lopsided' galaxies. A
galaxy is said to be lopsided if it displays a global
non-axisymmetric spatial distribution of type $m=1$ where $m$ is the
azimuthal wavenumber, or a $cos \phi$  distribution where $\phi$ is
the  azimuthal angle in the plane of the disk.

Surprisingly no further systematic work was done on this topic till
mid-1990's. Since then there has been a resurgence in this field.
The lopsided distribution has now been detected and studied also in the old
stellar component as traced in the near-IR starting
with the observations of Block et al. (1994) and Rix \& Zaritsky
(1995). This exciting new development of the imaging studies of
spiral galaxies in the near-IR K-band (2.2 $\mu$) was made possible
by the development of the NICMOS 3 array. The dust extinction
effects are negligible in the near-IR, hence these studies reveal the
spatial distribution of the underlying old stellar population, which constitute the main mass
component of the disk. These
observations detected a non-axisymmetric $m = 1 $ distribution of surface
density of old stars in the inner/optical region of the disk. Rix
\& Zaritsky (1995) define $A_1$, the fractional amplitude of
the first azimuthal fourier component ($m = 1$) of surface
brightness, to be the quantitative measure of disk lopsidedness. They
 find that $A_1 $ increases with radius. The average value
measured between 1.5-2.5 disk scalelengths is large $\geq 0.1$, and
$30 \% $ of the galaxies studied show a higher lopsidedness
(Zaritsky \& Rix 1997). A similar high average value of disk
lopsidedness was confirmed in a recent Fourier-analysis study of a
much larger sample of 149 galaxies (Bournaud et al. 2005 b).

The above analysis shows that nearly one third of the 149 galaxies
exhibit 10 \% or more asymmetry in the amplitude of the $m=1$
Fourier component. Thus, {\it lopsided distribution in the disk is a
general phenomenon}, and is stronger at larger radii. Hence it is
important to understand the origin and dynamics of the 
lopsided distribution in spiral galaxies.

The lopsided distribution in the HI gas has been mapped spatially
(Haynes et al. 1998), and also mapped kinematically for a few
galaxies (Schoenmakers, Franx \& de Zeeuw 1997, Swaters et al. 1999), and by
global velocity profiles for a much larger sample (Richter \& Sancisi
1994). Such an asymmetry has also been detected in dwarf galaxies
(Swaters et al. 2002), and also in the star-forming regions in
irregular galaxies (Heller et al. 2000). The asymmetry may affect
all scales in a galaxy. While the large-scale lopsidedness is more
conspicuous, the off-centering of nuclei is now often discovered at
high spatial resolution. A prototype of this $m=1$ nuclear
distribution is the inner region of M31, where the central black
hole is clearly off-centered with respect to its nuclear stellar
disk (e.g., Tremaine 2001). This frequent 
nuclear $m=1$  perturbation must play a central
role in the fueling of the active galactic nucleus (AGN) in a galaxy.

The origin and the evolution of lopsidedness are not yet
well-understood, though a beginning has been made to address these
problems theoretically. Like any other non-axisymmetric perturbation, the 
lopsided distribution would also tend to get wound up by the differential 
rotation in the galactic disk within a few dynamical timescales.
Since a large fraction of galaxies exhibit lopsidedness, it must be either 
a long-lived phenomenon or generated frequently.
 Tidal interaction (Beale \& Davies 1969),
and satellite galaxy accretion (Zaritsky \& Rix 1997) have been
suggested as the origin of the disk lopsidedness, these can occur frequently.
 Weinberg (1995) has shown that
the tidal interaction between the Galaxy and the Large Magellanic Cloud
(LMC) leads to a
lopsided distortion of the Galaxy halo at resonance points between
the LMC and the halo orbit frequencies, which in turn causes a
lopsided distribution in the disk of the Galaxy. Since galaxy
interactions are now known to be common, the origin of disk
lopsidedness as attributed to the disk response to the tidal distortion in a
halo has been proposed and studied by Jog (1997, 2002), and Schoenmakers et al. (1997). 
Some other possible mechanisms that have been suggested involve
  an off-center disk in a halo as in a dwarf galaxy (Levine \&
Sparke 1998), or gas accretion (Bournaud et al. 2005 b), or treating it as a
global, long-lived mode (Saha, Combes \& Jog 2007). 

The $m=1$ distribution in the inner regions of some galaxies such as
M 31 has been modeled through analytical work and numerical
simulations (Tremaine 1995, Statler 2001, Bacon et al. 2001, de
Oliveira \& Combes 2008). According to the various physical
conditions in galaxy nuclei (such as the mass of the bulge, the mass
of the nuclear disk and that of the central black hole, the presence
of gas, etc.), an $m=1$ mode is unstable, or an $m=1$ excitation is
very slowly damped and can persist for several hundreds of dynamical
times. Such long-lived lopsided distribution is also seen in the centers of advanced
mergers of galaxies (Jog \& Maybhate 2006).
Due to their persistence, the lopsided modes could play a
significant role in the evolution of the central regions of
galaxies, especially in the fueling of a central AGN.

An m=1 perturbation in a
 disk leads to a shift in the center of mass in the disk, and this
then acts as an indirect force on the original center of the
disk. The disk is inherently supportive of an m=1
mode, which is a particular feature only of a lopsided mode. This 
results in long-lasting global lopsided modes. While the $m=2$ case 
corresponding to the two-armed spiral pattern or bars has been studied 
extensively, the $m=1$ mode has not received comparable
attention in the literature so far. This has to be redressed: first, 
m=1 is common and the amplitude is even larger than for m=2 (Jarrett 
et al. 2003), second, the m=1 modes do not have an Inner Lindblad Resonance,
or ILR (e.g., Block et al. 1994),
and hence can allow transport of matter in the inner regions, and third, these 
appear to be long-lived.

The existence of long-lived lopsided modes is expected to have a significant impact
on the dynamics of the galaxy, the star formation in it, and on the nuclear fueling etc.
In the tidal picture, the disk lopsidedness can be used
as a diagnostic to study the lopsidedness of the dark matter halo
(Jog 1999). Similarly, higher-order ($m=2$) disk asymmetry can allow
one to study the ellipticity of the dark matter halo (Jog 2000, 
Andersen \& Bershady 2002).
 Thus in addition to being interesting and challenging in
itself, a study of disk lopsidedness can yield information about a
number of other interesting properties of galaxies. Extra-planar gas
and lopsidedness are frequently correlated (Sancisi et al 2008).

Recently, the lopsided distribution in galaxies in different environments
such as groups and clusters and centers of mergers has been studied.
These show different properties, such as the higher observed lopsided amplitudes in
the group galaxies (Angiras et al. 2006). In future studies, these can act as 
important tracers of the dynamics of disks and dark matter halos in these settings.

In Section 2, we discuss the observational properties of
lopsidedness as seen in HI and old stars, in the main disk as well
as in the central region of galaxies. The various theoretical models
proposed in the literature and their comparison is given in Section
3. The dynamics of lopsidedness in the central region of the
near-Keplerian case region is discussed in Section 4. Section 5
discusses the observations and dynamical implications of lopsidedness 
in galaxies in a different setting, namely in groups, clusters and in 
centers of mergers.
Section 6 discusses several related points including the 
comparison between m=1 and 2 cases, the deduction of the halo asymmetry etc.
Section 7 gives the effect
of lopsidedness on the galaxy. Finally, Section 8 gives a brief
summary and future directions for this field.

\section{Observations of lopsidedness in galactic disks}

While lopsidedness is seen to be a ubiquitous phenomenon (Section
1), various methods are used in the literature to define the
asymmetry in disk galaxies. We list and compare below the details,
such as the definitions, the size of the sample studied, the tracer
used (near-IR radiation from old stars, and 21 cm emission from the
HI gas etc.) There is no consensus so far as to what is the
definition for lopsidedness as well as what constitutes lopsidedness
(or the threshold). Obviously this has to be done if say the percentage
of galaxies showing lopsidedness as per the different criteria are
to be compared. At the end of Section 2.1, we recommend the use of a standard
criterion for lopsidedness as well as the threshold that could be adopted 
by future workers. 

The disk lopsidedness in spiral galaxies has been studied in two
different tracers- HI, the atomic hydrogen gas studied in the outer
parts of disks, and the near-IR radiation from the inner/optical
disks which traces the old stellar component of the disks. The
lopsidedness was historically first detected in the HI, which we now
realize is due to a higher lopsided amplitude at large radii. Hence
we follow the same, chronological order in the summary of observations given next.

\subsection{Morphological lopsidedness}

\subsubsection{Morphological lopsidedness in HI gas}

The asymmetric distribution of HI in M101 was noted in the early
HI observations (Beale \& Davies 1969). This phenomenon was first
highlighted by Baldwin et al. (1980), who studied galaxies with highly asymmetric spatial
extent of atomic hydrogen gas distributions in the outer regions in
the two halves of a galaxy, such as M101 and NGC 2841 (see Fig. 1 here).
 This paper mentions the asymmetric distribution of light and HI in
spiral galaxies such as M101. Quantitatively, they looked at the HI
distribution in the four prototypical galaxies, namely, M101, NGC
891, NGC 4565, NGC 2841. They defined a galaxy to be "lopsided" 
in which the galaxy is more extended on one side than the other, and
where the projected density of HI on the two sides of the 
galaxy is at least 2:1 , and in which the asymmetry persists over a 
large range in radius and position angle. All these lopsided galaxies 
were also noted to have large-scale velocity perturbations. 
For a quantitative measurement of lopsidedness the edge-on systems 
cannot be used.
The cut-off in inclination used for the near-IR and HI studies is given in Sections 2.1.2 and 5.1 respectively.

\bigskip

\begin{figure}[h]
\centering
\includegraphics[height=2.5in, width=2.5in]{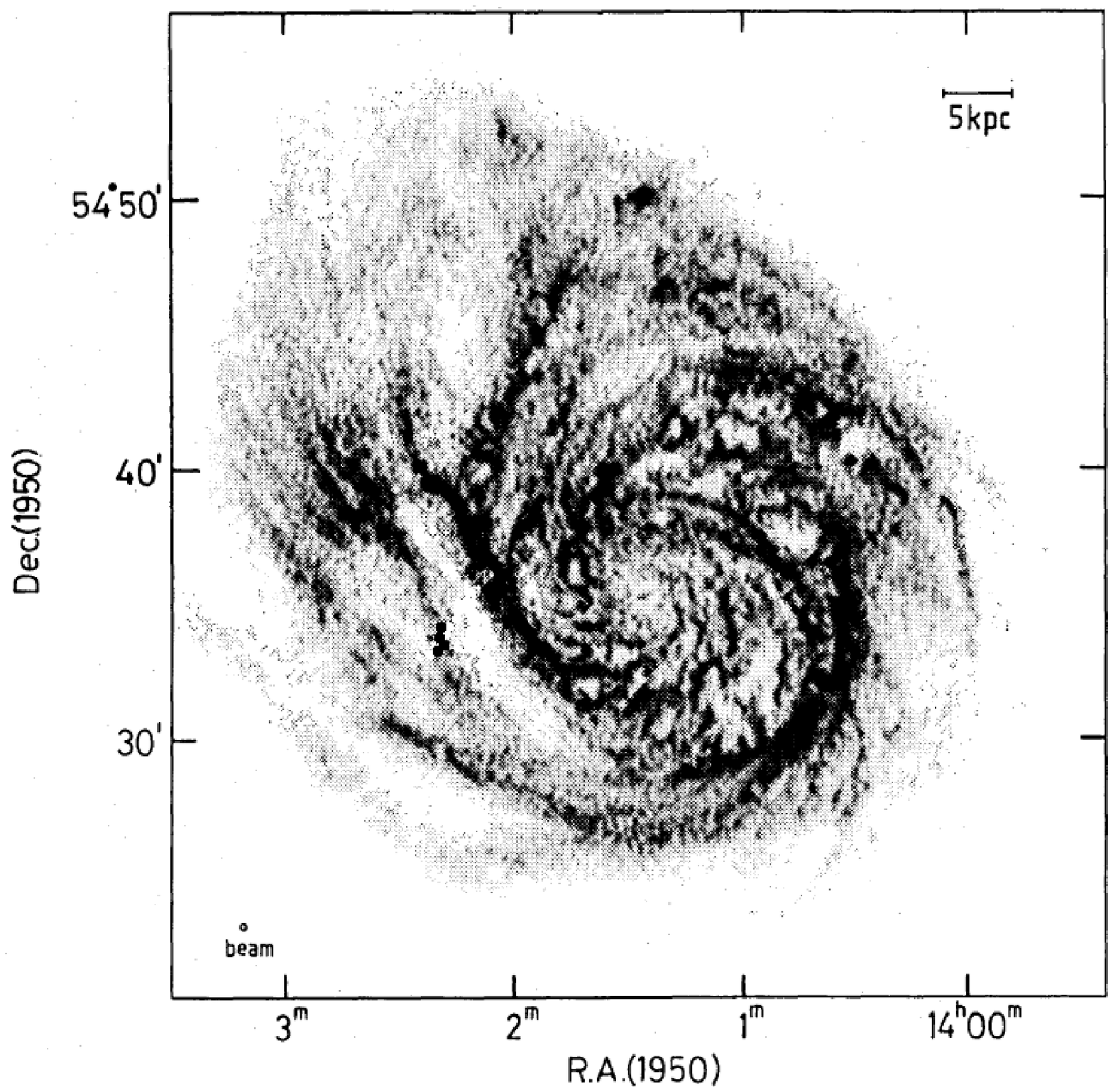}
\medskip 
\includegraphics[height=2.5in, width= 2.3in]{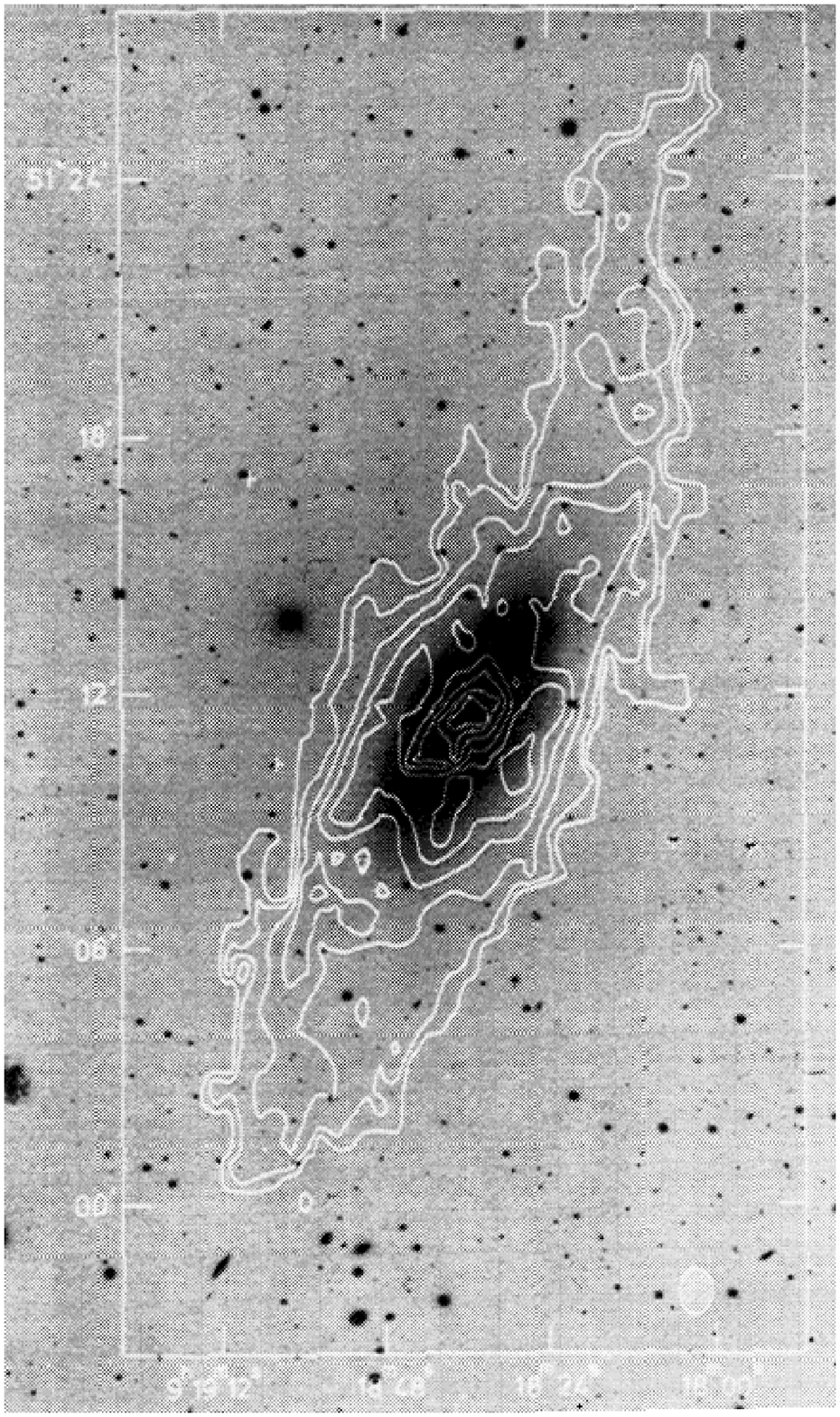} 
\bigskip
\includegraphics[height=2.5in, width=2.5in]{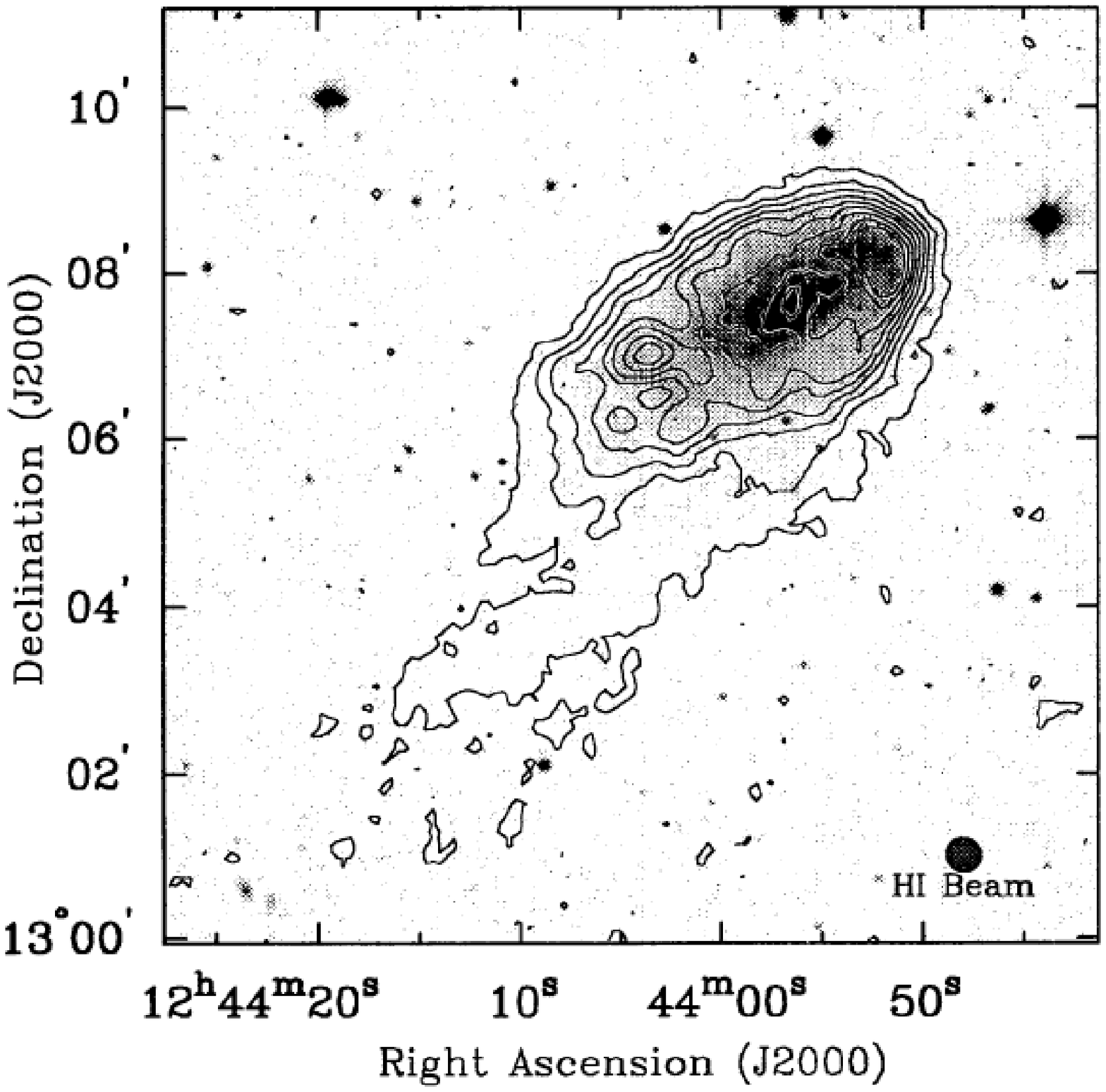}
\medskip
\includegraphics[height=2.6in, width=2.7in]{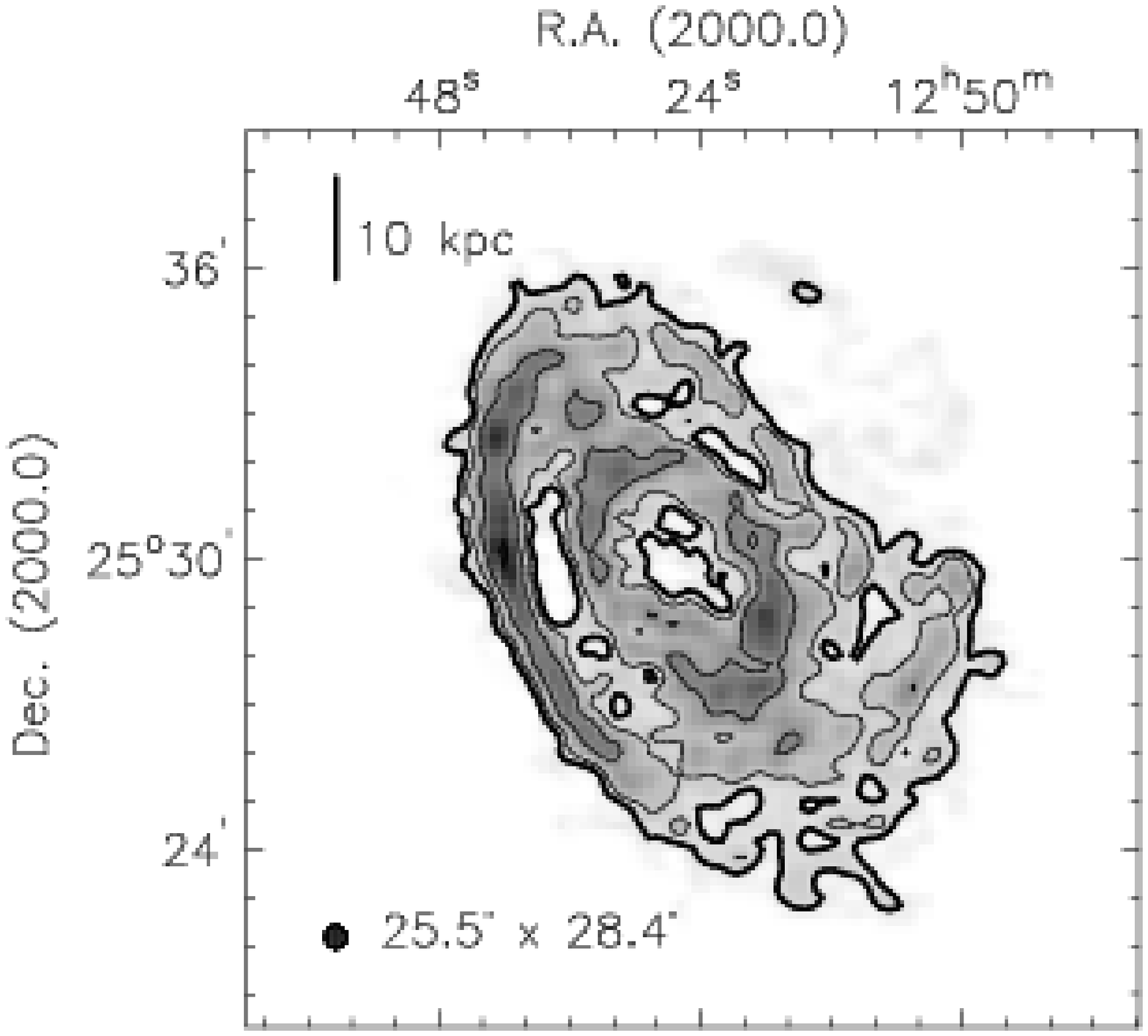}
\bigskip
\caption{Galaxies showing an asymmetry in the spatial extent of 2:1 or more in the HI distribution : 
M101 (top left, where the HI intensity is plotted here as gray scale) 
and NGC 2841 (top right: here the HI contours are superimposed
on an optical image),  These galaxies were termed "lopsided" galaxies
by Baldwin, Lynden-Bell, \& Sancisi (1980). These figures are from Braun (1995) and Bosma (1981) respectively.
Other typical examples are NGC 4654 (lower left: where HI contours are superimposed
on an optical image) and UGC7989 (lower right, showing contours and grey scale of the
HI intensity), from Phookun \& Mundy (1995),
and Noordermeer et al. (2005) respectively.} 
 \label{fig1}
\end{figure}

\bigskip

There was no further work on this topic for over a
decade. Further HI mapping of a few individual galaxies such
as NGC 4254 was done by Phookun, Vogel, \& Mundy (1993) which
stressed the obvious spatial asymmetry but they did not 
measure the lopsidedness. This paper studied the special case of
the not-so-common one-armed galaxies such as NGC 4254 where the phase 
varies with radius (see Section 2.3).

Richter \& Sancisi (1994) collected data from single-dish HI
surveys done using different radio telescopes, for a large sample of about 1700 galaxies. They found that 50 \% 
of galaxies studied show strong
lopsidedness in their global HI velocity profiles. This could be either due to spatial
lopsided distribution and/or lopsided kinematics. But since a
large fraction of their sample shows this effect, they concluded that it must reflect an asymmetry 
in the gas density, as is confirmed by the correlation between the spatial and global HI velocity asymmetries in some galaxies like NGC 891, see Fig. 2 here. They argued that the asymmetry in HI represents a large-scale 
structural property of the galaxies. The 
criteria they used to decide the asymmetry are: (1). significant flux difference 
between the two horns ( $> 20
\%$ or $> 8$ sigma) (2). Total flux difference ($> 55 : 45 \%$)
between the low and the high velocity halves (3). Width differences
in the two horns ($>$ 4 velocity channels or 50 km s$^{-1}$). One
word of caution is that it is not clear from their paper if these
three give consistent identification of a galaxy as being lopsided
or not.

\bigskip

\begin{figure}[h]
\centering
\includegraphics[height=5.0in]{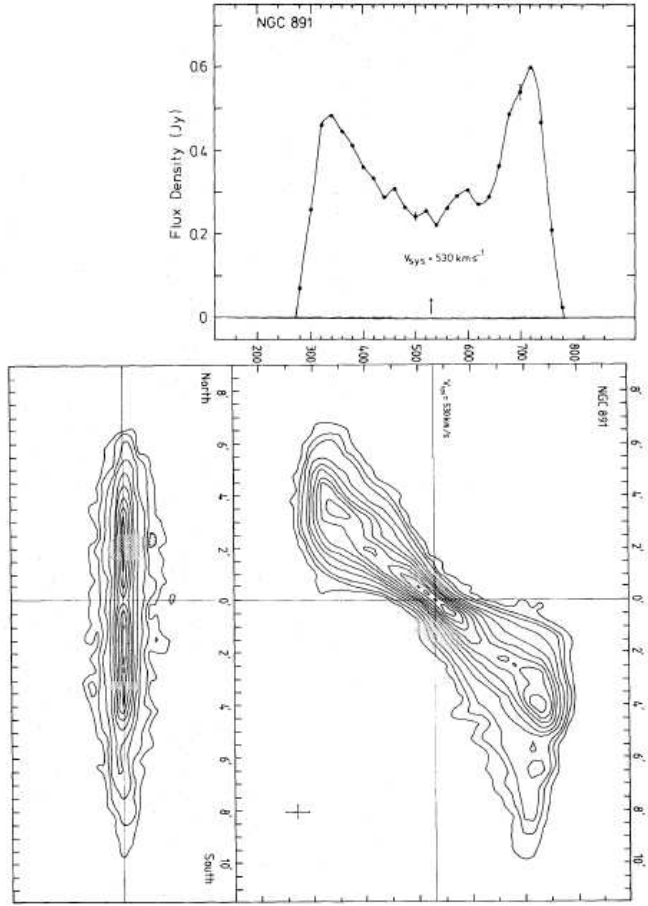} 
\caption{Asymmetric HI surface density plot of NGC 891 (contours at the bottom left);
position-velocity diagram of the same galaxy (contours at the bottom right);
and the global velocity profile in NGC 891 (spectrum at the top), from Richter \& Sancisi (1994).}
 \label{fig2}
\end{figure}
 
\bigskip

The global velocity tracer is likely to underestimate the asymmetry 
fraction as for example if the galaxy were to be viewed along the major 
axis as pointed out by Bournaud et al. (2005 b), or for a face-on galaxy 
as noted by Kamphuis (1993). 

The comparison of asymmetry in the stars as detected in the near-IR
and the HI asymmetry in surface density using the second criterion
of Richter \& Sancisi (1994) shows a similar behaviour in a sample
of 76 galaxies (Fig. 6 of Bournaud et al. 2005 b). However,  the asymmetry 
is quantitatively larger and more frequent in HI than in stars. 
This result supports the conjecture by Richter \& Sancisi (1994) that the asymmetry 
in the global velocity profiles is a good tracer of the disk mass asymmetry. 
While making this comparison, it should be noted, however, that the HI asymmetry 
is seen more in the outer radial range while the asymmetry in the near-IR is seen 
in the inner region of a galactic disk.

Haynes et al. (1998) studied the global HI profiles of 103 fairly
isolated spirals. 52 of the 78 or $\sim 75 \% $ galaxies showed
statistically significant global profile asymmetry of  1.05. Many
show large asymmetry: 35 have asymmetry $>$ 1.1, 20 have $ >$ 1.15,
and 11 have $ >$ 1.2 .

The atomic hydrogen gas is an ideal tracer for a quantitative study of 
lopsidedness in galaxies since the HI gas extends farther out than the 
stars. The typical radial extent of HI is 2-3 times that of the 
stars (e.g. Giovanelli \& Haynes 1988), and the amplitude of asymmetry 
increases with radius (Rix \& Zaritsky 1995) as seen for the stars.
However, surprisingly, a quantitative measurement of HI spatial lopsidedness has 
not been done until recently. In a first such study, the two-dimensional maps of the surface 
density of HI have been Fourier-analyzed recently
to obtain the m=1 Fourier amplitudes and phases for a sample of 18
galaxies in the Eridanus group (Angiras et al. 2006)- see Section 5.2 for
details. Such analysis needs to be done for nearby, large-angular size 
galaxies using sensitive data which will
allow a more detailed measurement of lopsidedness in nearby galaxies.
A study along these lines is now underway using the data from WHISP,
the Westerbork observations of neutral Hydrogen in Irregular
and SPiral galaxies  (Manthey et al. 2008).

The molecular hydrogen gas also shows a lopsided distribution 
in some galaxies, with more spatial extent along one half of a galaxy as in 
NGC 4565 (Richmond \& Knapp 1986), IC 342 (Sage \& Solomon 1991), NGC 628 
(Adler \& Liszt 1989) and M51 (Lord \& Young 1990).
However, this effect is not common in most cases and that can be understood 
as follows. The lopsidedness appears to be more strongly seen in the outer parts 
of a galaxy and the amplitude increases with radius as seen in stars (Rix \& 
Zaritsky 1995),and also in HI gas (Angiras et al. 2006). Theoretically it has 
been shown that the disk lopsidedness if arising due to a response to a distorted 
halo with a constant amplitude can only occur in regions beyond $\sim 2$ disk scalelengths (Jog 2000). In
 most galaxies the molecular gas in constrained to lie within two disk 
scalelengths or half the optical radius (Young 1990). 
Hence we can see that in most galaxies, there is no molecular gas in the regions 
where disk lopsidedness in stars or HI is seen. This is why the molecular gas being 
in the inner parts of the galactic disk does not display lopsidedness in most cases.

\subsubsection{Morphological lopsidedness in old stars}

The near-IR traces the emission from the old stars since dust
extinction does not significantly affect the emission in the
near-IR. The systematic study of this topic was started in the
1990's when a few individual galaxies such as NGC 2997 and NGC 1637 
were mapped in the near-IR by Block et al (1994).
They noted the m=1 asymmetry in these but did not study it
quantitatively. The pioneering quantitative work in this field was done by Rix \&
Zaritsky (1995) who measured the asymmetry in
the near-IR for a sample of 18 galaxies.  In each galaxy,
A$_1$, the fractional amplitude for the m=1 mode  normalized by an
azimuthal average (or m=0), was given at the outermost point measured in the
disk, i.e. at 2.5 exponential disk scalelengths. This distance is
set by the noise due to the sky background in the near-IR. The mean
value is 0.14 , and 30 \% have A$_1$ values more than 0.20 which were
defined by them to be lopsided galaxies. A typical example is shown in Fig. 3.

\bigskip

\begin{figure}[h]
\centering
\includegraphics[height=5.0in,width=4.0in]{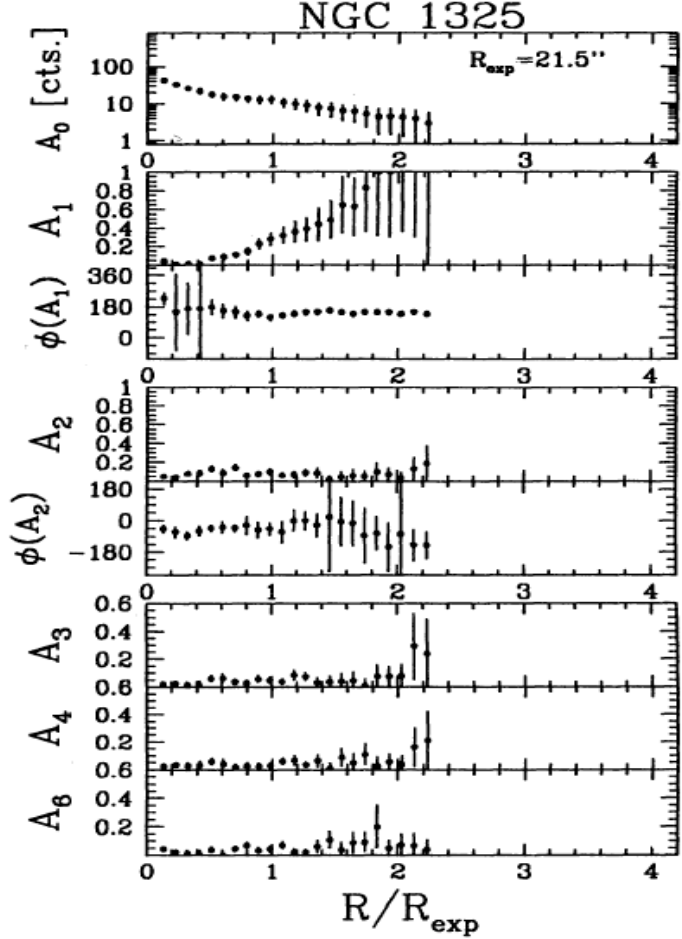}
\caption{The values of the various fractional Fourier amplitudes and phases vs. 
radius in units of the disk scalelengths for NGC 1325 (from Rix \& Zaritsky 1995). 
The amplitude scale in the lower three panels has been expanded by a factor of 5/3
for clarity, since these higher $m$ components have smaller amplitudes . The lopsided 
amplitude A$_1$ increases with radius, and the phase is 
nearly constant with radius.} 
 \label{fig3}
\end{figure}

\bigskip

This study was extended for a sample of 60 galaxies by Zaritsky \&
Rix (1997). They carried out the Fourier analysis of the near-IR surface brightness 
between the radial range of 1.5-2.5 R$_{exp}$, where  R$_{exp}$ is
the characteristic scale of the exponential light distribution.
 The normalised $m=1$ Fourier amplitude A$_1$ is  a more 
representative indicator of disk lopsidedness. 

It was shown that 30 \% of the galaxies have A$_1 > $ 0.2,
which was taken to define the threshold lopsidedness as in the
previous work. Rudnick \& Rix (1998) studied 54 early-type galaxies
(S0-Sab) in R- band and found that 20 \% show A$_1$ values measured
between the radial range of 1.5-2.5 R$_{exp}$ to be $ >$ 0.19.

A similar measurement has recently been done on a much larger sample of
149 galaxies from the OSU (Ohio State University) database in the near-IR (Bournaud et al.
2005 b), see Fig. 4. This confirms the earlier studies but for a larger sample,
and gives a smaller value of lopsided amplitude, namely
 $\sim 30 \% $ show lopsideness amplitude of
0.1 or larger when measured over the radial range of 1.5-2.5 R$_{exp}$. 
The galaxies with inclination angles of $< 70^{0}$ were used for this study.
Since this is a large, unbiased sample it can be taken
to give definitive values, and in particular the mean amplitude 
of 0.1 can be taken as the threshold value for deciding if a galaxy is lopsided. 
The lopsidedness also shows an 
increasing value with radius, as seen in the Rix \& Zaritsky (1995) study.

\bigskip

\begin {figure} [h]
 \centering
  \includegraphics[height=2.5in,width=3.0in]{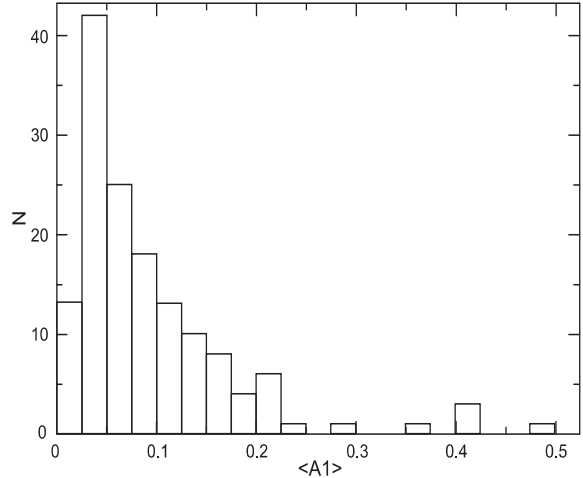}
   \caption{The histogram showing the distribution of lopsidedness measured in 149 galaxies at inclination of $<70^{0}$ from the OSU sample (Bournaud et al. 2005 b). The typical normalized lopsided amplitude A$_1$ measured over the radial range between 1.5 to 2.5 disk scalelengths is $\sim 0.1$. Thus most spiral galaxies show significant lopsidedness.}
 \label{fig4} 
 \end{figure}

\bigskip

In the Fourier decomposition studies the determination of the center is a tricky issue, and the same center has to be used
for all the subsequent annular radial bins, otherwise during the optimization procedure, a different center could
get chosen and will give a null measurement for the lopsidedness. This is applicable for the lopsidedness
analysis for HI (Angiras et al. 2006, 2007), and also for centers of mergers (Jog \& Maybhate 2006).  These two cases are discussed respectively in Sections 5.1 and 5.3.

The large number of galaxies used allows for a study of the variation with 
type. It is found that late-type galaxies are more prone to lopsidedness, 
in that a higher fraction of late-type galaxies are lopsided, and they show a 
higher value of the amplitude of lopsidedness (Bournaud et al. 2005 b), see Fig. 5. This is similar to what was 
found earlier for the variation with galaxy type for the HI asymmetries 
(Matthews, van Driel, \& Gallagher 1998). These samples largely consist of 
 field galaxies, while the group galaxies show a reverse trend with galaxy 
type (Angiras et al. 2006,  2007) implying a different mechanism for the origin 
of lopsidedness in these two settings, see Section 5.1 .

\bigskip

\begin{figure} [h]
\centering
 \includegraphics[height=2.5in,width=3.0in]{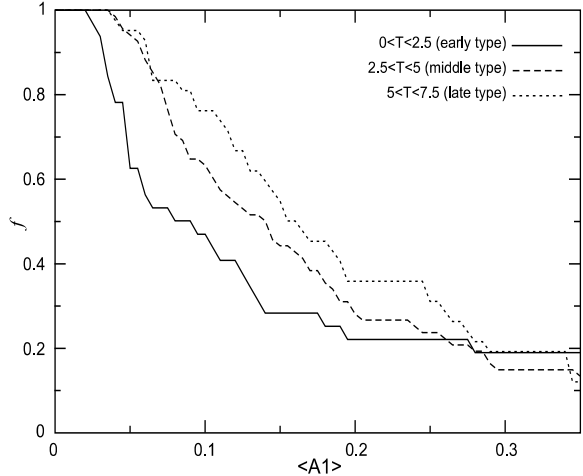} 
 \caption{The plot of the cumulative function of $<A_1>$ for three groups of Hubble types of spiral 
galaxies: the early-types ($0 < T < 2.5$), the intermediate types ($2.5 < T < 5$ ), and the late-types 
($5 < T < 7.5 $), where $T$ denoted the Hubble type of a galaxy (taken from Bournaud et al. 2005 b). The late-type galaxies are more lopsided than the early-type galaxies.}
 \label{fig5}
\end{figure}

\bigskip

While an axisymmetric galaxy disk gives rise only to
radial gravity forces, and therefore no torque, any
asymmetry in the disk, either $m=1$, $m=2$ or higher,
gives rise to tangential forces in addition
to radial forces, and then to a gravity torque.
  From the near-infrared images, representative
of old stars, and thus of most of the mass, it is possible to compute
the gravitational forces experienced in a galactic disk.
The computation of the gravitational torque presents important
complementary information, namely it gives the overall average
strength of the perturbation potential as shown for m=2 (Block et
al. 2002), and for m=1 (Bournaud et al. 2005 b), whereas the Fourier amplitudes give values 
which are weighted more locally. The gravitational potential is derived from the 
near-infrared image, see Bournaud et al. (2005 b) for the details of this method.
The m=1 component of the gravitational torque, Q$_1$ between 1.5-2.5 disk scalelengths 
is obtained, the histogram showing its distribution is plotted in Fig. 6 , which is similar to that 
for the lopsided amplitude $A_1$ (Fig. 4) as expected. 

\bigskip

\begin{figure}[h]
\centering
 \includegraphics[height=2.5in,width=3.0in]{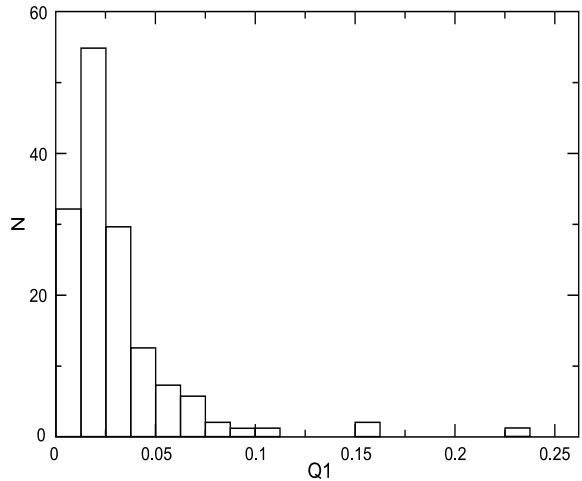} 
 \caption{The histogram of Q$_1$, the m=1 Fourier amplitude in the gravitational potential, 
 for the OSU sample of galaxies, from Bournaud et al. 2005 b.}
 \label{fig6}
\end{figure}

\bigskip

An even larger sample based on the SDSS
data has now been Fourier-analyzed by Reichard et al. (2008),
and they also get a similar average value of lopsidedness in spiral galaxies, see Fig. 7.
However, they use the surface density data between 50\% and 90\% light radii, 
so a clear one-to-one quantitative comparison of the values of lopsidedness
from this work as compared to the previous papers in the literature discussed above is not possible.
This work confirms that galaxies with low concentration, and low stellar mass density (or the late-type spirals)
are more likely to show lopsidedness, as shown earlier for HI by Matthews et al. (1998) and for stars by Bournaud et al. (2005 b).

\bigskip

\begin{figure}[h]
\centering
\includegraphics[height=2.5in,width=3.0in]{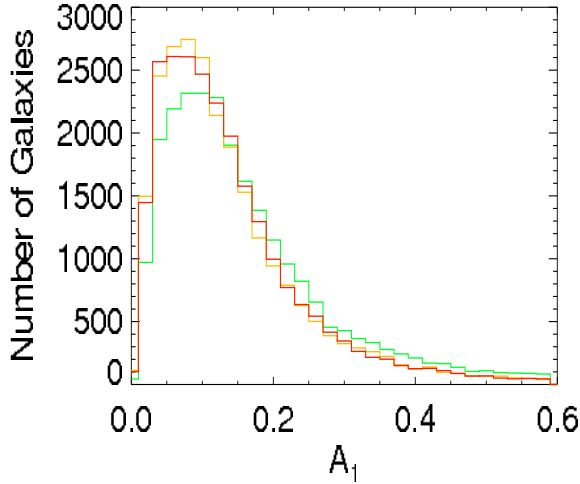}
\caption{The histogram of number of galaxies vs. A$_1$ values for the SDSS sample, from Reichard et al. (2008). The histogram gives similar values to the earlier studies by Rix \& Zaritsky (1995) and Bournaud et al. (2005 b)}
 \label{fig7}
\end{figure}

\bigskip

Another approach to measure the asymmetry involves the wedge method
(Kornreich et al. 1998, 2002) where the flux within
circular sectors or wedges arranged symmetrically with respect to the
galactic disk centre are compared. While it is easier to measure this
than the Fourier amplitudes, it gives only average values. An extreme
limit of the wedge method is applied by Abraham et al. (1996) and
Conselice et al. (2000). They define the rotation asymmetry
parameter as the ratio of fractional difference between the two
sides, so that 0 corresponds to a symmetric galaxy and 0.5 for the
most asymmetric galaxy. This is a more global or average definition
of the disk asymmetry and is suitable for studying highly disturbed
galaxies, and not those showing a systematic variation as in a
lopsided distribution. Such highly disturbed systems are seen at high
redshifts, for which this criterion was used first. 

An interesting point to note is that spurious signs of asymmetry
 arising due to dust extinction (Rudnick \& Rix 1998)
and that arising due to the pointing error of single-dish telescope
(Richter \& Sancisi 1994) were checked and ruled out. Conversely, a galaxy 
could look more symmetric in the global velocity profile than it is, if the 
galaxy is seen face-on. In that case even though the morphology is asymmetric 
-as in HI in NGC 628, the global velocity profile is symmetric, and hence the 
galaxy would appear to be kinematically symmetric- see Kamphuis (1993).

Based on the above discussion of the various methods, we recommend that the 
future users adopt the fractional 
Fourier amplitude A$_1$ as the standard criterion for lopsidedness. This is because it
gives a quantitative measurement, is well-defined, and can be measured easily 
as a function of radius in a galaxy, and thus allows an easy comparison of its value 
between galaxies and at different radii. The threshold value that could be adopted could be 
the average value of 0.1 seen in the field galaxies in the intermediate radial
range of 1.5-2.5 R$_{exp}$ (Bournaud et al. 2005 b), so that 
galaxies showing a higher value can be taken to be lopsided. A uniform criterion
will enable the comparison of amplitudes of lopsidedness in different galaxies, 
and also allow a comparison of the fraction of galaxies deduced to be lopsided in different studies.

\subsection{Kinematical lopsidedness}

\begin{figure}[h]
\centering
\includegraphics[height=1.8in,width=4.0in]{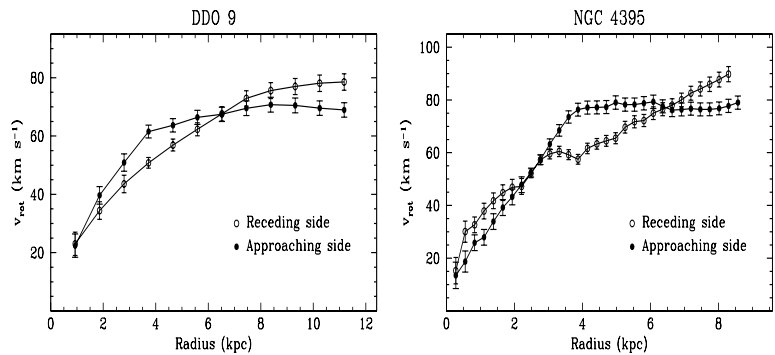}
\caption{The rotation curve in DDO 9 and in NGC 4395 is asymmetric on two sides of 
the galaxy, from Swaters et al. (1999).} 
 \label{fig8}
\end{figure}

The lopsidedness or a (cos $\phi$) asymmetry is often also observed
in the kinematics of the galaxies. This could be as obvious as the
asymmetry in the rotation curves on the two halves of a galactic
disk, as is shown in Fig \ref{fig8} (Swaters et al. 1999, Sofue \& Rubin 2001), or more subtle as
in the asymmetry in the velocity fields (Schoenmakers et al. 1997). Often the optical centres are distinctly separated
spatially from the kinematical centers as in M33, M 31, and
especially in dwarf galaxies as pointed out by Miller \& Smith
(1992). The rotation curve asymmetry is also seen 
as traced in the optical for stars (Sofue \& Rubin 2001). The
detailed 2-D velocity fields were so far mainly observed for HI as
in the interferometric data (see e.g. Schoenmakers et al. 1997, Haynes
et al. 1998). Now such information is beginning to be
available for the bright stellar tracers as in H$_\alpha$ emission
from HII regions (Chemin et al. 2006, Andersen et al. 2006), however since the filling factor
of this hot, ionized gas is small, it is not an ideal tracer for a 
quantitative study of disk lopsidedness.

Schoenmakers et al. (1997) use the kinematical observational data in
HI on two galaxies- NGC 2403 and NGC 3198, and deduce the upper
limit on the asymmetry in the m=2 potential to be $<$ a few percents.
However, this method gives the result up to the sine of the viewing
angle. Kinematic asymmetry in individual galaxies such as NGC 7479
has been studied and modeled as a merger with a small satellite
galaxy (Laine \& Heller 1999).

The rotation curve is asymmetric in the  two halves of a galaxy or
on the two sides of the major axis as shown for DDO 9 and NGC 4395 by
Swaters et al. (1999), see Figure 8. However, they do not make a more detailed
quantitative measurement of the asymmetry. Swaters (1999) in his study
of dwarf galaxies showed that 50\% of galaxies studied show
lopsidedness in their kinematics. Schoenmakers (2000) applied his calculations on kinematical 
lopsidedness in galactic disks to five galaxies
in the Sculptor group and found that all five show significant
lopsidedness. A similar result has been found for 
the 18 galaxies studied in the Ursa Major cluster (Angiras et al. 2007).
The frequency of asymmetry and its magnitude is higher in galaxies in
groups - see section 5.2 for details.

\bigskip

\bigskip

A galaxy which shows spatial asymmetry would naturally show
kinematical asymmetry (e.g., Jog 1997) except in the rare cases of 
face-on galaxies as discussed above where the galaxy can show asymmetry in the morphology 
but not in the kinematics. However, the papers which study lopsidedness do not
always mention it. On the contrary, in the past, several papers
have made a distinction between the spatial or morphological
lopsidedness and kinematical lopsidedness (e.g. Swaters et al. 1999,
Noordermeer, Sparke \& Levine 2001) and have even claimed (Kornreich et al.
2002) that the velocity asymmetry is not always correlated with the
spatial asymmetry. However, in contrast, it has been argued that the
two have to be causally connected in most cases (Jog 2002),
especially if the lopsidedness arises due to the disk response to a
tidal encounter. 

An important point to remember is that the same tracer (stars or HI)
should be considered to see if a galaxy showing spatial lopsidedness
is also kinematically lopsided or not, and vice versa. This is
because the HI asymmetry is higher and is seen in the outer parts of the
galaxy while the asymmetry in the near-IR is more concentrated in
the inner regions. This criterion is not always followed (see e.g., 
Kornreich et al 2002). Thus the often-seen 
assertion in the literature that the spatial asymmetry is not correlated with kinematic
asymmetry is not meaningful, when the authors compare the spatial
asymmetry in the optical with the kinematical asymmetry in the HI.

\subsection {Phase of the disk lopsidedness}

\bigskip

The phase of the lopsided distribution provides an important clue to its
physical origin, but surprisingly this has not been noted or used 
much in the literature. Interestingly, the phase is nearly constant with
radius in the data of Rix \& Zaritsky (1995), as noted by Jog
(1997). This is also confirmed in the study of a larger sample
of 60 galaxies by Zaritsky \& Rix (1997), (Zaritsky 2005, personal
communication), and also for the sample of 54 early-type galaxies studied
by Rudnick \& Rix (1998). A nearly constant phase with radius was
later confirmed for a larger sample of 149 mostly field galaxies
(Bournaud et al. 2005 b), and also for the 18 galaxies in the Eridanus
group (Angiras et al. 2006). The latter case is illustrated in Fig. 9. 
This points to the lopsidedness as a global $m=1$ mode, and this idea 
has been used as a starting point to develop a theoretical model 
(Saha et al. 2007).
There are a few galaxies which do show a systematic radial variation in phase, 
as in M51 (Rix \& Rieke 1993), which therefore appear as one-armed spirals.

\bigskip

\begin {figure} [h]
 \centering
\includegraphics[height=1.25in,width=2.5in]{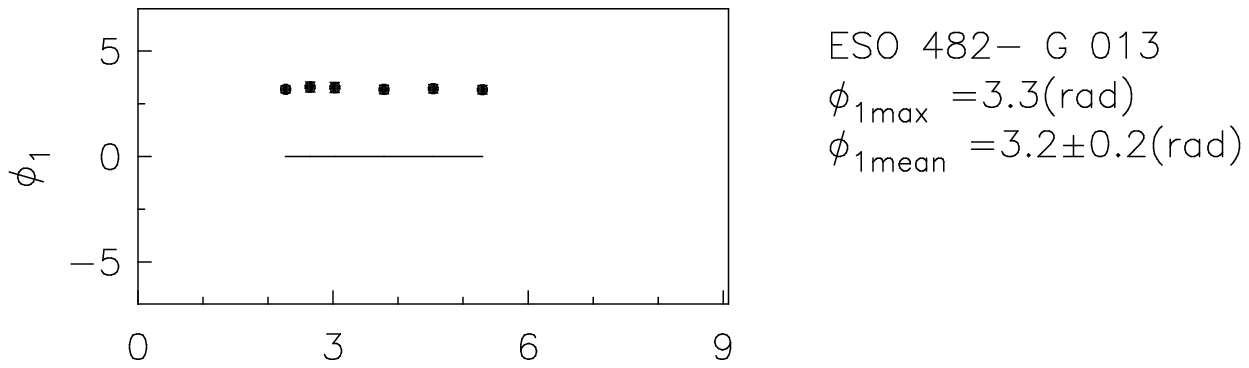}
\medskip 
\includegraphics[height=1.25in,width=2.5in]{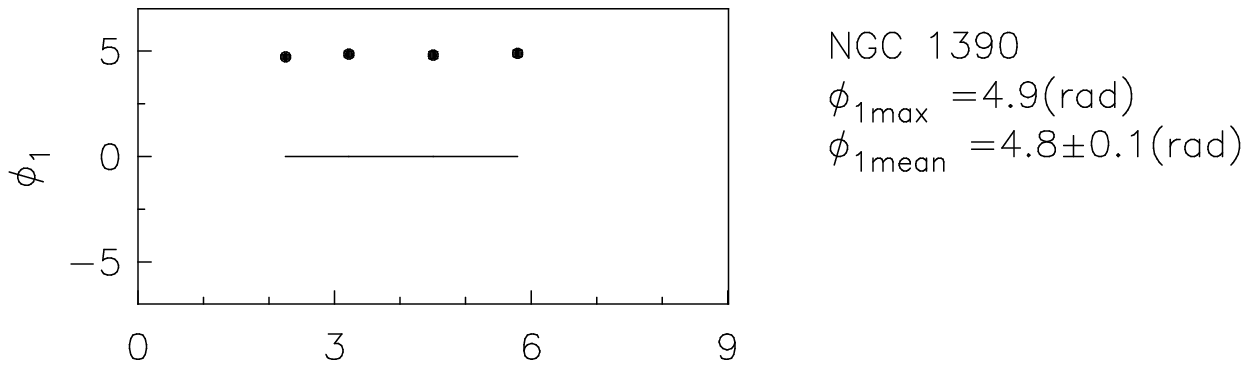} 
\bigskip
   \caption{The plot of the phase of the m=1 Fourier component vs. 
radius (given in terms of the disk scalelength) for the HI surface density 
for two galaxies ESO 482 - G 013 and NGC 1390 in the Eridanus group of galaxies, from Angiras et al. 
(2006). Note that the phase is nearly constant with radius indicating 
a global lopsided mode.}
\label{fig9}
  \end{figure}

\bigskip

In contrast, the central regions of mergers of galaxies, show highly fluctuating 
phase for the central lopsidedness (Jog \& Maybhate 2006). This may indicate an unrelaxed state, 
which is not surprising 
given that the mergers represent very different systems than the individual spirals
mainly discussed here.

\bigskip

\subsection{Observations of off-centered nuclear disks}

 A certain number of galaxies are observed to have an off-centered nuclear disk, and 
more generally an $m=1$ perturbation affecting more particularly the nuclear region.
 Our own Galaxy is a good example, since the molecular gas observations have
revealed that the molecular nuclear disk has three quarters of its mass at positive
longitude, which is obvious in the central position-velocity diagram
(the parallelogram, from Bally et al 1988). 
The asymmetry appears to be mainly a gas phenomenon, since the off-centreing
is not obvious in the near-infrared images (e.g. Alard 2001, Rodriguez-Fernandez \& Combes
2008). The gas is not only off-centered but also located in an inclined
and warped plane (Liszt 2006, Marshall et al 2008).
An $m=1$ perturbation is superposed on the
$m=2$ bar instability. The most nearby giant spiral galaxy, M31, has also revealed 
an $m=1$ perturbed nuclear disk in its stellar distribution (Lauer et al 1993, Bacon
et al 1994). The spatial amplitude of the perturbation is quite small, a few parsecs,
and this suggests that this nuclear lopsidedness could be quite frequent in galaxies. However,
 it is difficult to perceive it due to a lack of resolution in more distant objects.
 Since M31 is the prototype of the $m=1$ nuclear disk, we will describe it in detail
in the next section. Some other examples have been detected, like NGC 4486B in the Virgo
cluster (Lauer et al 1996), but the pertubation must then be much more extended,
and that phenomenon is rare.

\bigskip

\begin{figure}[h]
  \centering
  \includegraphics[width=8cm]{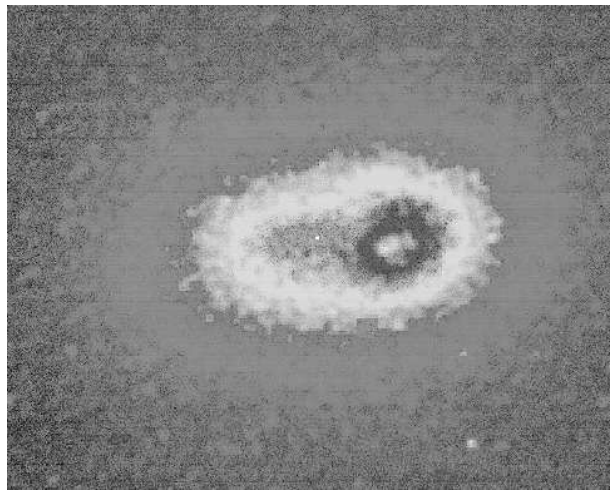}
   \caption{HST WFPC2 V-band image of M31. The surface brightness contributed by the UV cluster 
coinciding with the component P2 has been clipped out. The white dot indicates the position of the black hole.
>From Kormendy \& Bender (1999).}
 \label{fig10}
  \end{figure}

\bigskip

\subsubsection{ The case of the M31 nuclear disk } 

The first images of M31 to reveal the asymmetrical nucleus were the
photographs at 0.2" resolution from the Strastoscope II by Light et al (1974). 
They first resolved the nucleus, and measured a core radius of 0.48" (1.8pc).
The total size of the nucleus is 4 arcsec (15pc).
They showed that the nucleus is elongated, with a low intensity extension 
outside the bright peak (cf Fig. 10); and they 
 considered the possibility of a dust lane
to mask the true center. Nieto et al (1986) confirmed this morphology
in the near-UV and also evoked dust. Later, it was clear that 
dust could not be the explanation of this peculiar morphology,
since the center was still offset from the bulge in the near-infrared
image (Mould et al 1989). As for the kinematics, it was already 
remarked by Lallemand et al. (1960) that the nucleus is rotating rapidly,
showing a very compact velocity curve, falling back to zero at a radius
of 2 arcsec. This was confirmed by Kormendy (1988) and Dressler \& Richstone
(1988), who concluded to the existence of a black hole in the center of M31,
of $\sim$ 10$^7$ M$_\odot$, with the assumption of spherical symmetry. 
Lauer et al (1993, 1998) revealed with HST that the asymmetrical 
nucleus can be split into two components, like a double nucleus,
with a bright peak (P1) offset by $\sim$ 0.5" from a secondary 
fainter peak (P2), nearly coinciding with the
bulge photometric centre, and the proposed location of the black hole
(e.g. Kormendy \& Bender 1999). It is well established now
from HST images from the far-UV to near-IR (King et al. 1995, Davidge et al. 1997)
that P1 has the same stellar population as the rest of the
nucleus, and that a nearly point-like
source produces a UV excess close to P2 (King et al. 1995 ).

2-D spectroscopy by Bacon et al (1994) revealed
that the stellar velocity field is roughly centred on P2,
but the peak in the velocity dispersion map is offset by
$\sim$ 0.7" on the anti-P1 side (Fig. 11). With HST spectroscopy
the velocity dispersion peak reaches a value of $440\pm70$~km s$^{-1}$.
and the rotational velocity has a strong gradient (Statler et al. 1999). 
The black hole mass required to explain these observations 
ranges from 3 to $10\, \times 10^7$~M$_\odot$.
The position of the black hole is assumed to coincide with the centre of the UV peak,
near P2, and possibly with the hard X-ray emission detected by the Chandra satellite
(Garcia et al. 2000). 

\bigskip

\begin{figure}[h]
  \centering
  \includegraphics[height=8cm,width=6cm]{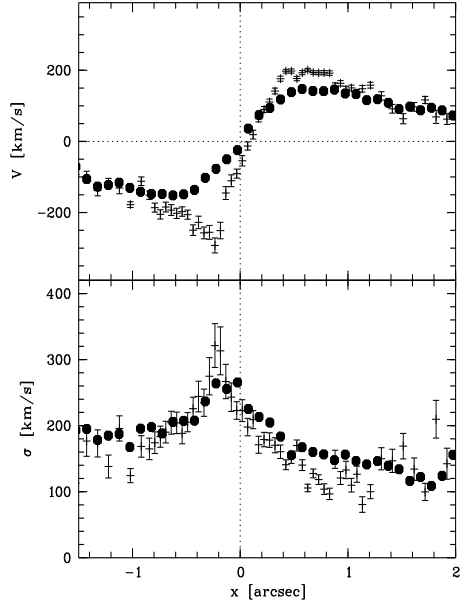}
   \caption{Velocity profile ({\bf top}) and velocity dispersion ({\bf bottom})
in the nucleus of M31. The crosses are from STIS (HST) and the filled circles
from OASIS (CFHT). The OASIS kinematics  have been averaged over a 
0.2'' wide slit (PA$=39^\circ$) - taken from Bacon et al. (2001).}
 \label{fig11}
\end{figure}

\bigskip

\subsubsection{Other Off-centered nuclei } 

\bigskip

It has been known from a long time that the nearby late-type spiral M33
has a nucleus displaced from the young population mass centroid,
by as much as 500pc (de Vaucouleurs \& Freeman 1970, Colin \& Athanassoula 1981).
This off-centreing is also associated with a more large-scale lopsidedness,
and can be explained kinematically by a displacement of the bulge with respect
to the disk. Such kind of off-centreing is a basic and characteristic property
of late-type Magellanic galaxies.
In NGC 2110, an active elliptical galaxy, Wilson \& Baldwin (1985) noticed a displacement of
the nucleus with respect to the mass center of 220pc, both in the light and kinematics.
 Many other active nuclei in elliptical galaxies
have been reported off-centered, but the presence of dust obscuration makes
its reality difficult to assert (e.g. Gonzalez-Serrano \& Carballo 2000,
where 9 galaxies out of a sample of 72 ellipticals are off-centered).

 Quite clear is the case of the double nucleus in the barred spiral M83 (Thatte et al
2000): near-infrared imaging and spectroscopy reveals, in spite of the high extinction,
 that the nucleus is displaced by 65pc from the barycenter of
the galaxy, or that there are two independent nuclei. Molecular gas with high velocity is associated with the
visible off-center nucleus, and this could be the remnant of a small
galaxy accreted by M83 (Sakamoto et al 2004).
In some cases what appeared to be a double nucleus could in fact be two regions of 
star formation in centers of mergers of galaxies as in Arp 220 (Downes \& Solomon 1998).

Recently, Lauer et al (2005) studied a sample of 77 early-type galaxies with HST/WFPC2 resolution,
and concluded that all galaxies with inner power-law profiles have nuclear disks,
which is not the case of galaxies with cores.  
They found 2 galaxies with central minima, likely to have a double nucleus (cf Lauer et al 2002), 
and 5 galaxies having an off-centered nucleus. This perturbation also appears
as a strong feature in the Fourier analysis (A$_1$ term).

Off-centering is also frequently observed in central kinematics, where the peak of the
velocity dispersion is displaced with respect to the light center
(Emsellem et al 2004, Batcheldor et al 2005). Decoupled nuclear disks, and 
off-centered kinematics are now clearly revealed by 2D spectroscopy.

\section{Theoretical models for the origin of
lopsidedness:}

The origin and the evolution of lopsidedness are not yet
well-understood and in fact not received much theoretical attention.
The lopsidedness is observed in a variety of tracers and settings. It is observed in old and young stars, and in HI and
H$_2$ gas, within the optical disk and far outside the optical disk
as seen in the tracer HI gas, and in field galaxies and in galaxies in groups.
Given all these parameters it is difficult to come up with a unique
physical mechanism for the generation or the maintenance of disk
lopsidedness. Indeed there could be multiple paths for it, such as
tidal interactions, gas accretion, or an internal instability. The
particular mechanism that dominates depends on the situation
concerned as can be seen from the following discussion. Towards the end of
this section (section 3.4), we give a summary of the most likely physical
mechanisms for the origin of disk lopsidedness in spiral galaxies.

We also stress that the standard m=2 case has been extensively
studied for years both analytically as well as by N-body
simulations, both in the context of central bars (e.g. Binney \&
Tremaine 1987, Combes 2008, Shlosman 2005) or as
two-armed spiral features (Rohlfs 1977, Toomre 1981). Thus the basic
physics of their growth and dynamics is fairly well-understood. In
contrast, the m=1 feature has just begun to be studied. We discuss
the physical differences between these two cases (m= 1 and 2), and 
later in Section 6 a comparison between their observed values is made.

\subsection{Kinematical model for the origin of
lopsidedness}

The simplest way to explain the observed lopsided disk distribution
is to start with a set of aligned orbits and see how long these will
take to get wound up, as was done by Baldwin et al. (1980), see Fig. 12. It is
well-known that the differential rotation in a galactic disk would
tend to smear any material feature over a few dynamical timescales.
Baldwin et al. (1980) showed that on taking account of the epicyclic
motion of the stars, the effective radial range over which the
differential motion affects the winding up is reduced by nearly a
factor of 2, and this helps in increasing the lifetime of the
feature. The net winding up time t$_{winding}$, for a feature
between two radii R$_1$ and R$_2$
is then given by t$_{winding}$ = 2 $\pi / \Delta (\Omega - \kappa)$,
where $\Omega$ and $\kappa$ are the angular speed of rotation and the 
epicyclic frequency respectively, and $\Delta$ denotes the difference 
taken at these two radii.
For a flat rotation curve with $\kappa = 1.414 \Omega$, this
gives t$_{winding}$ = 5 [2 $\pi / \Delta \Omega] $ , which is about
5 times longer than the usual winding up time for material arms.

\bigskip

\begin {figure} [h]
 \centering
\includegraphics[angle=-90,width=.45\textwidth]{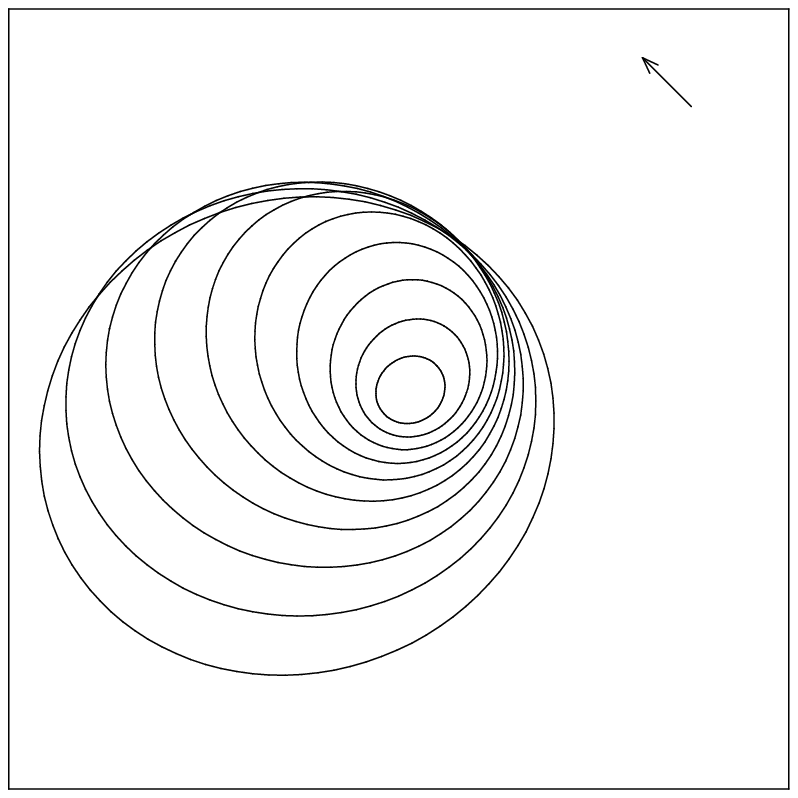}
\includegraphics[angle=-90,width=.5\textwidth]{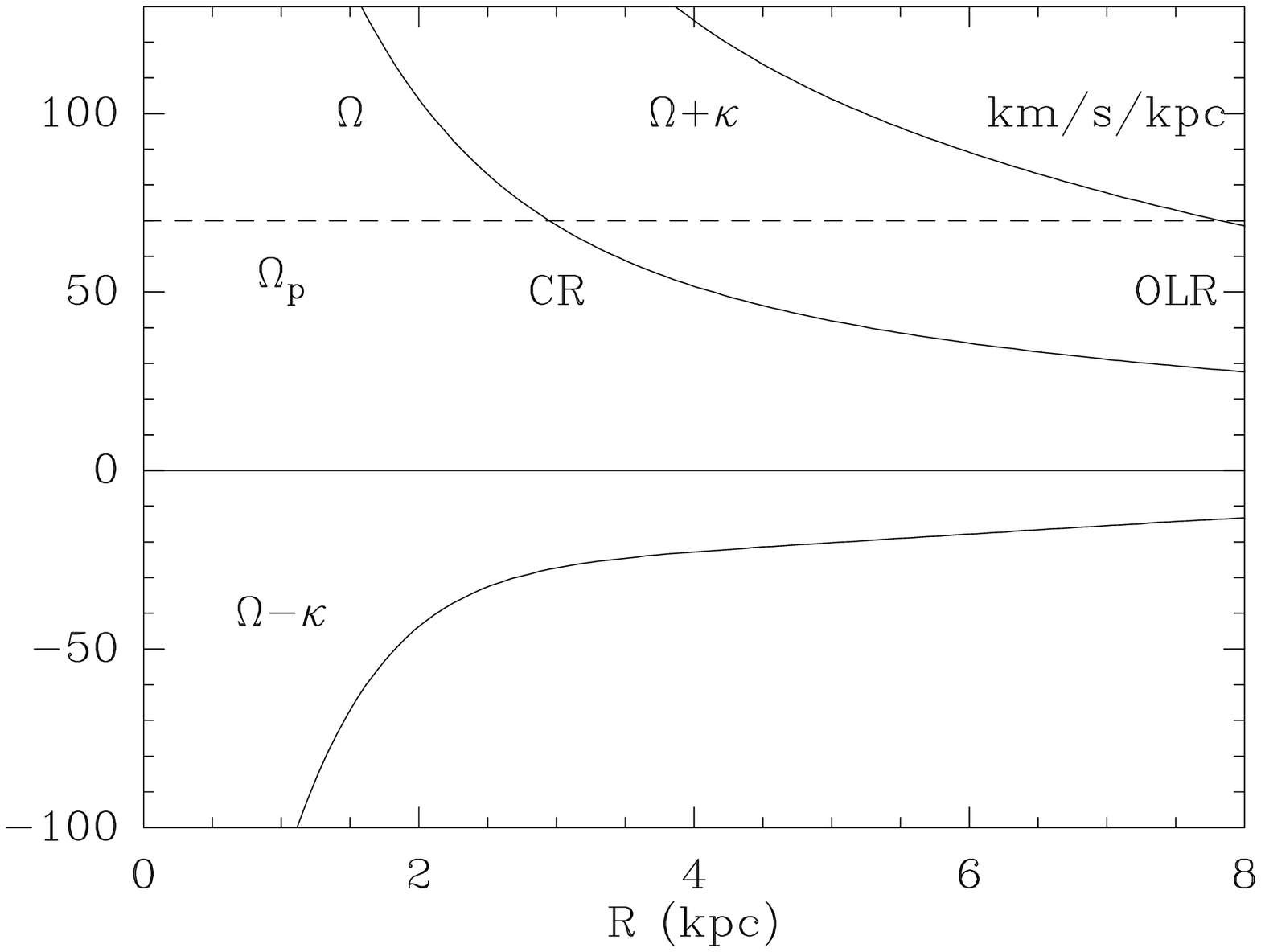}
\bigskip
\caption{{\it Left}: Pattern of lopsided elliptical orbits whose major
axes were aligned to begin,
following Baldwin et al. (1980).
The orbits tend to get misaligned with time due to the differential
rotation in the disk.
{\it Right}: Frequencies $\Omega$, $\Omega - \kappa$ and $\Omega + \kappa$
in a typical galaxy disk. In the case of a prograde
mode, a possible pattern speed $\Omega_p$ is indicated,
allowing Corotation (CR) and Outer Lindblad Resonance (OLR)  (taken from Combes 2000).
}
 \label{fig12}
\end{figure}

\bigskip

\bigskip

For the outer parts of a galaxy say at $\sim$ 15 kpc , this is
$\sim$ 2 Gyr, which is a few times larger than the local dynamical
timescale but is still less that the lifetime of the Galaxy. Hence
they argued that the lopsidedness cannot be primordial, and must be
generated repeatedly in the disk. Further, this model cannot explain why
 many isolated galaxies such as M101 show lopsidedness, which is
the puzzle they had started out to solve.

Further work on the kinematical origin of lopsidedness due to
 a cooperation of orbital streams of stars near resonance was
done by Earn \& Lynden-Bell (1996), with similar results for its
lifetime as discussed above. Apart from the short winding time, a
generic problem with the kinematical model is that it is not clear
what gives rise to the aligned orbits in the first place.

\subsection{Dynamical models for the origin of
lopsidedness}

The set of dynamical models discussed next deal with a more physical
origin for the lopsidedness. The most commonly proposed models
include tidal encounters, gas accretion, instability in a
counter-rotating disk, an off-centered disk in the halo, and ram
pressure stripping in a cluster. Of these the most promising are the
first two, with the tidal encounters being the dominant mechanism
for group galaxies.
In addition to these externally triggered processes, the disk
lopsidedness could also arise as a global m=1
instability in self-gravitating disks.

\subsubsection{Tidal encounters, and disk response to distorted halo}

One of the earliest ideas suggested for
generating lopsidedness in a galaxy was due to a tidal encounter as
applied to M101 (Beale \& Davies 1969), although no details were
worked out. It is easy to see that a perturbation as say due to a
tidal encounter between two galaxies with an arbitrary orientation
can generate a force term of
type cos $\phi$, which can then generate lopsidedness in the galaxy 
(Combes et al. 2004, see Chapter 7.1.1).
 The other modes will also be
generated but generally m=1 and 2 are observed to be the strongest 
(Rix \& Zaritsky 1995). Also see Section 6 for a discussion the 
observed relative strengths of the $m$ = 1 and 2 modes.

In addition to the above direct triggering of lopsidedness as a disk
response to the tidal force, it can also be generated more
indirectly due to the response of the disk to the distorted halo
which feels a stronger effect of the interaction (Weinberg 1995, Jog
1997, Schoenmakers et al. 1997). The details of lopsidedness thus
induced will be summarised in this section. The effect can be seen as 
in the spatial or surface density distribution as well as in the kinematics.

A generally stronger perturbation resulting from the infall of a
satellite galaxy can also result in the disk lopsidedness as shown 
in the N-body simulation study  by Walker et al. (1996). Zaritsky \& Rix
(1997) further used this idea to constrain the rate of infall of
satellites onto a galaxy from the fraction of galaxies showing
lopsidedness and the star formation, both triggered by
the satellite infall. Note that this gives an upper limit on the
satellite infall rate if there are other mechanisms which
also give rise to the disk lopsidedness; such as gas accretion as 
discussed in Section 3.2.2.

Since the dark matter halo is more extended than the disk, during a
tidal encounter the halo is expected to experience a stronger tidal
perturbation on general grounds. Weinberg (1995, 1998) has studied
this case numerically for the specific case of the interaction
between the LMC and the Galaxy. The live halo shows a strong
lopsided response at the resonance points as set by the orbit of the
LMC with respect to the Galaxy. The disk response to
this distorted halo is shown to be much stronger than the direct lopsidedness
triggered in it due to the encounter. A detailed numerical
evolution of a galactic disk and the halo perturbed by an impulsive
perturbation (Kornreich et al. 1998) shows that m=1 can grow as a
free sloshing libration but can only last for a dynamical timescale.
On the other hand, a number of encounters covering a large parameter 
space for the galaxy mass ratios, orbits, disk inclination were studied 
by numerical simulations by Bournaud et al. (2005 b). They conclude that the 
resulting lopsidedness can have 
an amplitude as high as the typical observed value $\sim 10 \% $, and it lasts 
for more than ten dynamical timescales $\sim$ 2 Gyr, after which the lopsidedness drops rapidly.

\bigskip

\noindent {\it 3.2.1.a $\:$ Orbits and isophotes in a perturbed
disk}

\bigskip

We next briefly summarise the results for the orbits, isophotes and
kinematics for a disk that is perturbed by a linear lopsided
perturbation potential, as say due to a
distorted halo (Rix \& Zaritsky (1995), 
Jog (1997, 2000), Schoenmakers et al (1997)) where the
halo distortion could be ascribed to a tidal encounter. These
equations of motion are general and are applicable irrespective
of the mechanism giving rise to the external potential to which the
disk responds. This simple model allows one to draw general
conclusions about the strength of the perturbation potential by a
comparison of results with observations as is shown next.

The details of this approach (Jog 2000) are summarised below. The
dynamics of particles on closed orbits in an axisymmetric disk
perturbed by a lopsided halo potential is treated. The cylindrical
coordinate system ($R , \phi $) in the galactic disk plane is used.

The unperturbed, axisymmetric potential in the disk plane,
$\psi_0(R)$, and the perturbation potential, $\psi_{lop} (R) $ are
defined respectively as:

$$\psi_0 (R) \: = \: {V_c}^2 ln R  \eqno(1)$$

$$\psi_{lop} (R) \: = \: {V_c}^2 \: \epsilon_{lop} \: cos \phi \eqno(2)$$

\noindent where $\psi_{0} (R) $ represents the typical region of flat
rotation, with $V_c$ being the constant rotational velocity, as seen
in a typical spiral galaxy. This is perturbed by a small, constant,
non-rotating, perturbation potential with a lopsided form as given
by $\psi_{lop} (R)$. Here $\epsilon_{lop}$ is a small perturbation
parameter, which is taken to be constant with radius for simplicity. That is,
 a halo with a constant lopsided distortion is assumed.
This is a simple model but it still allows one to
understand the resulting orbits, isophotes and the kinematics in a lopsided galaxy.

Consider a circular orbit at $R_0$. The coupled equations of motion
for the perturbed quantities $\delta R$ and $\delta \phi$ are solved
together using the first-order epicyclic theory. The resulting
solutions for the perturbed motion are:

$$R \: = \: R_0 \: ( 1 \: - 2 \:
   \epsilon_{lop} \: cos \phi) \: \: \: \: ; \: \: \: \: \: \: \: V_R \:
   = \: 2
   \: V_c \: \: \epsilon_{lop} \: sin \phi \: \: \: \: ; \:
  \: \: \: \: \: \: V_{\phi}
   \: = V_c \: ( 1 +  3 \: \epsilon_{lop} \: cos \phi )  \eqno(3)$$

\noindent Thus, an orbit is elongated along $\phi = 180^0$, that is
along the minimum of the lopsided potential, and it is shortened
along the opposite direction. This result was also noted by Earn \&
Lynden-Bell (1996).

The loop orbits discussed here are not valid for radii
much smaller than the disk scalelength (Rix \& Zaritsky 1995), but
this does not affect the applicability of this analysis to galactic
disks because the disk lopsidedness is typically observed only
at radii beyond 1.5 disk scalelengths.

The other signatures of the effect of the lopsided perturbation
potential are its effect on the isophotes and the kinematics in the disk, discussed next.
Since the imaging observations give information on the isophotes
rather than orbits, it is necessary to also obtain the isophotal
shapes in an exponential galactic disk in a lopsided potential. For
an exponential disk, the above approach gives:

 $$ A_1 \: = \:  \frac {\epsilon_{iso}}{2} \:
          \frac {R}{R_{exp}}  \eqno (4) $$

\noindent where $\epsilon_{iso}$ is the ellipticity of the isophote
at $R$, and $R_{exp}$ is the exponential disk scale length. Note
that here the radii measuring the minimum and maximum extents of an
isophote are along the same axis - unlike in the standard
definition of ellipticity where these two are along directions that
are normal to each other. {\it The resulting isophotes have an
egg-shaped oval appearance}, as observed say in M 101. Thus, the
azimuthal asymmetry in the surface density or the fractional Fourier amplitude for m=1, as denoted by $A_1$
manifests itself as an elongation of an isophote, and both represent
the same underlying phenomenon.

The effective surface density in a self-gravitating, exponential galactic disk may be 
written as:

$$ \mu (R, \phi) \: = \: \mu_0 \: exp [ - \frac {R}{R_{exp}} ( 1- \frac{\epsilon_{iso}}{2} 
\: cos \phi ) ]  \eqno (5) $$

For a particular isophote, the term in the square bracket is a constant and hence this formally
defines the parametric form of an isophote. Thus the minimum radius of an isophote occurs along
$\phi = 180^{0}$ while the maximum occurs along $\phi = 0^0$; while the opposite is true for
an individual orbit. Thus the isophotes are elongated in a direction opposite to an orbit, and the elongation
is along the same direction where the maximum effective surface density occurs-
this will be true for any self-gravitating system (Jog 1997).

To obtain the lopsided potential in terms of the observed lopsided Fourier amplitude,
 the equations of perturbed motion (eq. [3])have to be solved with
 the equation of continuity, and the effective surface density (eq. [5]]). Now, the equation
of continuity is given by:

$$ \frac{\partial}{ \partial R} \: [R \mu (R, \phi) \: V_R (\phi) ] \:
 + \: \frac {\partial}{\partial \phi} [\mu (R, \phi) \: V_{\phi} (\phi) ] = 0 \eqno (6) $$

Solving these, and combining with eq.[4], yields the
following important results (valid for $ R \geq R_{exp}$):

$$ \epsilon_{lop} \: = \:  \frac {A_1 } {\left (
  {2 R} /{R_{exp}} \right ) - 1 } \eqno (7)  $$

\noindent and,

$$ \epsilon_{iso} / \epsilon_{lop} \: = \: 4 (1 - \frac{R_{exp}}{2 R})  \eqno (8) $$

\noindent Thus, the ellipticity of isophotal contours $\epsilon_{iso}$ is
{\it higher by at least a factor of 4}
compared to $\epsilon_{lop}$. Thus even a $\sim $ few \% asymmetry
in the halo potential leads to a large $\sim 10 \% $ spatial
lopsidedness in the disk. This makes the detection of lopsidedness
easier, and it explains why a large fraction of spiral galaxies is
observed to be lopsided.

For the typical observed values of $A_1 \geq 0.1 $ at $R/
R_{exp} = 1.5-2.5 $ (see Section 2.2), the typical $\epsilon_{lop}
\sim 0.03$ (from eqs.[3], and [4]), or $\sim 0.05$ in view of the
negative disk response discussed next. Thus, from the observed disk
lopsidedness, we obtain a value of the {\it halo lopsidedness}. In
the limiting case of high observed $A_1 \sim 0.3 - 0.4 $, the
resulting $\epsilon_{lop}$ is still small $\leq 0.1 $, this is due
to the high ratio of $\epsilon_{iso} / \epsilon_{lop}$ (eq. 8). This is an
interesting physical result, because it means that despite the
visual asymmetry, such galaxies are dynamically robust.

The above results for orbits and isophotes are shown to be
applicable for both stars and gas in the same region of the galaxy
since they respond to the same lopsided potential and have
comparable exponential disk scale lengths (Jog 1997). This was
confirmed by a detailed comparison of the Fourier analysis of the
two-dimensional HI data and the 2MASS near-IR representing stars for
a few galaxies in the Eridanus group (Angiras et al. 2006)
 and in Ursa Major (Angiras et al. 2007). The two tracers show comparable lopsided 
amplitudes (see Fig. \ref{fig25}). This is true even though the HI gas 
in these group galaxies obeys a gaussian rather than an exponential radial distribution.

In the above analysis, the phase is taken to be constant with radius and hence set equal to zero.
 When the phase of the potential
varies with radius, the resulting isophotes show a prominent
one-arm, as observed in M51 and NGC 2997.

\bigskip

\noindent {\it 3.2.1.b $\:$ Kinematics in a perturbed disk}
 
\bigskip

The kinematics in the disk perturbed by a lopsided potential is also
strongly affected. The net rotational velocity, $V_{\phi}$, (see eq.
[3]), is a maximum at $\phi = 0^0 $, and it is a minimum along the
opposite direction. This results in distinctly {\it non-axisymmetric
rotation curves} in the two halves of a galaxy (Jog 1997), with the
maximum difference between the rotational velocities $\sim 10 \%$ or
20-30 km s$^{-1}$ for the typical observed $A_1$ values (Jog
2002). This naturally explains the observed asymmetry in rotation
curves of galaxies such as M 101. The observers most often give an azimuthally averaged
data, thereby the precious information on the kinematical asymmetry is lost. It is strongly recommended (see Jog 2002)
that the observational papers give, when possible, a
 full azimuthal plot of the rotation velocity 
or at the very least the average taken in each hemisphere separately. The latter is done in many 
papers- see e.g. Begeman 1987, which can be used to deduce the kinematical asymmetry in galaxies. 
 Such asymmetry has also been
studied by Swaters et al. (1999) from their kinematical data in HI
on DDO 9 and NGC 4395. They show that the rotation curve rises more
steeply in one half of the galaxy than in the other.

The asymmetry in the velocity fields resulting from the disk response to a
lopsided halo perturbation has also been studied by Schoenmakers
(1999). The results obtained are applied to analyze the
kinematical data from a few spiral galaxies (Schoenmakers et al.
1997, Swaters et al. 1999). Schoenmakers et al (1997) show that the
Fourier amplitudes $m + 1$ and $ m - 1$ of the velocity field are affected when the perturbation
potential of type $m$ is considered. By comparing the observed values for
m = 2 with the calculated values they obtain an upper limit
(uncertain up to the sine of the inclination angle) for the lopsided
perturbation potential.

The approach described in this section assumes a simplified perturbation 
lopsided potential with a constant amplitude, also only closed orbits 
are considered for simplicity. The orbits would change slightly and 
would not be closed if the random motion of the particles is taken 
into account. However, this does not affect the isophotal shapes - see 
Rix \& Zaritsky (1995).

\bigskip

\noindent {\it 3.2.1.c $\:$ Radius for the onset of disk lopsidedness}

\bigskip

The lopsided distribution in a disk cannot self-support itself because
the potential corresponding to it opposes the perturbation potential, as discussed next.
The effective disk surface density or the
 disk density response is shown to be a maximum along $\phi =
0^0$, that is along the maximum of the lopsided potential (Jog 1997), see eq.(5) above.
 This has interesting and subtle dynamical consequences (Jog 1999). This can be seen from
the self-gravitational potential corresponding to the
non-axisymmetric disk response, which is obtained by inversion of Poisson
equation for a thin disk using the Henkel transforms of the
potential-density pairs. This response potential is shown to oppose the imposed
lopsided potential. This may seem counter-intuitive but it arises due to the self-gravity of the disk. Thus in the inner parts of the disk, the disk resists any imposed perturbation potential. 

 A self-consistent calculation shows that the net
lopsided distribution in the disk is only important beyond 1.8 disk
scale lengths and its magnitude increases with radius. This indicates
the increasing dynamical importance of halo over disk at large
radii. The negative disk response decreases the imposed lopsided
potential by a factor of $\sim 0.5 - 0.7 $ (Jog 2000). The above
radial dependence agrees well with the onset of
lopsidedness as seen in the near-IR observations of Rix \& Zaritsky
(1995). On taking account of this effect, a given observed lopsided amplitude 
corresponds to the deduced perturbation potential to be higher by a factor of $\sim 1.3 - 1.4$.

The radius of onset of lopsidedness and the reduction in the imposed potential depend on the form of 
the perturbation potential which was taken to be constant for simplicity in Jog (1999). A more realistic case with a radially varying perturbation potential as in a tidal encounter
with amplitude decreasing at low radii, will result in the highest decrease at lowest radii
(Pranav \& Jog 2008). In this case the actual value of reduction will decide
the radius beyond which net disk lopsidedness is seen.

This idea of negative disk response is a general result and is 
applicable for any gravitating system which is perturbed by an external mechanism. Although it is
shown here for a $cos \phi$ perturbation resulting from a tidal
perturbation, it is applicable for any linear perturbation
of the system.
A similar study for the onset of warps (which can be represented by an m=1 mode 
along the vertical direction) has been done (Saha \& Jog 2006). This shows the onset of 
warps from a radius of 4-5 disk scalelengths, in 
good agreement with observations (Briggs 1990). The disk self-gravity 
is more important along the z-direction for a thin disk and hence the 
disk is able to resist vertical distortion till a larger radius than the planar distortion.

\bigskip

\noindent {\it 3.2.1.d $\:$ Comparison of $A_1$ vs. tidal parameter}

\bigskip

Since tidal encounters (e.g. Beale \& Davis 1969, Weinberg 1995) and satellite 
accretion (Zaritsky \& Rix 1997) have often been suggested as the mechanism for 
the origin of the disk lopsidedness, it is instructive to 
check how the observed amplitude $A_1$ for lopsidedness varies with 
the tidal parameter $T_p$ as was done by Bournaud et al. (2005 b). The tidal 
parameter $T_p$ was calculated in each case as:

$$ T_p \: = \: log \: {\Sigma}_i {\large[} (\frac {M_i}{M_0}) (\frac {R_0}{D_i})^3 {\large]}  \eqno(9) $$

where the sum is computed over the companions for a galaxy of target mass $M_0$ and radius $R_0$ 
and $M_i$ is the mass of the companion at a projected distance $D_i$ on the sky. 
The summation is over neighbours within 2 degrees on the sky and within 500 km s$^{-1}$
velocity range of the test galaxy. 

\bigskip

\begin{figure}[h]
\centering
\includegraphics[height=2.5in,width=2.5in]{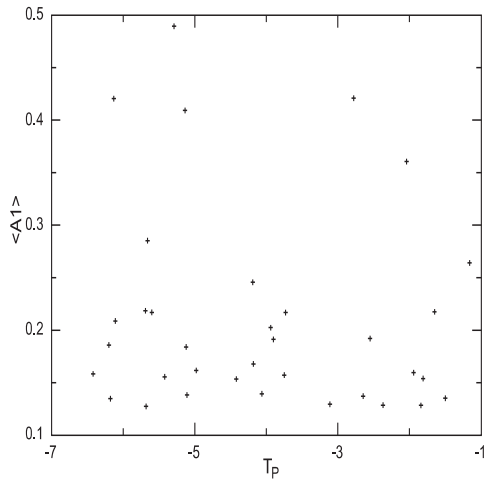} 
\caption{Plot of A$_1$ vs. tidal parameter for the 35 strongly lopsided galaxies in the OSU sample (taken from Bournaud et al. 2005 b). There is no correlation between these quantities, in particular the isolated galaxies with high A$_1$ (top l.h.s. corner of this plot) cannot be explained by a recent tidal interaction.}
 \label{fig13}
\end{figure}

\bigskip

The result is plotted in Figure 13, for the 35 most lopsided galaxies which are at an inclination of $<$ 70$^0$. Surprisingly, this does not show a correlation between the lopsided amplitude and the strength of the tidal parameter. In particular it is hard to explain the galaxies with high $A_1$ and low tidal parameter (in the top l.h.s. of this figure) in the tidal picture. On the other hand, this still does not rule out tidal encounters as the origin for lopsidedness if it is long-lived, or if it arises due to a satellite merger (Walker et al 1996, Bournaud et al. 2005 b).

In order to check the typical values of lopsided amplitudes generated in tidal encounters and satellite accretion, 
N-body simulations for a number of encounters and satellite mergers were studied.
 They included the tidal encounters 
between nearly equal-mass galaxies (up to the ratio of 4:1) and mergers of small-mass galaxies (in the ratio of 
5:1-20:1) (Bournaud et al. 2005 b).
It was found that tidal interactions can indeed generate
 a fairly large amplitude similar to or higher than the average value of $\sim 0.1$. However, the amplitude
then drops rapidly and hence can be seen only for $< 2$ Gyr. 
A typical example of an encounter between a 2:1 mass ratio is shown in Figure 14.
Thus tidal encounters cannot explain the high amplitude of lopsidedness seen in several isolated galaxies such as NGC 1637,
although it could arise due to a recent satellite accretion. 
While a satellite accretion of mass ratio 7:1-10:1 can result in a strong lopsidedness, it can
also thicken the disk
more than is observed (Bournaud, Combes, \& Jog 2004 , Bournaud et al. 2005 a). The thickening of disks can  set a limit on
the rate of satellite mergers (Toth \& Ostriker 1992).
It needs to be checked if a small-mass satellite falling onto a galactic disk
 can generate the right amplitude distribution of lopsidedness without thickening 
the disk, and further if there exist satellites in sufficient numbers to fall in at a steady rate
to repeatedly generate lopsidedness as required by the observed high fraction of lopsided galaxies.

The mechanism for origin of lopsidedness involving a tidal encounter has been explored by Mapelli et al. (2008) 
in the context of NGC 891. 
They show that the lopsidedness seen in the atomic hydrogen gas in NGC 891 
is due to the fly-by encounter with its neighbour UGC 1807. They argue that this is a preferred mechanism over gas accretion from cosmological filaments or that due to ram pressure from the intergalactic medium.

\bigskip

\begin{figure}[h]
\centering
\includegraphics[height=4.0in,width=3.5in]{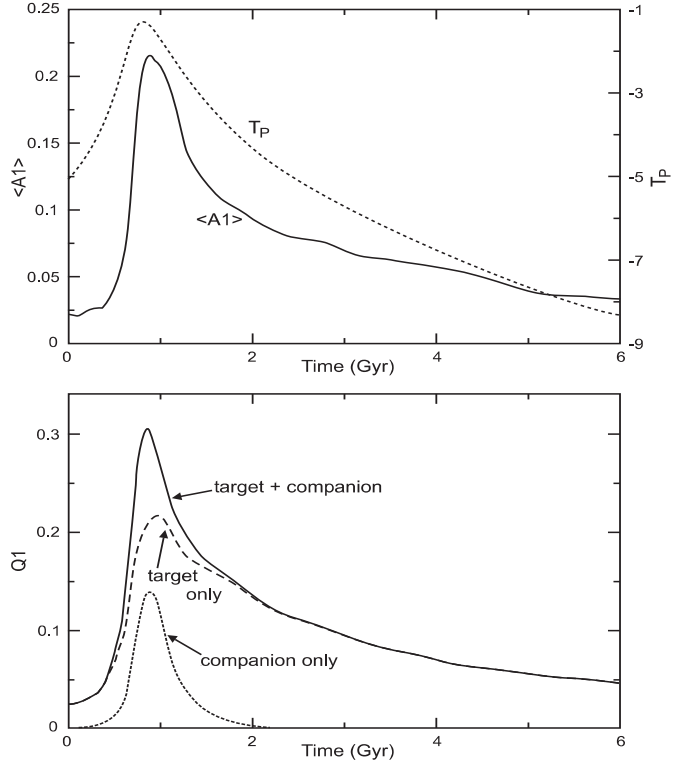} 
\caption{Plot of A$_1$ vs. time in the middle radial range of 1.5-2.5 disk scalelengths, generated in a distant interaction between galaxies of mass ratio 2:1 (taken from Bournaud et al. 2005 b). The peak value of A$_1$ is large $\sim 0.2$, higher than the average value seen in the OSU field galaxies sample, but it drops rapidly to 0.05 in a few Gyr. The lower panel shows the same in terms of Q$_1$, the cumulative potential from the disk.}
 \label{fig14}
\end{figure}

\bigskip

Further, a number of statistical features for the field galaxies
show that a tidal encounter cannot be the primary mechanism for the
origin of the disk lopsidedness- for example, the lopsidedness is higher 
for late-type galaxies whereas tidal encounters and mergers would tend to 
lead to the secular evolution of a galaxy towards early-type galaxies. Thus 
if tidal interactions were the primary mechanism for generating lospidedness, 
then the early-type galaxies should show a higher amplitude of lopsidedness. 
This is opposite to what is sen in the field galaxies (Bournaud et al. 2005 b). 
Thus other mechanisms such as gas accretion (Section 3.2.2) could be important 
in generating the lopsidedness in field galaxies.

In the group galaxies, on the other hand, the tidal interactions being more frequent, 
 play a dominant role in generating lopsidedness.
This is evident from the fact that the early-type galaxies show higher lopsided amplitudes 
as seen in the Eridanus group galaxies (Angiras et al. 2006) and
less strongly in the Ursa Major group of galaxies (Angiras et al. 
2007). The details are given in Section 5.

\subsubsection{Gas Accretion, and other mechanisms}

The intergalactic gas accretion was proposed as a qualitative idea
to explain the m=1 asymmetry in NGC 4254 by Phookun et al.(1993).
They proposed that the lopsidedness could arise due to the subsequent swing amplification
 in stars and gas (as in Jog 1992).
An extensive study of origin of lopsidedness via N-body simulations
(Bournaud et al. 2005 b) shows that while tidal encounters can explain
the observed amplitudes of disk lopsidedness, these cannot explain
the various observed statistical properties such as the correlation
between A$_1$ and A$_2$, and the higher lopsidedness seen for the
late-type field galaxies. In order to do this, one needs to take account
of gas accretion from outside the galaxy.

There is growing evidence that galaxies steadily accrete gas from
the external regions, as seen from the cosmological models (Semelin
\& Combes 2005), and also observed in nearby galaxies (Sancisi et al. 2008). 
Thus it is natural to see how this affects the mass
distribution in a disk. While the details of gas infall are not yet
well-understood, it is plausible that a galaxy may undergo an
asymmetric gas infall on the two sides from say two different
external filaments. An application of this idea showed that the gas
infall and the resulting star formation can well reproduce the
striking asymmetry observed in NGC 1367 (Bournaud et al. 2005 b), 
see Fig. 15 here.

\bigskip

\begin{figure}[h]
  \centering
  \includegraphics[width=10cm]{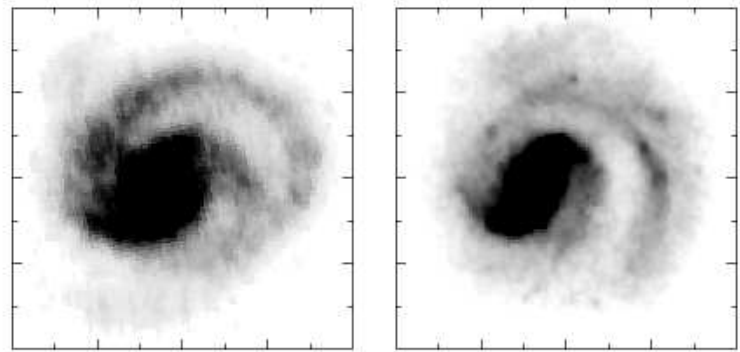}
   \caption{NGC data - near-IR data (left panel), and the same 
from N-body simulations with gas accretion in two streamers at a rate 
so as to double the mass of the galaxy in a Hubble time (right panel), taken from Bournaud et al. (2005 b).}
 \label{fig15}
\end{figure}

\bigskip

A word of caution  is that if
 gas accretion is the mechanism for generation of lopsidedness,
one would expect to see asymmetry in the gas velocity fields whereas these
are smoothly continuous as pointed out by Baldwin et al. (1980).
In the case of galaxies in groups, in any case,
the tidal interactions
may play a dominant role, see Section 5.2
for the details.

Other physical models have also been developed in the literature but
these are probably not as widely applicable due to the specific
parameters or conditions chosen, as discussed below. This includes a
proposed model where the growth of m=1 is treated as an instability
in a self-gravitating disk (Lovelace et al 1999). This results in strongly unstable
eccentric motions but only within the central disk scalelength. Here a linear analysis is used to treat slowly growing mode, and the pattern speed could be either positive or negative.
In another model where the disk is off-centered with respect to the
halo up to a maximum distance of the core radius (Levine \& Sparke 1999) also shows lopsidedness. However, it is seen
 only in the
inner regions within the core radius of the halo where the halo density and hence the rotation speed is constant.
Both these papers do not yield higher lopsidedness
in the outer parts, and this result contradicts the
observations which preferentially show lopsidedness in the outer
parts of a galaxy. The kinematical and morphological asymmetries
resulting from the latter model are shown to be not always
correlated (Noordermeer et al. 2001). This is somewhat
unexpected - since on general physical grounds the two would be
expected to be related causally (see Section 2.2). In any case, the
restricted set of initial conditions required for this model makes
it applicable only to a few galaxies such as the dwarf galaxies.

A self-consistent model for m=1 in a self-gravitating disk has been
proposed by Syer \& Tremaine (1996) in the so-called razor-thin
disks. However, the density response due to an imposed perturbation
opposes the perturbing force (see Jog 1999) and hence the disk
asymmetry cannot result from the orbital asymmetry as was also argued by
Kuijken (1993) and Earn \& Lynden-Bell (1996). Moreover, the model by Syer \& Tremaine (1996)
gives the density response to be maximum along the perturbed force,
which is in contrast to the result by Rix \& Zaritsky (1995), and
Jog (1999).

\subsubsection{ Lopsidedness as an instability}

An obvious possible explanation for the origin of the lopsided mode is
that it arises due to gravitational instability in the disk. For example, this was
proposed and studied for the gas by Junqueira \& Combes (1996). A
similar model for the disk in a dark matter halo perturbed by a
satellite was studied by Chan \& Junqueira (2003), however these
models generate lopsidedness only in the inner regions in contrast
to the observed trends. An internal mechanism based on the
non-linear coupling between m=2 (bars or spiral arms) and m=3 and
m=1 has been proposed by Masset \& Tagger (1997) which gives rise to
the excitation of m=1 modes in the central regions.

Lopsided instabilites have been shown to develop in counter-rotating
stellar disks which have a high fraction of retrograde orbits
(Hozumi \& Fujiwara 1989, Sellwood \& Valluri 1997, Comins et al 1997,
Dury et al 2008).
Galaxies where the gas participates to the counter-rotation are 
quite often observed to develop
$m=1$ perturbations (e.g. Garcia-Burillo et al 2000, 2003). 
However, since counter-rotation is rarely seen in stellar disks
(Kuijken, Fisher, \& Merrifield 1996, Kannappan \& Fabricant 2001, McDermid et al 2006),
this cannot be the primary mechanism for the generation of
lopsidedness in disks.

In a recent work, the self-gravity of a slowly-rotating global m=1 mode has been shown to 
lead to a long-lasting lopsided mode in a purely exponential 
disk as in a spiral galaxy (Saha, Combes, \& Jog 2007). This model was 
motivated by the fact that the observations show that the lopsidedness 
has a constant phase with radius which indicates a global mode.
Further, it was noted that m=1 is unique in that the centre of mass of 
the disturbed galaxy is shifted away from the original centre of mass 
and thus acts as a restroing force on the latter. Thus the system can 
self-support the m=1 mode for a long time especially when one takes 
account of the self-gravity of the global mode.

Using the linearized fluid equations
and the softened self-gravity of the perturbation, a self-consistent
quadratic eigenvalue equation is derived for the lopsided
perturbation in an exponential galactic disk and solved.
Fig. 16 shows the resulting isodensity contours, clearly the centres of isocontours are
progressively more disturbed in the outer parts.

\bigskip

\begin{figure}[h]
\centering
{\rotatebox{270}{\includegraphics[height=2.3in]{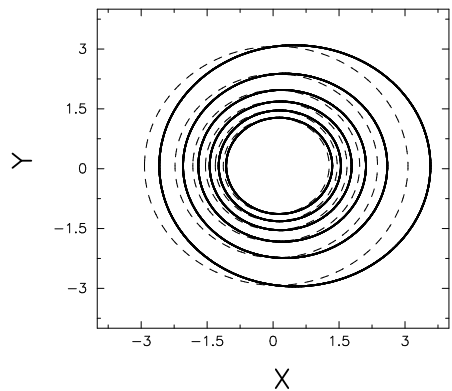}}}
\bigskip
\caption {Contours of constant surface density for a global m=1 mode, taken from Saha et al. (2007). Here the x and y axes
are given in units of the disk scalelength. The maximum surface density occurs at (0,0). The outer
contours show a progressive deviation from the undisturbed circular distribution, indicating
a more lopsided distribution in the outer parts- as observed.}
 \label{fig16}
\end{figure}

\bigskip

The self-gravity of the mode results in a significant
reduction in the differential precession, by a factor of $\sim 10$ compared to the free
precession. This leads to 
persistent m=1 modes, as shown in Fig. 17.

\bigskip

\begin{figure}[h]
\centering
{\rotatebox{270}{\includegraphics[height=2.3in]{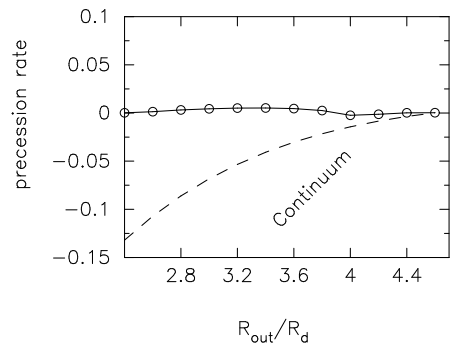}}} 
\bigskip
\caption{The precession rate for the global lopsided mode vs. the size of the disk in units of the disk scalelength
in a galactic disk (shown as the line with circles)
is very low, thus the mode is long-lived. The dashed line denotes the free precession ($\kappa - \Omega$).
This is taken from Saha et al. (2007).}
 \label{fig17}
\end{figure}
\bigskip

N-body simulations are performed to test the
growth of lopsidedness in a pure stellar disk, which confirm these results (see Fig. 18). Both
approaches are compared and interpreted in terms of slowly
growing instabilities on timescales of $\sim$ a few Gyr, with almost
zero pattern speed.

\bigskip
\begin{figure}[h]
\centering
{\rotatebox{270}{\includegraphics[height=5.2in]{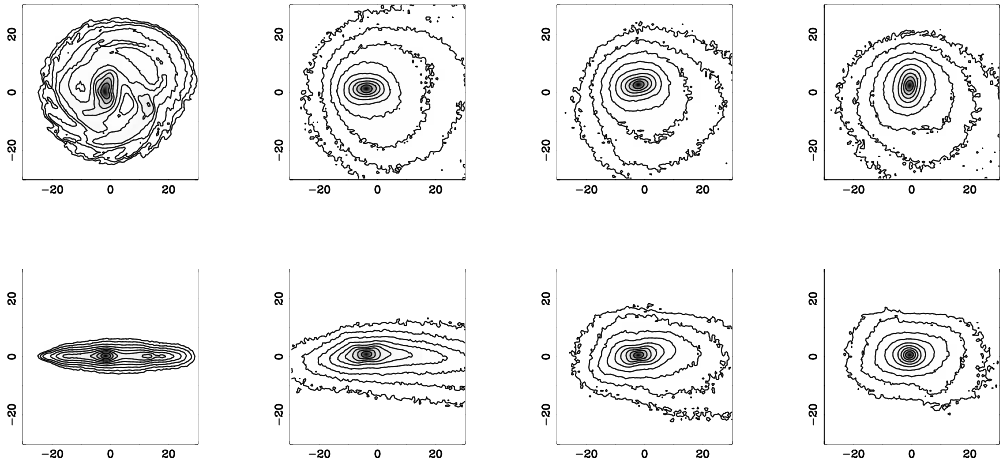}}}
\bigskip
\caption{The isodensity contours in a purely stellar exponential galactic disk, given in a logarithmic 
scale of the surface density of the stellar disk, face-on (top panel) and edge-on (lower panel) views
at four different epochs: T =0, 4.8, 9.6, 14.4 Gyr, from left to right, taken from 
Saha et al. (2007). The global lopsided mode is long-lived 
and lasts for $\sim 14$ Gyr.} 
 \label{fig18}
\end{figure}

\bigskip

Though this is a somewhat idealized approach, it is precisely this
that has allowed the authors to focus on the basic dynamics of the excitation and
growth of m=1 modes. For example, here the only important input
parameters are the softening and the Toomre Q parameter. A smaller
value of Toomre Q results in a fast initial growth of the mode but
later as the Q increases due to the heating in the system, the mode
is self-regulated and has a nearly constant amplitude A$_1$ that is
long-lived. The softening acts as an indicator of the coherence in
the mode, and a higher value results in a faster growth rate of the
modes.
These global modes are precessing remarkably slowly, and therefore are
relatively long-lived.
Numerical analysis of the eigen modes of a cold thin disk
shows that, if treated as modes of zero pattern speed, warps and lopsidedness 
are fundamentally similar in nature (Saha 2008).

Such small pattern speed is also in agreement with the work of Ideta (2002) 
who showed that the rotating m=1 mode in a live halo would be damped very rapidly
by the density wake induced in the halo, see Fig. 19.
Hence he argued that the m=1 modes must be non-rotating at a rate smaller than
1 km s$^{-1}$ kpc$^{-1}$. This would give
  the damping time to be comparale to the Hubble time, which 
can explain the high frequency of lopsidedness seen in spiral galaxies.

\bigskip

\begin{figure}[h]
\centering
\includegraphics[height=2.8in]{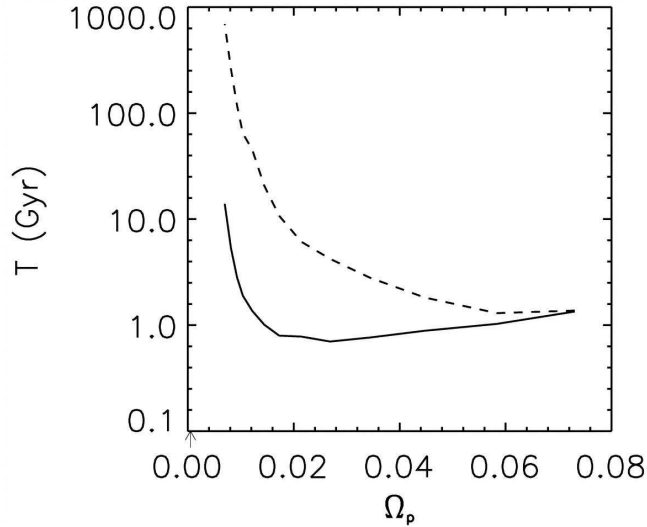} 
\caption{The damping time of the lopsided mode vs. the pattern speed, for a self-gravitating case (solid line) and without self-gravity (dashed line) respectively, taken from Ideta (2002). Damping times $\sim$ Hubble time imply a very low pattern speed of the lopsided mode.}
 \label{fig19}
\end{figure}

\bigskip

In a recent paper Dury et al. (2008) have studied a galactic disk in an inert halo by N-body simulations and argued that that a rotating m=1 mode can occur as a result of the swing amplification (Toomre 1981).

\subsubsection {Effect of inclusion of rotation and a live halo}

The work by Saha et al. (2007) shows that a gravitating disk is susceptible to the 
growth of m=1 modes, irrespective of the origin of these modes. These 
could be triggered as a response to a distorted halo, or by gas accretion. 
We caution, however, that Saha et al. (2007) treat a simple, specialized 
case of slow rotating modes in a pure exponential disk.
A more realistic treatment should include a live or a responsive halo in addition to the 
disk. Here more parameters enter the picture such as the relative mass of the halo and the
disk, and ratio of the core radius of the halo to the disk
scalelength etc. The inclusion of a live halo is expected to further
support the persistent m=1 modes, in analogy with what was shown for bars by Athanassoula (2002) and for a general, local non-axisymmetric feature by Fuchs (2004). In contrast, other papers suggest that a responsive halo tends to damp a feature in the disk by the wake created 
in the halo. This was
 shown for disk lopsidednesss (m=1 in the plane) by Ideta (2002), and shown
in the case of warps treated as an m=1 mode normal to the galactic plane (Nelson \& Tremaine 1995).
This issue needs to be clarified by further dynamical studies of a global m=1 mode of an arbitrary
pattern speed in a galactic disk. This study should include
 a live halo, and have a high resolution (since poor resolution may smoothen the response and give a spuriously long-lasting mode), with the aim to
 check the lifetime of such a mode.

A lopsided  mode in a collisionless spherical dark matter 
halo is shown to be long-lasting or slowly damped compared to the dynamical timescales (Weinberg 1994, 
Vesperini \& Weinberg 2000). However it is not clear if
this is due to the fact that the halo is supported by random motion
and hence a perturbation in it is long-lived. On the other hand,
  in a spiral
galaxy, the presence of differential rotation puts a limit on any
material feature due to the precession rate.
In contrast, the numerical simulations of perturbations triggered in 
a galaxy with live halo, due to the tidal encounter of nearly equal-mass galaxies 
(up to the ratio of 4:1) and mergers of small-mass galaxies (in the ratio of 
5:1-20:1) show that the lopsidedness thus generated, although long-lived compared to the dynamical timescales, does not last beyond $\sim 2$ Gyr (Bournaud et al. 2005 b).
The crucial factor that decides the lifetime of the lopsided mode could be its pattern speed, with the small speed cases lasting for a long-time close to the Hubble time (Ideta 2002, Saha et al. 2007). The off-centering of the mass distribution is less pronounced in the case of high rotation speed. This could lead to a short lifetime of the m=1 mode as pointed out by Ideta (2002). In this case the restoring term in the equations of motion (Saha et al. 2007, eq. 18) is less pronounced and hence the m=1 mode lasts for a shorter time. Alternatively, it could be that the wake generated in a halo could be larger for a higher pattern speed, and hence
the halo tends to dampen such modes. This may be the reason why the lopsidedness generated in a disk with a live halo due to a tidal encounter or a minor merger (Bournaud et al. 2005 b) lasts for $< 2 $ Gyr.

While slow m=1 modes are shown to be long-lived, the uniqueness of this solution is still not established,
or that this is 
what explains the observed lopsidedness. Also it is not clear what would excite such slow modes.
 The generating mechanism may have a strong bearing on the resulting pattern speed: a tidal encounter is expected to 
result in lopsidedness with a high pattern speed $\sim$ the relative velocity over the impact parameter
as argued by Ideta (2002). Gas accretion, on the other hand, 
may not easily give a global m=1 mode, while observations show the mode to be global.
 Further numerical simulations should check if a satellite
accretion can give rise to a slow, global mode.
 An actual measurement of the pattern speed of the lopsided
mode in a real galaxy will help settle this issue, and we urge observers to take up this important measurement.

\subsection{Comparison between origin of m=1 and m=2; stars and gas}

As discussed at the beginning of Section 3, the m = 2 case is fairly
well-understood, while the m=1 case has only begun to get attention
from theorists. There are several differences in the dynamics and
evolution of the two features, and also as applied to stars or gas.
First, the presence of dark matter halo is likely to have a
substantial role to play in the origin and evolution of lopsidedness
in a disk. This is especially true
 in the outer parts of a disk since the disk
lopsidedness is observed to increase with the radial distance 
 where the halo is more important. In the inner regions of a
galaxy, on the other hand, the inclusion of the bulge is likely to
play an important role in stabilizing the m=1 mode (de Oliveira \&
Combes 2008).

 Further, the m=1 mode generally has no ILR (Inner Lindblad Resonance, e.g., Block et al.
1994) hence its evolution differs from that of m=2. The m=1 mode does not
get damped easily due to the angular momentum transport occurring at
the resonance points as in the case of m=2 (Lynden-Bell \& Kalnajs 1972). In
case of m=2 this causes an absorption of the wave at the ILR and
thus a break in the feedback loop. But in absence of an ILR for m=1,
this break does not arise and this helps in sustaining the global
m=1 mode for a long time. In this picture, the gas being cold
behaves in a different way. The absorption at the resonance point is
only partial for gas and hence even m=2 can be sustained in gas
despite the presence of an ILR. This helps support the generation of
m=2 and higher order modes in the presence of gas.

\subsection{ A summary of the various mechanisms}

Of the various mechanisms proposed so far, the most promising ones,
as judged by the resulting agreement with the observations, are
those involving tidal encounters and gas accretion.

An m=1 perturbation in a disk leads to a shift in the centre of mass
in the disk, and this then  acts as an indirect force on the
original centre of the disk. The disk is thus shown to naturally
support an m=1 mode, and as pointed out above, this is a
characteristic property valid only of a lopsided mode. This basic
physics is sometimes clouded over because of the additional effects
introduced due to the inclusion of the dark matter halo, the bulge,
and the gas as in a real galaxy. Further, depending on whether the
halo is live or rigid, and whether it is pinned or not, and
whether the pattern is rotationg or stationary can lead to
additional complexities. 
Also, other features like the wandering of
the centre (Miller \& Smith 1992) may introduce an m=1 mode. In
short, there seem many paths to get m=1 in a galaxy, and therefore
it is important to identify which is the most applicable one in a
real galaxy.

\noindent {\it Long-term maintenance of disk lopsidedness}

It has been realized from the beginning that the lopsided modes
should be fairly long-lived (e.g., Baldwin et al. 1980) or excited
frequently. This is needed 
in order to explain the high fraction of galaxies showing
lopsidedness, and also the strong lopsidedness seen in isolated
galaxies like M101. It has been noted that since m=1
does not have an ILR, it should be the preferred mode in the
galactic disk (Section 3.3). This, however, does not say anything
directly about its lifetime.
The persistence of lopsided mode is still an open question, as shown by the discussion below.

While tidal encounters can generate
the right lopsided amplitudes, these are not correlated with the
strength of a tidal encounter (Bournaud et al. 2005 b), or with the
presence of nearby neighbors (Wilcots \& Prescott 2004). This could
be explained if the disk lopsidedness once generated either directly
in the disk, or as a response to a long-lived halo distortion, were
long-lived, so that there is no clear correlation with a tidal
encounter. N-body simulations with a live halo show the resulting m=1 
modes in the disk to last for $\sim 2-3$ Gyr, which is much smaller than the Hubble time, thus these
need to be triggered again.
A tidal encounter will typically generate a fast mode
which is expected to be not long-lived (see the discussion in Section 3.2.3).

While the satellite accretion of mass ratio 7:1-10:1 can result in a strong lopsidedness, it can
also thicken the disk more than is observed (Bournaud et al. 2004). Also, it has a short lifetime of $< 2$ Gyr (Bournaud et al. 2005 b).
It further needs to be checked if a smaller-mass satellite 
falling onto a galactic disk
 can generate the right amplitude distribution of lopsidedness without thickening 
the disk, and  if there are adequate number of such satellites that can fall in at a steady rate.

The pattern speed is expected to have a significant effect in determining the 
lifetime of a lopsided mode with a slow patttern being long-lived. It is not 
clear if a live halo will help or hinder the long-term sustenance of an m=1 mode,
as discussed in Section 3.2.3.

Future work needs to study the long-term maintenance of m=1 modes when generated
by accretion of a low-mass satellite. A similar study needs to be done for the case of
 gas accretion and to see if the latter gives a global mode.
The observations of group galaxies with their generally stronger and frequent triggering of lopsidedness (see Section 5)
can act as a constraint on any generating mechanism proposed for the field galaxies.

\section{Lopsidedness in the central region}

\subsection{Stability of central nuclear disks}

It is now well established that all galaxies with bulges
or spheroids host a massive central black hole (Gebhardt et al. 2000).
 The central region, or nuclear disk, in galaxies therefore have
a gravitational potential very close to the Keplerian. 
Some similarities exist with proto-planetary systems (Rauch \& Tremaine 1996), 
or with the formation of new stars through accretion disks
(Adams et al. 1989, Alexander et al. 2007).
The nearly Keplerian potential, with the angular
velocity $\Omega \sim r^{-3/2}$ favors eccentric orbits and $m=1$ 
modes, instead of $\Omega \sim r^{-1}$
of galactic disks which favor $m=2$ perturbations.

Nearly keplerian disks have the particular property that
the orbit precession rate is almost
zero ($\Omega \sim \kappa$). If the apsides are aligned at a given time,
they will stay so in a $\Omega_p \sim 0$ mode. The self-gravity of the disk
makes $\kappa > \Omega$, and the orbits differentially precess
at a rate ($\Omega - \kappa) < $ 0. However, if the disk self-gravity is not
large, a small density perturbation could be sufficient to counteract the
small differential precession. Goldreich \& Tremaine (1979)
showed that in the case of Uranian rings, the self-gravity could
provide the slight impulse to equalize the precession rates, and
align the apsides.
Two kinds of waves could propagate in such disks-
slow stable modes, and unstable rapid waves,
growing on a dynamical time-scale.

 The density wave theory (e. g., Lin \& Shu 1964) predicts
that in a self-gravitating stellar disk, global spiral modes can develop
only between the radial range delimited by the Lindblad resonances,
i.e. for $m^2 (\Omega - \Omega_p)^2 < \kappa^2$. 
Only in gaseous disks, where the pressure forces dominate,
 acoustic waves can propagate outside this range.
Considering the $m=1$ waves, for a pure keplerian potential
(neglecting the self-gravity of the disk), $\Omega = \kappa$,
and the perturbations are neutral. If there is some self-gravity
in the disk, then ($\Omega - \kappa) < 0$, there is only an outer
Lindblad resonance and a corotation, but no inner resonance
(for prograde modes with $\Omega_p >0$), and therefore the radial range for
the development of $m=1$ perturbations is quite large. 

\bigskip

\begin{figure}[h]
  \centering
  \includegraphics[height=8cm,width=8cm]{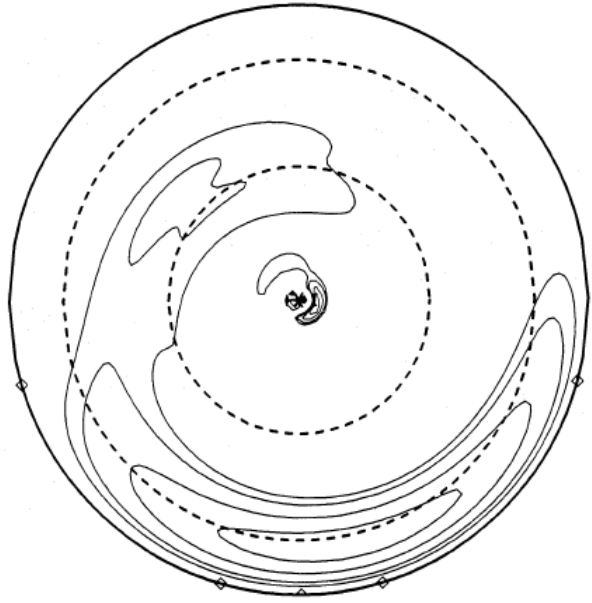}
   \caption{Isodensity contours for the lowest order $m=1$ mode, in a disk of mass equal to the
central point mass. Note the shape in alternating ``bananas'' instead 
of a continuous one-arm spiral. The dashed lines show the locations
of the corotation and the outer Lindblad resonance.
>From Adams et al. (1989).}
 \label{fig20}
  \end{figure}

\bigskip

The study of WKB modes in a non self-gravitating gaseous disk,
with truncation radii at the inner and outer boundary, was
first done by Kato (1983), who found a region of trapped one-arm waves,
with quite low pattern frequency, much lower than the orbital frequency.
 Adams et al (1989) considered the influence of self-gravity,
in order to apply to young stellar objects, which can have accretion
disks with masses of the same order as the central stellar mass.
 The self-gravity of such a disk is sufficient to modify
the precession rate of nearly keplerian orbits ($\Omega - \kappa$)
to a common and coherent value. They found a pattern speed which 
is of the same order as the angular velocity in the disk, and the unstable waves
develop with a growth time comparable to the dynamical time-scale.
 The special shape of modes is shown in Fig. 20.

Crucial to the development of the perturbations, a special characteristic of the $m=1$ mode is to 
shift the gravity center of the system from the dominant central mass (the black hole for instance),
also see Section 3.2.3.
In the reference frame of the black hole, this implies the introduction of an inertial force, 
which by reference to celestial mechanics, is called the indirect term. 
This term is the mediator of angular momentum exchange
between the disk inside and outside corotation and the central
mass.  The coupling with the outer Lindblad resonance
provides the amplification that is usually provided by the corotation
in $m=2$ modes. A feedback cycle has been proposed by
Shu et al (1990), and called SLING (Stimulation by the Long-range 
Interaction of Newtonian Gravity). 
The modes depend strongly on the outer disk boundary conditions,
since a reflection of short waves is assumed there, in the 
4-waves feedback cycle. This cycle is only possible with a
gaseous component, since the short waves are not
absorbed at the resonance but cross the OLR (Outer Lindblad Resonance).

\bigskip

\subsubsection{Slow stable modes, damping slowly}

\bigskip

Another possibility to explain central lopsidedness
is to exploit the slow modes, that are stable, but can be 
long-lived, excited by some external mechanism, such as
the accretion of a globular cluster or a giant molecular cloud.
The precession rate of eccentric orbits is $\Omega - \kappa$ =0 
in the potential of a point mass M$_{BH}$, and slightly negative in 
the presence of a small disk of mass M$_d$, lighter than the
central point mass, with amplitude varying as
$({\rm{M}_d}/{\sqrt{\rm{M}_{BH}}})$ or 
$\propto ({\rm{M}_d}/{{\rm{M}_{BH}}}) \Omega$. If self-gravity has a large enough role, and in
particular, if the disk is cold enough and its Jeans length smaller than the disk radius,
 $m=1$ density waves can propagate; their dispersion relation has been studied
in the tight-winding limit or WKB approximation (Lee \& Goodman 1999, Tremaine 2001). In the linear
approximation, the pattern speed for the wavelength $\lambda = 2 \pi / k$ is in
first approximation for $({\rm{M}_d}/{\rm{M}_{BH}}) << 1$, as given by:
$$
\Omega_p = \Omega - \kappa + {{\pi G \Sigma_d |k|}\over{\Omega}}
F({{k^2c^2}\over{\Omega^2}})
  \eqno (10) $$
where $\Sigma_d$ is the surface density of the disk, and $F$ the usual reduction
factor that takes into account the velocity dispersion $c$ of the stellar disk, and
its corresponding velocity distribution (e.g., Tremaine 2001). The pattern speed
then remains of the order of $({\rm{M}_d}/{\rm{M}_{BH}}) \Omega$, for a sufficiently
cold disk, and is much smaller than the orbital frequency. These slow waves
exist whenever the thin-disk Jeans length $\lambda_J = {c^2}/{G\Sigma_d}$
is lower than $4 r$, while the Toomre parameter $Q$ is less relevant
(Lee \& Goodman 1999).

The study by Tremaine (2001) shows that disks orbiting a central mass
support slow $m=1$ modes, which are all stable. Their frequencies are
proportional to the strength of collective effects, which is
either self-gravity, or velocity dispersion (or pressure in
fluid disks). The latter phenomenon can be simulated
with softened gravity. There are then two kinds of slow modes:
the g-modes (where self-gravity is dominating, and softening
is unimportant), which are long waves, with kr $<<$ 1, with 
negative frequency $\omega <0$; and the p-modes, which depend on
the softening $b$, which can have both short and long waves
(kr $\sim b$), and with positive frequency $\omega >0$; as the
softening increases, the amplitude of the mode decreases, as well as
$\omega$.

In numerical N-body simulations of nuclear stellar disks,
Jacobs \& Sellwood (2001) have reported the presence of a slowly
decaying prograde $m=1$ mode in annular disks around a slightly softened
point mass, but only for disk masses less than 10\% of the central mass
concentration. This confirms the existence of a persistent
slow mode, with positive $\omega$ increasing
with the mass of the disk, and decreasing with the amplitude
of the perturbation.

 Touma (2002) has computed the normal modes of a series of
N rings in a thin disk, through linearized dynamics, and
using the Laplace-Lagrange secular theory of planetary motions 
(valid for small eccentricities). The gravity is
softened to mimic a hot stellar disk, and varies as the velocity dispersion.
 The modes are stable when all rings are prograde, but a fraction of only 5\%
of counter-rotating rings is sufficient to make unstable modes appear.
Sambhus \& Sridhar (2002) built a model of the M31 nucleus
with counter-rotating orbits in a razor thin nucleus,
 and checked that this amount of
counter-rotation could be compatible with observations.
The counter-rotating stars could come from a past accreted
system, like a globular cluster.

Bacon et al (2001) explored by N-body simulations
the possibility of stable $m=1$ mode to explain the M31
eccentric nuclear disk. They found that for a disk mass
accounting for $\sim$ 20 -- 40\% of the total central mass,
self-gravity is sufficient
to counteract the differential precession of the disk.
An external perturbation
can excite this mode, and it is then long-lasting, over
100 Myr, or 3000 rotation periods. 
The prograde mode found in the simulations compares
well with the $p$-modes of Tremaine (2001).
There is a remarkable agreement between the observed
($\sim 3km s^{-1} pc^{-1} $) and predicted value
of the pattern speed, in spite of
all approximations, and although the WKB approximation is not
satisfied.

Although the slow modes are long-lived, their exciting mechanisms
should be found, to explain the high frequency of the phenomenon.
The dynamical friction of the $m=1$ wave on the stellar bulge has been
proposed by Tremaine (1995) as an amplification mechanism,
if its pattern speed is sufficiently positive. This amplification results
from the fact that the friction decreases the energy less than the angular momentum.
The orbits with less and less angular momentum are more and more eccentric, and the
$m=1$ mode develops. Although the efficiency of the mechanism
has not been proven, it should not apply for the slow modes considered here,
in a slightly rotating bulge.
An external perturbation is more likely to trigger the
$m=1$ perturbation.  There is the possibility of
infalling of globular clusters, through dynamical friction,
a mechanism explored in the next section. Also 
interstellar gas clouds should be continuously
infalling onto the center, since within 10-100 pc of 
M31 nucleus, dust lanes, and CO molecular clouds are observed
(Melchior et al. 2000). The interval between two
such external perturbations (either passage of a globular cluster,
or a molecular cloud) in M31 is of the same order of magnitude,
so that the external perturbations are an attractive mechanism.

 Each episode of $m=1$ waves will heat the disk somewhat,
but the instability is not very sensitive to the initial radial
velocity dispersion. Over several 10$^8$ yr periods, the nuclear
disk could be replenished by fresh gas from the large-scale M31
disk and subsequent star formation.
The hypothesis of cold gas accretion from the disk of
M31 itself, has then not only the advantage to trigger the
$m=1$ perturbation, but also 
to explain the maintenance of a rather thin and cold nuclear disk. 

\bigskip

\begin{figure}[h]
  \centering
  \includegraphics[height=5.5in]{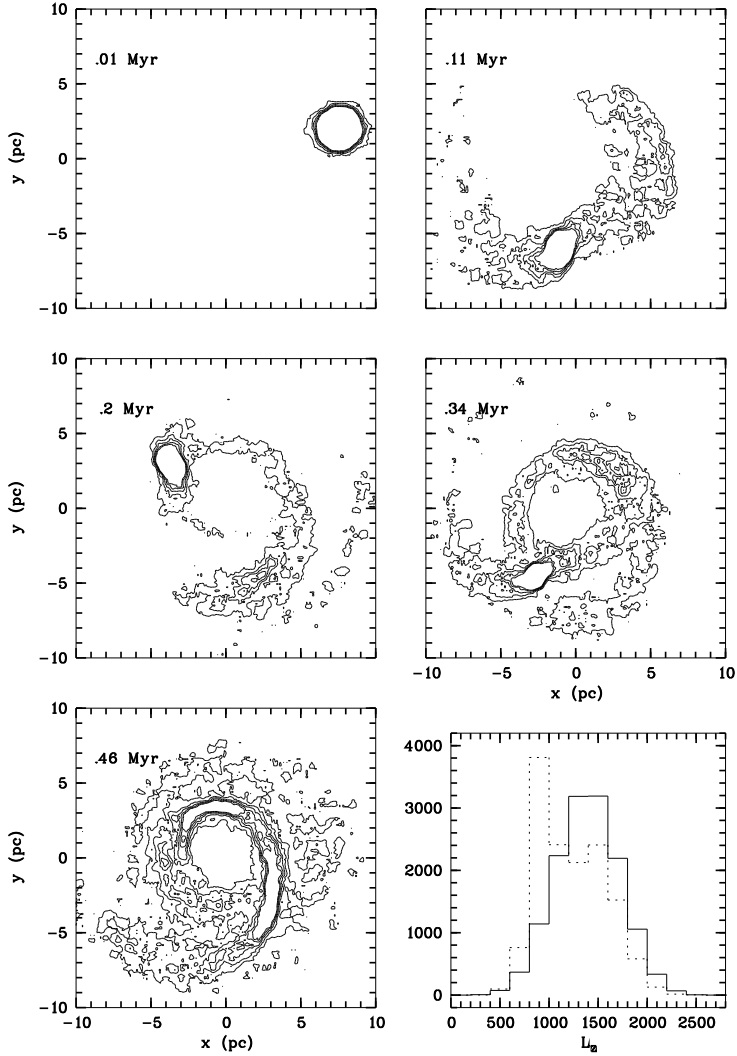}
   \caption{Simulation of a globular cluster infall towards a
galaxy nucleus containing a massive black hole, fit to the 
M31 characteristics. The panels show the face-on view contours
from the cluster stars, from the start to 0.46 Myr.
The bottom right panel shows the histogram of the angular momentum
for the cluster particles at the beginning (solid line) and at the end 
(dashed line) of the simulation. From Emsellem \& Combes (1997).}
 \label{fig21}
  \end{figure}

\bigskip

\subsection{Double nuclei by infalling bodies}

 One solution to the double nuclei problem is to assume that dense stellar
systems, like globular clusters, or a dwarf satellite, or even black holes, are
regularly infalling into the galaxy center, and are responsible for the 
observed morphology. Although the events may be relatively rare,
their actual frequency is not well-known, and the question remains
open as to a possible fit to the observations.
 
The typical dynamical friction time-scale, for an object of mass $M$ at about 10pc
from the center, in a spiral galaxy with a bulge of $\sim$ 10$^{10}$ M$_\odot$,
typical of an Sb galaxy like M31, is $\sim$ 10$^7 (10^6M_\odot/M)$ yr,
and it could be much smaller inside. Although short with respect to galactic time-scales,
this is much larger than the orbital time of 3 $\times$10$^5$ yr at this radius.
Tremaine et al (1975) precisely proposed that the central stellar nuclei
in spiral galaxies are the results of many globular clusters infalling
by dynamical friction. Typically nuclear stellar systems of 10$^7$-10$^8$ 
M$_\odot$ would require the infall of a hundred globular clusters.
In this frame, the frequency of the event is relatively large, and 
could be compatible with the observations.

N-body simulations of globular clusters or dwarf galaxies infalling through the 
gravitational field of a disk, a bulge, and/or a central black hole, have been
carried out by many authors (e.g. Charlton \& Laguna 1995, 
Johnston et al 1999, Combes et al 1999, Bekki 2000b), and applied
to explain the M31 nucleus morphology
(Emsellem \& Combes 1997, Quillen \& Hubbard 2003). 
N-body simulations demonstrate that the infalling system is destroyed
by the tidal forces of the black hole at the right distance of the nucleus
(about 3pc). The debris then rotate around the nucleus in eccentric orbits,
and form an eccentric disk. Several hypotheses can then be explored:
either the infalling system is alone able to form the nuclear disk (Bekki 2000a),
but then its mass corresponds more to the core of a dwarf galaxy having
merged recently with the big primary (a quite rare event). Or the 
infall of the system excites an eccentric mode in the nuclear disk
(Bacon et al 2001), as proposed by Tremaine (1995).
Also, the presence of the infalling system not yet diluted
in the nuclear disk increase the asymmetry.
The details of the dynamics, and in particular the inclination 
of the nuclear disk with respect to the main disk of M31,
or the shift of the velocity dispersion peak from the black hole position,
due to the systematic rotation of the luminosity peak of the disk,
are all explained by the model by Emsellem \& Combes (1997), see Fig. 21.

 The nature of the infalling system is constrained by the 
present metallicity and colors of the nuclear disk, especially
if the assumption is made that the infalling system is the first one
and forms totally the nuclear disk (Bekki 2000a). 
The observed colors of the double luminosity peaks 
in M31 are quite similar to the nuclear disk
ones, and different from the bulge, so the hypothesis that the 
nuclear disk is formed from the infalling systems themselves is possible.
The hypothesis of globular clusters is more likely, in the sense
that the probability to observe it is larger, the friction time-scale being longer, 
and it requires at least 25 globular clusters to form the nuclear disk.
The hypothesis of a dwarf galaxy merger does not correspond to the
quite un-perturbed state of the M31 disk. 
There is some evidence of a past merger around M31,
in the shape of an extended stellar disk, loops and shells
(Ibata et al 2001, Irwin et al 2005). However, the time to form these
stellar streams is much longer than the time-scale
for the galaxy core to infall to the center.

It is interesting to discuss in this context
the case of the double nucleus in NGC 4486B, which is thought
to be similar to the M31 case, but with larger masses, and larger separation
(Lauer et al 1996). NGC 4486B is a compact elliptical galaxy,
in the outer envelope of M87 in the Virgo cluster. The double 
nuclei are separated by 12pc, and produce two almost equal luminosity peaks 
at similar distance from the photo-center, so creating almost no
lopsidedness in projection. If explained by a nuclear disk, it is
of very small eccentricity. The hypothesis of a past merger as
the origin of the two stellar nuclei is weaker in this case, since the 
environment of the cluster center does not favor mergers.

The idea of infalling systems can be generalized
to all galaxy mergers as the origin of lopsidedness.
In particular, mergers of galaxies could
naturally form eccentric disks. The disk would come from the disruption 
by tidal forces of one of the stellar core. The presence of massive
black holes in each spiral galaxy with bulge, strongly supports
this scenario. The end steps of the merger process would
form a binary black hole. The destruction of nuclear stellar systems
by the tidal forces of the black holes led Merritt \& Cruz (2001)
to suggest that the existence of low-density cores (and not cusps)
in giant galaxies is the consequence of mergers.
Further simulations of mergers with central black holes 
should be performed to explore the formation of
eccentric disks. 

Somewhat larger-scale asymmetry on scales $\sim$ 1 kpc is also seen in mergers
of galaxies and is deduced to be long-lived, as discussed in Section 5.3.

\subsection{Core wandering}

\begin{figure}[h]
  \centering
  \includegraphics[height=6cm,width=8cm]{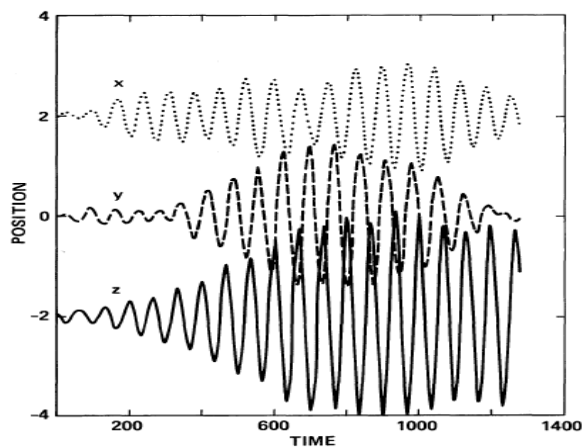}
   \caption{Evolution with time of the three coordinates
of the ``nucleus'' consisting in the central 1024 particles, selected from an N-body
simulation of 100 352 particles. The saturation of the amplitude of oscillations
is visible. From Miller \& Smith (1992).}
 \label{fig22}
  \end{figure}

Some of the nuclear lopsidedness might also be explained through
a special oscillation of the central black hole, called "core wandering".
This name came from the physics of globular cluster, that was observed
to reveal slow oscillation, with a time-scale larger than the crossing 
time in numerical simulations (Makino \& Sugimoto 1987).
Miller \& Smith (1992) showed by a large series of numerical simulations
that a massive nucleus cannot coincide with the mass centroid of its
galaxy in a stable way. The type of instability, where the motion of the nucleus
implies potential distortions in the center, which trap more particles,
is overstable, and reaches a saturation limit. The phenomenon is local, 
and the time-scale of the oscillation of the nucleus is of the same order as the
central dynamical period (cf Figure \ref{fig22}).
This core wandering appears physical, and not the consequence of
a $N^{-1/2}$ random noise oscillation, as tested by simulations with
highly varying particle number. In that case, the perturbation amplitude
that starts the growth is indeed depending on $N$, but not the limiting
amplitude, nor the growth rate, which is always a few dynamical times.
The $N^{-1/2}$ phenomenon can be hard to distinguish in small-N
simulations, and in globular clusters (Sweatman 1993), but this is
not the case for galaxies.

The stochastic part of
the core wandering phenomenon has been modeled by Chatterjee et al (2002), 
separating the force on the central mass in the collective action of the stellar
system in which it is embedded, and the fluctuating stochastic force provided
by individual stellar encounters. This second force produces a Brownian
motion of the central point mass. These motions occur on a time scale
much shorter than the time-scale of evolution of the stellar system.

As for the coherent modes of the stellar system coupled to the central mass,
the growing oscillation looks like a density wave, and the saturation
amplitude of the motion reaches a galaxy core radius.
This unstable phenomenon involves the nucleus, even if there is no
central black hole in the center, and is also observed in dynamical
friction experiments, when the decay of a satellite is studied (e.g.
Bontekoe \& van Albada 1987). The limiting amplitude of the nucleus
oscillation is then reached from above.
This kind of oscillation is also observed in spherical galaxies,
and continues to develop for a Hubble time (Miller \& Smith 1994).

\bigskip

\bigskip

Taga \& Iye (1998a) studied the oscillations of a central massive
black hole in a rotating galaxy, and also confirmed that the
phenomenon is not an $N^{-1/2}$ random noise effect, but
a true physical phenomenon. They
found by N-body simulations that a massive central body
can undertake long-lasting oscillations, but only when its mass is lower
than 10\% of the disk mass. This produces disk oscillations around it.
 Crucial must be the rotation of the pattern around the center, since 
it does not vary its amplitude, when the number of particles change.
The disk oscillations occur only when the black hole is allowed to move,
when the black hole is artificially nailed down
to the center, the disk oscillations vanish.
 They also conclude that the mechanism at the origin of this
instability is a density wave, with a fixed pattern as a function 
of radius.

When the disk around the central mass is fluid, an instability
akin to the one proposed by Shu et al (1990) is possible, with a feedback
provided by reflection on a sharp edge in the outer disk.
Heemskerk et al (1992) estimate that the edge effect is artificial,
and studied instead an $m=1$ instability arriving only when the masses
of the central object and the disk are comparable. There can
be angular momentum exchange between the mass and the disk.
To simplify their model, they considered only a gaseous disk with a central gap.
This instability is confirmed by Woodward et al (1994), and requires
the coupling with the central mass, which is displaced from the center,
and moves along a smooth, tightly wound, spiral trajectory.
Taga \& Iye (1998b) found that the sharp edge condition is not
necessary, and that an eccentric instability develops in a 
stellar disk, provided that the central mass is smaller than the
disk mass (about 10\%), and it is mobile. The eccentric instability then develops
a one-arm spiral, with an amplitude that is stronger than when the central mass is fixed to the center.
The mechanism is strongly dependent on the softening used,
and should be local to the central parts.

\subsection{Other mechanisms}

If a normal disk around a black hole is not spontaneously unstable
to $m=1$ perturbations, the modifications of the stellar distribution function $F$,
and in particular the depletion in low-angular momentum orbits, leading to the
empty loss-cone phenomenon, can provide the source of instability. If the derivative
of $F$ with respect to $J$ is positive, then spherical near-Keplerian systems are neutrally
stable (allowing the displacement of the nucleus
with respect to central mass), and the flattened non-rotating systems 
are unstable to $m=1$ modes (Tremaine 2005).

Peiris \& Tremaine (2003) construct eccentric disk models to represent the
double nucleus in M31, and claim that the inner nuclear disk must be inclined
by at least 20 degrees with respect to the main disk of the galaxy to represent the data.
Although their model is only dynamical, and does not include all the physics, with 
self-gravity, etc. , this suggests that the 
lopsidedness might be related to a warp or a misalignment. The latter could be 
the source of dynamical friction against the bulge, and relatively rapid alignment
should ensue. A possibility is that the bulge is itself misaligned with the main
galaxy disk, which could be due to a recent galaxy interaction (e.g. Ibata et al 2001,
Block et al 2006).

Salow \& Statler (2001, 2004) compute a more sophisticated model, including
self-gravity. They populate quasi-periodic orbits for stars, in the rotating frame
with a constant precession speed. The eccentricity of the orbits change sign with
radius, so that the apocenter of the orbits change phase in the plane
of the nuclear disk (cf fig 23). The resulting best fit model is similar to that
obtained with N-body simulations of a strong $m=1$ mode in a cold thin disk with a central
black hole (Bacon et al 2001). In particular the pericenters of the orbits in the inner
and outer disks are in phase opposition. The precession rate, however, is rapid
($\Omega$ = 36.5 km s$^{-1}$ pc$^{-1}$).

\bigskip

\begin{figure}[h]
 \centering
  \includegraphics[height=4cm]{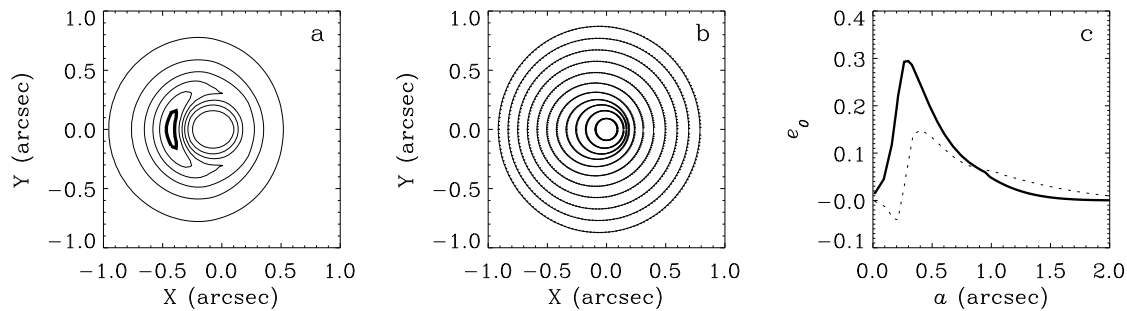}
\bigskip
  \caption{(a) Density contours of the best-fit model of the M31 eccentric
disk, the central black hole being at (X,Y)= (0,0); (b) uniformly precessing orbits in the total potential.
(c) the radial variation of the eccentricity $e$ of orbits, as a function of semi-major axis $a$: {\it dotted line}--
the orbit model, with eccentricities changing sign with radius; {\it full line} -- as a consequence of disk self-gravity,
the eccentricity does not change sign any longer. 
>From Salow \& Statler (2004).}
 \label{fig23}
  \end{figure}

\bigskip

\section{Lopsidedness in galaxies in groups, 
clusters and mergers}

\subsection {Lopsidedness in galaxies in groups}

Tidal encounters between galaxies in groups are more probable 
given the higher number density and the consequent frequent 
interaction between
galaxies in groups. Further, these have relative velocities similar to the field
galaxies, hence the lopsidedness arising due to a 
response triggered by tidal encounters is more likely to
occur in these. This is in fact borne out by the observations of
Hickson group galaxies (Rubin et al. 1991) where a large fraction
($> 50 \%$) of galaxies show lopsided rotation curves. This is much
higher than the case of field spiral galaxies where only $\sim$ 25
\% show asymmetric rotation curves (Rubin et al. 1999, Sofue \&
Rubin 2001). The observation of rotation curves of 30 galaxies in
20 Hickson groups (Nishiura et al. 2000) confirms this higher
frequency. In another study involving HI observations, all the five
major spirals in the nearby Sculptor group of galaxies (Schoenmakers
1999) show kinematical lopsidedness, and two show morphological
elongation or asymmetry.
A multi-wavelength study of two group galaxies NGC 1961 and NGC 2276
shows evidence for lopsidedness which has been attributed to tidal interactions
(Davis et al. 1997).

The two-dimensional maps of HI have been Fourier-analyzed recently
to obtain the m=1 Fourier amplitudes and phases for a sample of 18
galaxies in the Eridanus group (Angiras et al. 2006). This is the 
first quantitative analysis of asymmetry in the surface density distribution of the HI gas, and is similar to the Fourier analysis now done routinely in the 
literature for the near-IR data representing old stars in galaxies. 

The group location of
this sample allows us to serendipitously study the lopsidedness in a group
setting. The galaxies studied show a higher magnitude of asymmetry than the
field galaxies, and also a higher fraction of galaxies show asymmetry. 
The average amplitude of lopsidedness measured for the
Eridanus group galaxies is nearly twice that in the field galaxies
over the same radial range of 1.5-2.5 disk scalelengths. Second, nearly 
30 \% of the sample galaxies show lopsidedness amplitudes three times larger than the field average of 0.1. 
Fig. 24 shows the results for two galaxies UGC 068 and NGC 1325, in the Eridanus group.

The asymmetry is measured in this case to over twice the radial distance that is
typically covered in the near-IR studies (e.g. Bournaud et al .
2005 b). This is because the tracer used here is HI which extends farther out than the stars, and because the sky
background limits the Fourier analysis in the near-IR to $\sim 2.5 $
disk scalelengths.

\bigskip

\begin{figure}[h]
\centering
\includegraphics[height=1.25in,width=2.5in]{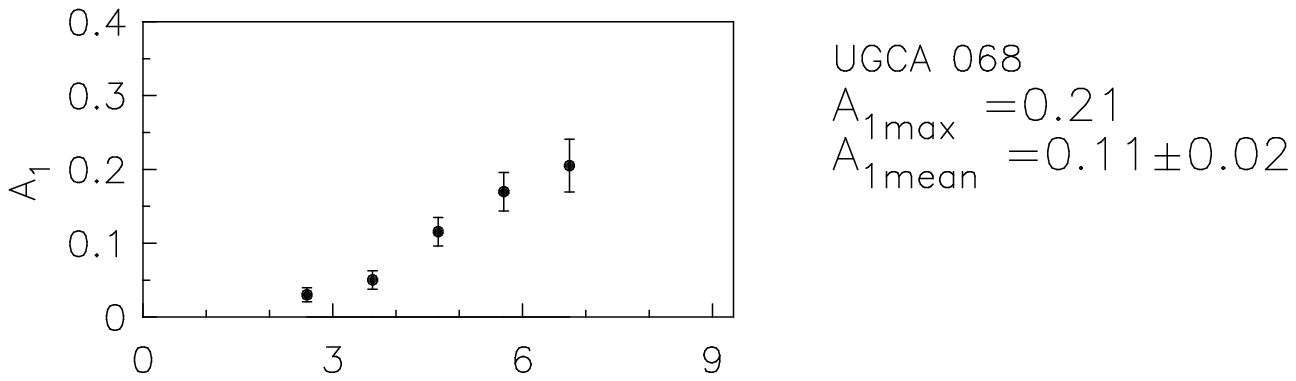}
\medskip 
\includegraphics[height=1.25in,width=2.5in]{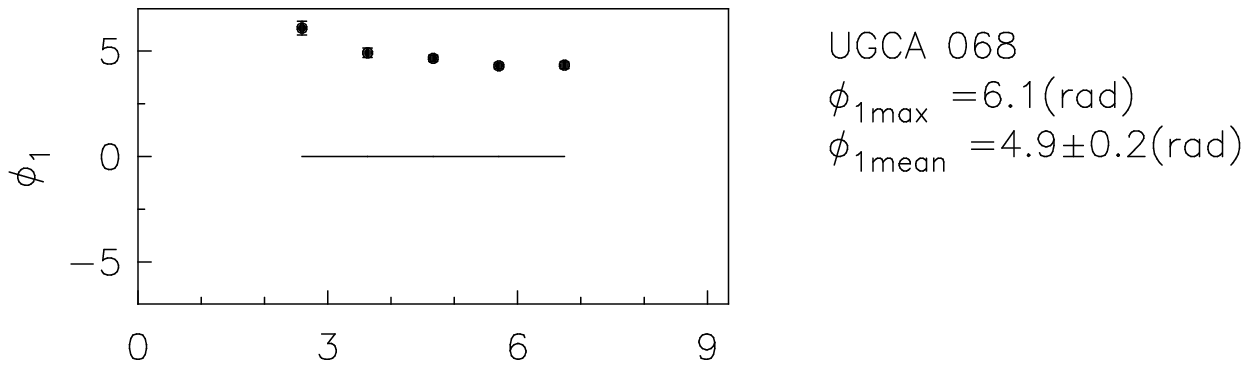} 
\bigskip
\includegraphics[height=1.25in,width=2.5in]{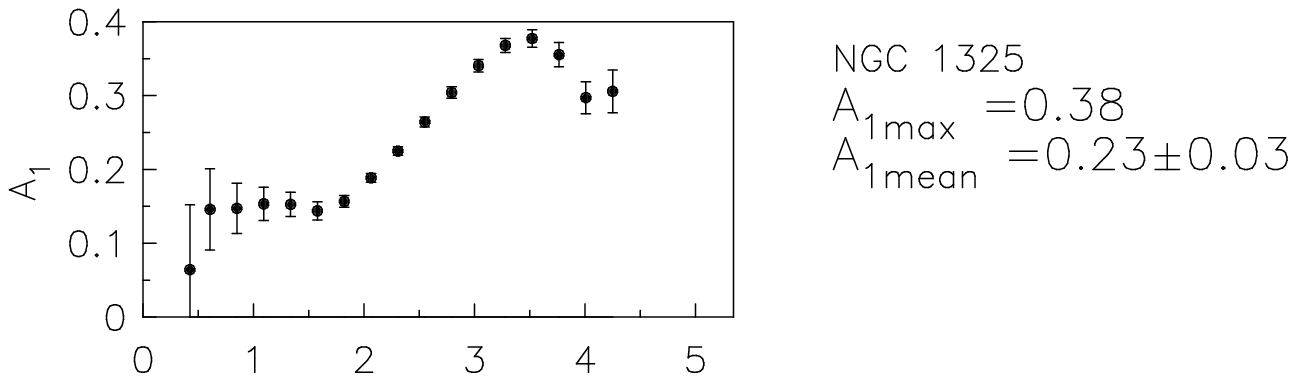}
\medskip
\includegraphics[height=1.25in,width=2.5in]{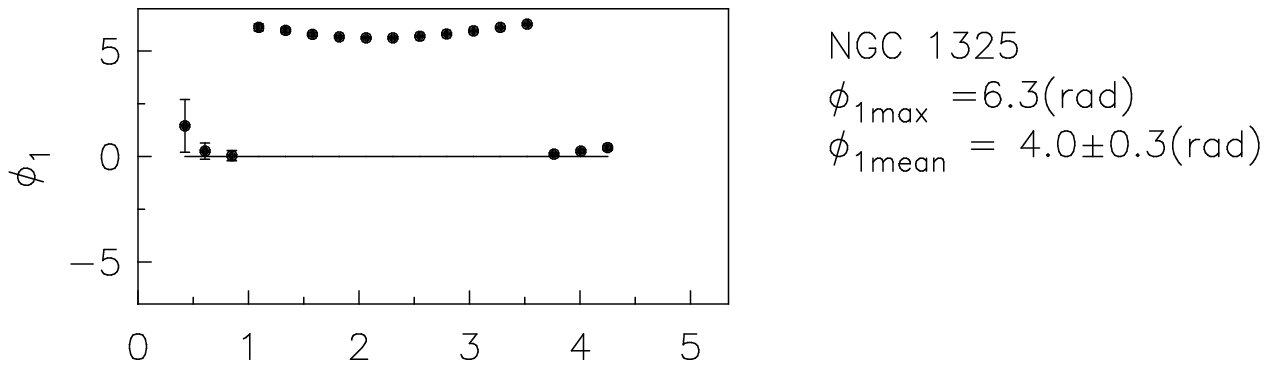}
\bigskip
\caption{The lopsided amplitude and phase of the HI surface density distribution 
versus radius for the two galaxies UGC068 and NGC 1325 in the Eridanus group, taken
from Angiras et al. (2006). The amplitude increases with radius and the 
phase is nearly constant indicating that m=1 is a global mode. The lopsidedness
in the HI can be measured until several disk scalelengths, more than twice the 
radial distance possible for the stars.} 
 \label{fig24}
\end{figure}

\bigskip

The A$_1$ values measured from the HI data and the R-band data are
available for four galaxies, and these were compared. It was found that in the 
radial region of overlap, the two tracers show a similar value of 
lopsidedness, see Fig. 25. This confirms that the origin of lopsidedness in HI is of 
a dynamical origin and not purely of a gas-dynamical process that only 
applies to the gas.

\bigskip

\begin{figure}[h]
\centering
\includegraphics[height=2.0in,width=2.5in]{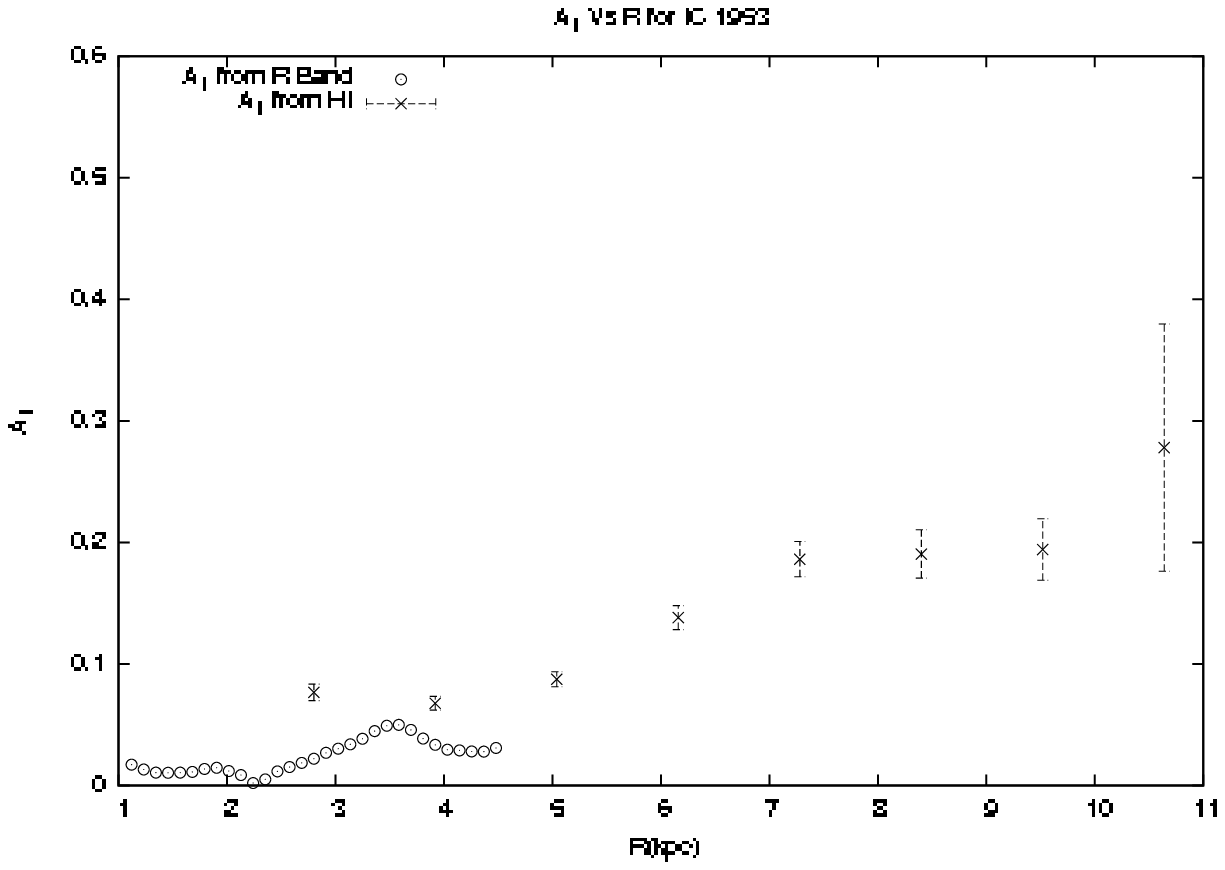}
\medskip
\includegraphics[height=2.0in,width=2.5in]{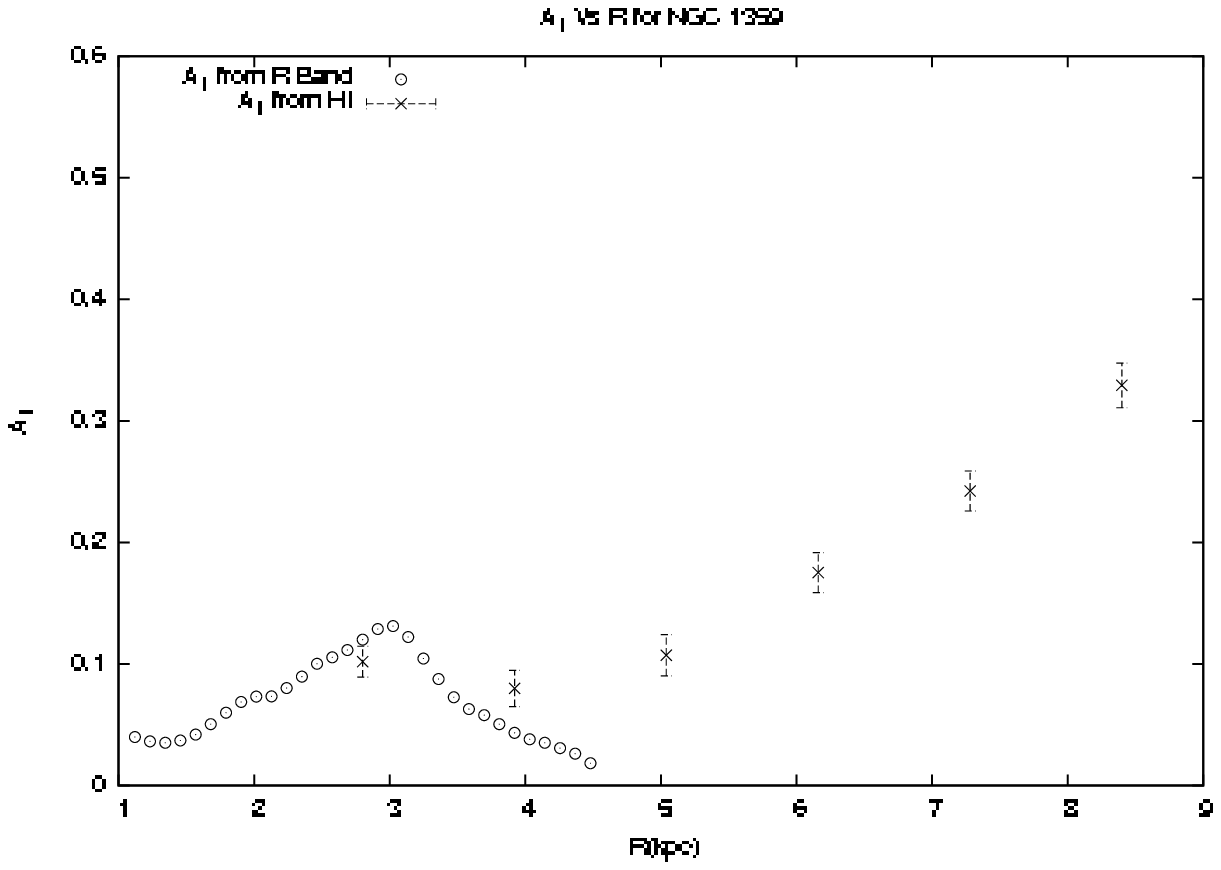}
\bigskip
\caption{A comparison of the lopsided amplitude A$_1$ obtained by 
analyzing the HI data and the R-band data for stars for NGC 1953 (left panel) and 
NGC 1359 (right panel), from Angiras et al. (2006). In the radial region 
of overlap the values are comparable, thus indicating the same origin 
for the lopsidedness both in stars and gas. The figures also strikingly
illustrate that HI is a much better tracer than stars for the study of 
lopsidedness  at large radii.}
 \label{fig25}
\end{figure}

\bigskip

A similar Fourier analysis has been done for a sample of 12 galaxies
in the Ursa Major (Angiras et al. 2007) by analyzing
the 2-D HI data (Verheijen 1997) available for these. The average value of lopsided amplitude
in the 1.5-2.5 disk scalelength region in this sample is smaller than in the Eridanus group, and is closer to the field
galaxies case, as shown in Fig. 26. This could reflect the different spatial distribution of galaxies in the two groups- the ones in the Ursa Major are distributed along a filament. 

\bigskip

\begin{figure}[h]
\centering
\includegraphics[height=2.5in,width=2.5in]{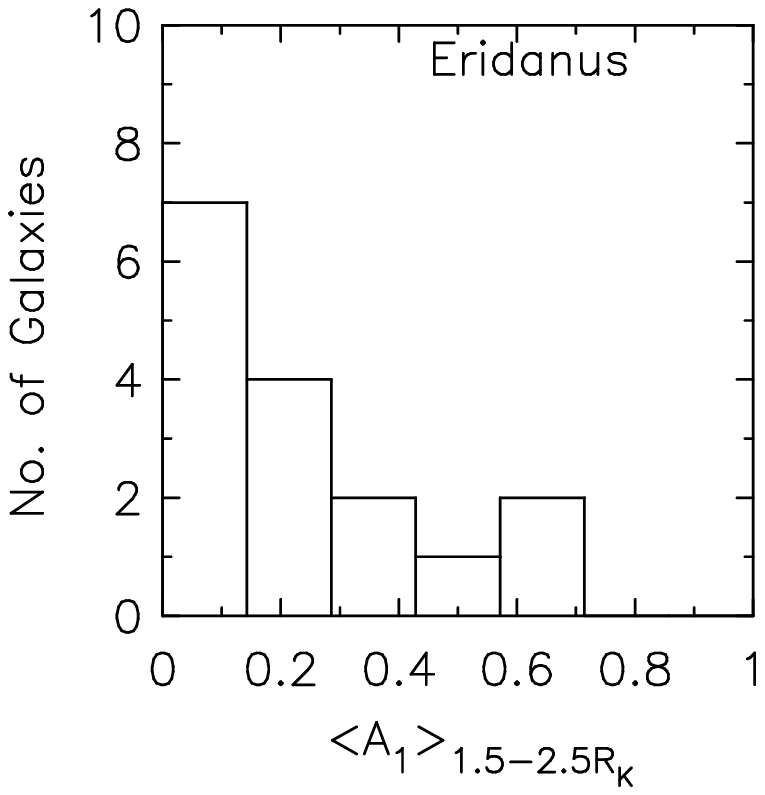}
\medskip 
\includegraphics[height=2.5in,width=2.5in]{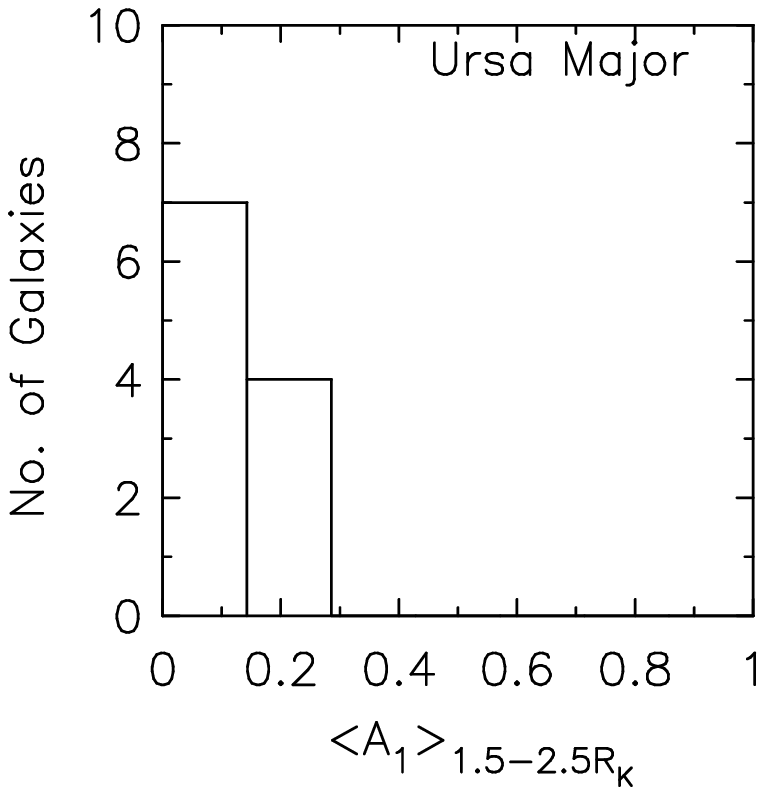} 
\caption{The histograms denoting the number of galaxies vs. the lopsided amplitude
measured in the 1.5-2.5 disk scalelength range for the Ursa Major Group
(left) and the Eridanus group (right) of galaxies, taken from Angiras et al. (2007).
The galaxies are more lopsided in the Eridanus group, indicating a substantial variation
between groups.}
 \label{fig26}
\end{figure}

\bigskip

An interesting characteristic of lopsidedness in group galaxies as noted 
by Angiras et al. (2006) is that the early-type galaxies show a higher quantitative 
lopsidedness than do the late-type galaxies, see Fig. 27. This is opposite to what is seen in the field galaxies (Bournaud et al. 2005b)
 and indicates that tidal interactions play a dominant role in generating lopsidedness
in the group galaxies. This is not surprising given the high concentration of galaxies in a group.
Tidal interactions would tend to cause a secular evolution of galaxies to an earlier type and hence
when these are the dominant mechanism for generating lopsidedness, one would expect higher values of 
lopsidedness for early-type galaxies, as argued by Bournaud et al. (2005 b).

\bigskip

\begin{figure}[h]
\centering
\includegraphics[height=2.0in,width=2.5in]{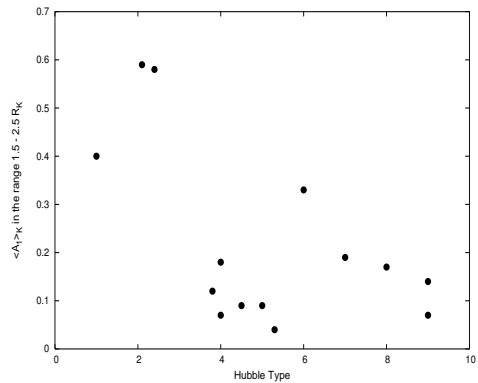} 
\caption { The lopsided amplitude measured between the radial range of 1.5-2.5 disk scalelengths vs. the galaxy type,
from Angiras et al. (2006). The early-type galaxies show a higher lopsidedness than the late-type galaxies.
This is opposite of what is seen for the field galaxies, see Fig. 5.}
 \label{fig27}
\end{figure}

\bigskip

A distinguishing feature of asymmetry in the group galaxies is that
the values of the asymmetry as measured by the mean fractional Fourier amplitudes A$_1$, A$_2$ and A$_3$ for the modes m=1,2 and 3
are found to be comparable (Angiras et al. 2007). Also,
 the derived perturbation potential parameters $\epsilon_1$, $\epsilon_2$ and $\epsilon_3$ are found to be comparable.
Although this last result depends on the model used, it reinforces the similar result obtained for the Fourier amplitudes which are directly observed and hence are model-independent (Section 3.1.2).
 This is in
contrast to the field galaxies where A$_1$ and A$_2$ are comparable and are generally stronger than A$_3$ and the other higher mode amplitudes (Rix \& Zaritsky 1995). 
 This indicates the importance of multiple, simultaneous
 tidal interactions that can occur under the special conditions of a group environment. This needs to be studied by future dynamical studies including by N-body simulations.
Since tidal interactions are frequent, occurring on a timescale of $< \sim 0.5 Mpc / 300$ km s$^{-1} \sim 3 Gyr$, the long-term maintenance of lopsidedness is not a problem in this environment.

\subsection{Lopsidedness in galaxies in clusters}

The galaxies in clusters may undergo even more frequent encounters
than in the groups because of the generally higher number density of
galaxies. However  the relative velocity between galaxies is higher in clusters and
hence the encounters are weaker in strength. The accumulation
of a large number of weak interactions  has been called
the galaxy harassment  (Lake, Kat z, \& Moore 1998). 
The asymmetry in NGC 4252 including the smoothly varying HI velocity field along the tidal tail, has been attributed to this effect 
(Haynes, Giovanelli \& Kent 2007). The
amplitude of lopsidedness generated due to tidal encounters is
expected to be weaker in this case because of the quick encounters.
 On the other hand, other dynamical processes such as asymmetry arising due
to ram pressure may be specifically applicable in a group or a
cluster setting. This may even be the dominant source of gas
asymmetry in these and can affect the HI gas lying on the outer
parts of a galactic disk. An interesting interplay of these various effects is possible as in NGC 4848 in the Coma cluster, which shows a lopsided distribution of the molecular gas (Vollmer et al. 2001). This has been explained by the interaction between the galactic gas, and the gas removed by ram pressure stripping around 4 $\times$ 10$^8$ yr ago which is now falling back.

Since the cluster galaxies often show HI deficiency especially in 
the outer parts where lopsidedness is generally more common, this 
could be a potential problem with detecting lopsidedness 
in cluster galaxies.

Despite this,  in some galaxies of  groups and 
clusters, the gas alone in known to show a strong asymmetry - as in NGC 4647 (Young 
et al. 2006).
This strong gas asymmetry has been frequently attributed to
ram pressure stripping.  However, the role of 
the gravitational potential asymmetry in
 lopsidedness cannot be neglected, even in these cases.

\subsection{Lopsidedness in centers of advanced mergers}

A fairly new regime that is just beginning to be explored is 
the asymmetry at the centers of advanced mergers of galaxies. 
 A recent systematic study was done by Jog \& Maybhate (2006), with a view to
understand the mass asymmetry and the relaxation in the central regions of mergers.

Interactions and mergers of galaxies are known to be common and significantly affect
their dynamics and evolution.
The outer regions of merger remnants covering a distance of $\sim $ few kpc to a few 10 kpc have 
been well-studied. These can be fit by an r$^{1/4}$ profile (class I), an outer exponential (class II) or a no-fit profile (class III)
(Chitre \& Jog 2002), and the first two can be explained as arising due to equal-mass mergers (e.g., Barnes 1992)
or unequal-mass mergers (Bournaud, Jog \& Combes 2005 a) respectively, while the third class
corresponds to younger remnants.
Despite 
its obvious importance for the evolution of the central regions, 
the luminosity distribution in the central regions of mergers was not 
studied systematically so far.
A few mergers where this has been studied,
such as NGC 3921 (Schweizer 1996), and Arp 163 (Chitre \& Jog 2002),
show wandering or meandering centres for the consecutive isophotes.

Jog \& Maybhate (2006) chose a sample of 12 advanced mergers which showed signs of recent interaction such as tidal tails or loops but had a merged, common center and covered all three classes discussed above; and the angular size of the galaxy was sufficiently large to allow the Fourier analysis by dividing the image into
a few radial bins. The sample was chosen so as to cover the three classes showing different remnant profiles
as described above.
 The K$_s$ band images from 2MASS were analyzed using the task ELLIPSE in STSDAS. The elliptical isophotes were fitted to galaxy images while allowing the center, ellipticity and the position angle to vary so as to get the best fit. 
Fig. 28 (top panel) shows the result for Arp 163. The
 isophotes are not concentric, instead the centers (X$_0$,Y$_0$) of consecutive isophotes 
show a wandering or sloshing behaviour, indicating an unrelaxed central region.

Another measure of asymmetry is the lopsidedness of the distribution, to obtain this a galaxy image was Fourier-analyzed with respect to a constant center and the amplitude A$_1$ and the phase p1 of the $m=1$ mode were plotted versus radius- as shown for Arp 163 in the lower panel of Fig. 28. During the Fourier analysis the center was kept fixed, for the reason as discussed
in Section 2.1.2 . The intensity and hence the mass distribution is highly lopsided with the fractional amplitude 
for the m=1 mode of $\sim 0.15$ within the central 5 kpc.

\bigskip

\begin{figure}[h]
 \includegraphics[width=12cm]{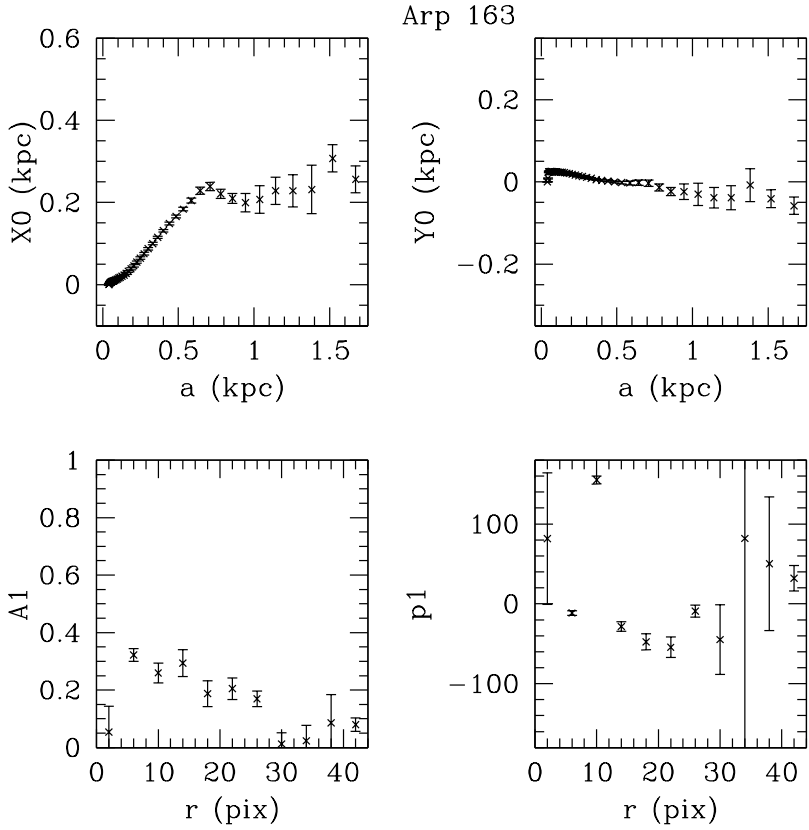} 
 \caption{Central region of the merger Arp 163 - The top panel shows the centers of the isophotes vs. the semi-major axis: the centers show a sloshing pattern indicating an unrelaxed behaviour. The lower panel shows the Fourier 
amplitude $A_1$ and the phase $p_1$ vs. radius for the lopsided mode ($m=1$): 
the lopsided amplitude is large and the phase fluctuates with radius. This is taken 
from Jog \& Maybhate (2006).}
 \label{fig28}
\end{figure}

\bigskip

All the sample galaxies show strong sloshing and lopsidedness in the central regions. The
asymmetry does not seem to significantly depend on the masses of progenitor galaxies - being similar for class I and II cases. However, it is higher for the mergers in early stages of relaxation (class III).
The corresponding values especially for the central lopsidedness are found to be smaller by a factor of few for a control sample of non-merger galaxies. This confirms that the high central asymmetry in mergers is truly due to the merger history.

The ages of remnants are deduced to be $\sim$ 1-2 Gyr as seen from the remnants with similar outer disturbed features in the N-body simulations of mergers (Bournaud et al. 2004), during which time these are likely to get chosen as our sample galaxies.

Thus the central asymmetry is long-lived, lasting for $\sim 100$ local dynamical timescales. Hence it can play an important role in the dynamical evolution of the 
central regions. First, it can help fuel the central AGN since it provides a means
of outward
transport of angular momentum. Second, it can lead to the secular growth of the bulges via 
the lopsided modes. These need to be studied in detail theoretically. Since this predicted evolution is due to the central asymmetry that is merger-driven, this process could be important in the hierarchical evolution of galaxies.

\section{Related topics}

\subsection {Relative strengths of lopsidedness (m=1) and
bars/spiral arms (m=2)}

Nearly all physical mechanisms, such as a tidal encounter or gas
accretion, that can give rise to an m=1 mode will also give rise to
an m=2 mode. Which of these two dominates depends on the detailed
parameters of the system. For example, for a distant encounter, the
m=2 mode is stronger. A prograde encounter is likely to preferentially 
generate the m=2 mode while a retrograde encounter favours the generation of a lopsided (m=1)
mode (Bournaud et al. 2005 b).

It is well-known that a two-armed spiral pattern is supported as a
kinematical feature over most of the galactic disk, or in the region
of nearly flat rotation curve in general, since the pattern speed
 given by $\Omega - \kappa / 2 $ is nearly constant in
this case (Lindblad 1959). The density wave theory of spiral
features is built around this idea (Rohlfs 1977).

The human eye/ brain likes to notice bisymmetry. This is one
reason why the two-armed spirals have received enormous amount
of attention in the theory of galactic structure and dynamics. This
is despite the fact that it has been known for a long time that a
higher $m$ value or a flocculent behaviour is more commonly observed in galaxies (see
e.g., Elmegreen \& Elmegreen 1982).

It should be noted that the early observational studies of spiral
structure such as the Hubble atlas (Sandage 1961) used blue filter
where the emission from the young stars stands out, and also the
dust extinction plays a major role in the galaxy image produced. In
contrast, the more recent studies using the near-IR band data, in
particular in the K$_s$ band avoid this problem and trace more
accurately the underlying old stellar mass
population (e.g., Block et al. 1994, Rix \& Zaritsky 1995).
Interestingly, in their recent study leading to the Large Galaxy
Catalog, Jarrett et al. (2003) show that m=1 is the most common mode
seen in the near-IR. They argue that, as noted by Block et al. (1994), there is theoretical reason to believe that m=1 modes should dominate the internal structure of spirals.

The question then is, why is there a prevalent notion even amongst
professional astronomers that lopsidedness is unusual in a galaxy
while a two-armed spiral is the norm, which is totally in contrast
to the observed case. We think one reason for this wrong notion
could be that the phase of lopsidedness is observed to be nearly
constant (see Section 2.3). Thus the resulting isophotal contours
are oval-shaped or egg-shaped, and their elongation is not striking
in the inner or optical parts of a galaxy, so an untrained eye may
 miss it. \footnote {On the other hand,
a little training or awareness allows one to detect lopsidedness
in galaxies easily, as the authors of this present article have
found !}
 If there were a strong radial dependence of the phase, then the resulting
one-armed structure as it occurs in M51 or NGC 4564 would be easy to see.

 The amplitude of lopsidedness is found to increase with radius for
the stars (Rix \& Zaritsky 1995) and for HI gas (Angiras et al.
2006, 2007). This is the reason why even a low-sensitivity HI map reveals
the lopsidedness easily - see e.g. the images of a face-on galaxy
like M101 or an edge-on galaxy like NGC 891 (Baldwin et al. 1980).
Thus it is not surprising that lopsidedness was first detected in
the HI maps. The fact that it is now found to be equally ubiquitous
in the older stars as well, is what makes its study even more
important and challenging. For example, it points to a basic
dynamical mechanism for the origin of lopsidedness, and not
something based purely on the gas dynamical processes.
An interesting feature regarding observational detection of lopsidedness,
first noted by Rix \& Zaritsky (1995), is that it does not get confused with the
inclination angle, unlike the A$_2$ values. 

\subsubsection {Observed amplitudes of m=1 and m=2 components}

The observed amplitudes A$_1$ and A$_2$ of m=1 and 2 respectively
for stars are generally comparable, and both are much larger than
the values for the higher $m$ modes for the field galaxies (Rix \& Zaritsky 1995, Bournaud et
al. 2005 b). Normally m=2 is taken to denote a bar or a spiral arm
 in the inner or outer regions respectively, or it can also denote
disk ellipticity. The amplitudes for m=2 have been measured for
larger samples (Laurikainen, Salo, \& Rautiainen 2002, Buta et al.
2005, Bournaud et al. 2005 b). The average values of A$_1$ and A$_2$
between 1.5 - 2.5 disk scalelengths have been measured for the 149
galaxies in the OSU catalog (see the Appendix in Bournaud et al.
2005 b), and the two generally show a positive correlation (see Fig. 8
in that paper). This cannot be explained if the tidal encounters were the main generating mechanism since m=2 spiral arms or bars are more easily triggered on 
direct orbits where $\Omega - {\kappa}/2$ is positive (e.g., Gerin et al. 1990). In contrast, a 
lopsided mode is more likely to be triggered or last longer in a retrograde orbit since the pattern speed
of m=1 asymmetries, $\Omega - \kappa$, is negative in a galactic disk.
The simulations by Thomasson et al. (1989) shows that retrograde tidal encounters between galaxies lead to the
formation of leading one-armed spiral arms, as in NGC 4622.
Observationally very few galaxies
show a leading arm. Pasha (1985) found that only 2 out of the 189 galaxies studied
show a leading arm (NGC 3786, NGC 5426), and each of these happen to be in a pair, which agrees with the work by Thomasson et al. (1989).
It is possible that the pattern speed of m=1 arms in other galaxies is small - this has to be checked observationally.
As discussed in Section 3.2.3 this can have a bearing on the mechanism for the origin of
lopsidedness. For example, if the pattern speed is small, the global modes are long-lived (Saha et al. 2007) and do not need to be triggered frequently. 

On the other hand, the galaxies in groups show a higher lopsidedness for the early-type galaxies, and show a comparable magnitude for all the lower modes, m=1,2, and 3. The frequent and even concurrent
tidal encounters in this setting are probably responsible for this (see Section 5.2).
 
The different modes could interact directly but this obvious line of research
 has not been followed up much. A
non-linear coupling between m=1,3 and m=2 was proposed by Masset \& Tagger (1997)
for the central regions. Even though the m=1 mode is largely seen in the outer galactic
disk while the bars and the spiral structure (m=2) are seen more in the inner parts
of a galactic disk, they can still have a dynamical effect on each other.
The heating due to a bar (m=2) can suppress the further 
growth of a lopsided mode, as was seen in the numerical simulations for a purely exponential
disk by Saha et al. (2007).
 Conversely, the presence of a lopsided mode can lead to a bar
dissolution as has been studied by Debattista \& Sambhus (2008).
This topic needs more study and could shed an important light on the dynamical evolution of a disk
due to various non-axisymmetric features.

\subsection{Asymmetry in the dark matter halo}

The study of disk asymmetry including lopsidedness has triggered many applications where the
asymmetry is used as a tool to gain information about the shape and the density distribution in the 
dark matter halo.

The dark matter halo is generally assumed to have a spherical shape, for the sake of simplicity.
This view has been challenged, and
there have been studies which have used various tracers such as the polar rings (Sackett \& Sparke 1990), 
warps (e.g. Ideta et al. 2000), gas flaring (Olling 1996, Becquaert \& Combes 1997, Narayan, Saha \& Jog 2005, Banerjee \& Jog 2008),
to deduce the shape of the dark matter halo in galaxies. A summary of topic is given in
 Natarajan (2002).

In the tidal picture of the origin of disk lopsidedness, the disk responds to a distorted 
dark matter halo. Thus we can use the observed disk asymmetry to deduce the asymmetry in the 
galactic plane of the dark matter halo, which is not visible directly.
As discussed in detail in Section 3.2.1, the observed disk lopsidedness
can be used to deduce the halo lopsidedness and indicates a few \% lopsidedness for the halo
in a typical galaxy. Similarly, on treating a self-consistent disk response, and using the
disk ellipticity, one can deduce the ellipticity of the dark matter halo (Jog 2000). 
The idea of negative disk response (Jog 1999, 2000) has been applied by Bailin et al. (2007) for a more 
realistic radial variation of the ellipticity of the potential to show that the disk response circularizes the net potential in the central region of a triaxial halo.

The amplitude of lopsidedness is higher in the group 
galaxies and can therefore imply a higher distortion of the halo, $\sim$ 10 \%
as shown for the case of the Eridanus group galaxies (Angiras et al. 2006).
The power in the various m modes is comparable (see Schoenmakers 2000,
Angiras et al. 2007).
The values of all three perturbation potentials derived $\epsilon_1, \epsilon_2 , 
\epsilon_3$ are comparable (Section 5.1). This can be an important 
clue to the mechanism for generating lopsidedness in groups, and perhaps indicates 
the importance of multiple, simultaneous tidal interactions that can occur under 
the special conditions of a group environment. 

The asymmetry in the dark matter halo of the Galaxy has been studied quantitatively as follows. The
recent survey of atomic hydrogen gas in the outer Galaxy (Levine et al. 2006) has revealed
a striking asymmetry in the thickness map of HI gas. The gas in the Northern part flares more with 
the thickness higher by a factor of $\sim 2$ compared to that in the South, at a galactocentric radial distance
of 30 kpc. This has been modeled by Saha et al (2008), who obtain the vertical scaleheight for the galactic disk in the 
gravitational field of the dark matter halo by solving the vertical force equation and the Poisson equation together. 
This model shows that the above asymmetry is best explained by a lopsided dark matter halo, with a small elliptical distortion that is out of phase with the lopsidedness.

The centres of dark matter halos are predicted to show lopsidedness as based on the N-body simulations
in the $\Lambda$CDM cosmology (Gao \& White 2006)
and the size of asymmetry is larger for larger size halos as in clusters of galaxies,
though these models need to be followed
by direct predictions which can be checked 
against observations as stressed by these authors.

The Fourier harmonic technique developed mainly to study the disk asymmetry (Jog 1997, Schoenmakers et al. 1997)
has now been applied to the kinematical data along the minor axis for the dwarf galaxy DDO 47 (Gentile et al. 2005).  This study 
has shown that the velocity dispersion components are too small to arise due to a cusp. Hence the galaxy was deduced to have a genuine core-like density distribution of the dark matter 
halo in the central regions.

The asymmetry in the halo is expected to be long-lived because of its
collisionless nature.  However, a finite pattern speed can reduce this life-time to be $\sim $ a few 
Gyr, or much less than the Hubble time (see the discussion in Section 3.2.3).

\subsection {Comparison with warps}

Spiral galaxies also show a bending of the midplane at large radii, this is known as the warp.
A warp is also a global feature of type m=1 except in the vertical direction (Binney \& Tremaine 1987).
Warps are extremely common and in fact all the main galaxies in the Local group, namely the Galaxy, M31, M33, LMC are warped. Warps are seen mainly in the HI gas, typically beyond a 4-5 disk scalelength radius (Briggs 1990)
but in many cases are also seen in stars in a radial region somewhat inside of this (Reshetnikov \& Combes 1998).

A tidal encounter in an arbitrary orientation can generate both lopsidedness as well as warps, as is known, see e.g. Weinberg (1995), and
Bournaud et al. (2004). Individual galaxies often exhibit both these phenomena, and galaxies with an intermediate angle of inclination allows both to be seen easily as in NGC 2841 (see Fig. 1b in the present paper).
Both the features share some common properties - namely they are seen preferentially
in the outer parts of a galactic disk, and their long-term maintenance against differential rotation is a problem. 
The disk self-gravity resists imposed perturbation in the inner parts, as denoted by the negative
 disk response (Section 3.2.1.c). The resulting net self-consistent disk response shows that 
the disk will exhibit lopsidedness only outside of $\sim 2$ disk scalelengths (Jog 1999, Jog 2000). A similar calculation for the perturbation along the vertical direction shows that the onset of warps in a galactic disk occurs only beyond 4-5 disk scalelengths (Saha \& Jog 2006).

We note, however, that warps and lopsided distribution are physically different features. First, in a lopsided distribution the centre of mass is shifted with respect to the original centre of mass of the galaxy. On the other hand, 
in a standard m=1 S-shaped warp, the mass distribution is symmetric with respect to 
the centre of mass and to the symmetry plane of $z = 0$. Second, the onset of warps is 
determined by a somewhat arbitrary threshold of a few degree lifting of the mid-plane away from its central value, which is set by the observational detection limits. On the other hand, a lopsided amplitude is well-determined quantitatively following the Fourier analysis- although the threshold value of what constitutes a lopsided galaxy is still fairly arbitrary (see Section 2.2).

\subsection{Implications for high redshift galaxies}

Lopsidedness is more likely at high redshift, since galaxy interactions are more frequent.
There is multiple observational evidence of more asymmetric
galaxies at high redshift (e.g. Simard et al 2002). 
Measuring the lopsidedness is however complex due to the clumpy/non-uniform
background distribution in a galaxy (e.g. Elmegreen et al 2004), and the overall low spatial resolution.

\section{Effect of lopsidedness on galaxy evolution}

A galactic disk is inherently susceptible to the formation of the
lopsided mode (Section 3). This mode can be long-lived as
evident from the high fraction of galaxies that are observed to show
lopsidedness. The dynamical origin and the evolution of these
features is a challenging problem as discussed here. But apart from that, does the
existence of lopsidedness affect the galaxy in any way? The answer
is a resounding yes.

The list of processes whereby a lopsided distribution can
affect the evolution of the galactic disk in a significant way include the following:

1. A lopsided distribution can help in the angular momentum transport
in the disk and can thus contribute to the secular evolution in the disk.
This is especially important  given the long-lived
nature of the lopsidedness. This can cause a redistribution of matter as studied
by Lynden-Bell \& Kalnajs (1972). This process is especially important at lower radii 
because the dynamical timescale is smaller. Hence the net dynamical evolution 
is likely to occur on timescales less than the age of the galaxy. The details for this process need to be worked out.

2. The fueling of the central active galactic nucleus (AGN) can occur due to the m=1 motion of the central black
hole (Section 4.2) and this could be more effective than the first process above. 

3. Lopsidedness could affect the details of galaxy formation. For example, the accretion of mass as mediated by
m=1 would be especially important for the highly disturbed high redshift galaxies.

4. The lopsided distribution results in an azimuthal asymmetry in star formation in a galactic disk.
In a lopsided potential, the effective disk surface density is shown
to be a maximum at $\phi = 0^0$, corresponding to an overdense
region, while there is an underdense region in the opposite
direction along $\phi = 180^0$. The fractional increase in surface
density at $\phi = 0^0$ is high $\sim 0.3 - 0.5 $ for strongly
lopsided galaxies (see Fig. 1, Jog 1997). Thus, the molecular gas in
the overdense region could become unstable and result in enhanced
star formation, as shown
for example for the parameters for M101. Further, the enhanced star
formation in the overdense region is argued to give rise to a
preferential formation of massive stars (Jog \& Solomon 1992). This will
result in more HII regions in the overdense region. This prediction is exactly in agreement
with observations of more HII regions seen along the SW in M101.

Lopsidedness has also been observed in the H$_{\alpha}$ emission
from the star-forming regions in dwarf irregular galaxies (Heller et
al. 2000). Such asymmetry is expected to be common in all galaxies
and we suggest future work in this area is necessary.

\section{Summary and Future Directions}

We have reviewed the spatial and
kinematical lopsidedness in a galaxy - both the observations and dynamics, as seen in the
various tracers - stars and gas, and in the inner and outer regions,
and in different settings- field and group environment.

The lopsidedness is shown to be a common phenomenon. Nearly 30 \% of
spiral galaxies show a 10\% fractional amplitude in m=1 or the first
Fourier mode. The amplitude can be higher and can go up to 30 \% in
strongly lopsided galaxies like M 101. In a group environment, this
effect is stronger: all the galaxies show lopsidedness and the
average amplitude of lopsidedness is nearly twice that in the field
case.

We recommend that the future users adopt the fractional 
Fourier amplitude A$_1$ as the standard criterion for lopsidedness.
Further, the threshold value that could be adopted could be 
the average value of 0.1 seen in the field galaxies in the intermediate radial
range of 1.5-2.5 R$_{exp}$ (Bournaud et al. 2005 b), so that 
galaxies showing a higher value can be taken to be lopsided. A uniform criterion
will enable the comparison of amplitudes of lopsidedness in different galaxies, 
and also allow a comparison of the fraction of galaxies deduced to be lopsided in different studies. 

A variety of physical mechanisms have been proposed to explain the
origin of the lopsidedness, of which the most promising are the ones
involving a tidal encounter, gas accretion, and gravitational instability. A unique feature of the m=1 perturbation
in the galactic disk is that it leads to a shift in the centre
of mass in the disk, and this further acts as an indirect force
on the original centre of the disk. The self-gravity of the perturbation 
decreases the precession rate by a factor of $\sim 10$ compared to free precession.
 The disk is thus shown to
naturally support a slowly-rotating, global, lopsided mode which is long-lived. However, the
uniqueness of this solution has been proven. Also it is not clear what would give rise to
such slow modes, except gas accretion. A
fast pattern speed as 
would occur for lopsidedness generated in a tidal encounter 
cannot yet be ruled out.

The high fraction of galaxies showing lopsidedness has still not been explained fully. The 
N-body simulations for a tidal interaction between galaxies with a live halo and gas (Bournaud et al. 2005 b) 
show that the
lifetime of the lopsided mode thus generated is a few Gyr. The other mechanism involving steady gas accretion 
would probably generate a one-armed lopsided mode which is not seen commonly.
Satellite accretion could generate the right amount of lopsidedness but would 
also thicken the disks more than is seen. It needs to be checked if a small-mass satellite 
falling in can generate the right amplitude distribution of lopsidedness without thickening 
the disk, and of course if there are such satellites available to fall in at a steady rate.
A measurement of pattern speed in real galaxies would be extremely useful in 
constraining the main mechanism for generating lopsidedness. For example, a tidal encounter is expected to give rise to a lopsided mode with a small but finite pattern speed.

In group galaxies, the ongoing continuous tidal interaction can help one get over the 
maintenance problem easily. This can explain why nearly all galaxies in a group are strongly
lopsided, and may perhaps explain why these show equal amplitudes of higher asymmetry modes. This needs to be confirmed by
detailed dynamical studies and simulations. 

Some of the open problems in this field include: a measurement of pattern speed of lopsided 
mode in real galaxies, the amplitude and thickening of the disk generated by an accretion of a 
low-mass satellite, and 
the study of origin and evolution of lopsidedness in galaxies in groups, and the evolution timescale 
of a m=1 mode in a collisionless dark matter halo. These deal with the origin and the evolution of lopsidedness
in galaxies. The related field of problems involving the study of the dark matter shape has much promise, and some work has been started along this direction with models to explain the observed HI thickness distribution.
 In the central regions of mergers of galaxies, the dynamics of the sloshing and lopsidedness seen on scales of $\sim 1$ kpc needs to be investigated. On further small scales, simulations with central black holes should be carried out to explore the formation of eccentric disks as in M31.

An even more interesting set of questions has to do with the 
effect the lopsidedness has on the further evolution of the galaxy. These include:
the angular momentum transport, the fuelling of the central active galactic nucleus, 
enhanced star formation in the overdense regions, and the early evolution of a galaxy
mediated by the m=1 mode for the studies of galaxies at  high 
redshift, and the coupling between the various modes (bars, lopsidedness etc) in a galactic
disk and its effect on the further dynamical evolution of a galaxy.

In summary, the study of asymmetry is a rich and a challenging
area in galactic structure and dynamics, with lots of open questions
- both observational and theoretical.

\bigskip

\noindent {\bf Acknowledgments:} We are happy to acknowledge the support of the Indo-French grant IFCPAR/2704-1 .

\bigskip

\bigskip
\centerline{\bf References}

\bigskip

\noindent Abraham, R.G., van den Bergh, S., Glazebrook, K., Ellis,
R.S., Santiago, B.X., Surma, P., \& Griffiths, R.E. 1996, ApJS, 107, 1

\noindent Adams, F.C., Ruden, S.P., \& Shu, F.H. 1989, ApJ 347, 959

\noindent Adler, D.S., \& Liszt, H.S. 1989, ApJ, 339, 836

\noindent Alexander, R.D., Armitage, P.J., Cuadra, J., \& Begelman, M.C. 2007, 
(arXiv 0711.0759)

\noindent Alard, C. 2001, A\&A, 379, L44

\noindent Andersen, D.R., \& Bershady, M.A. 2002, in "Disks of Galaxies: Kinematics, Dynamics and Peturbations", ASP Conference Proceedings, Vol. 275. Eds. E. Athanassoula, A. Bosma, and R. Mujica (San Francisco: Astronomical Society of the Pacific), p.39

\noindent Andersen, D.R., Bershady, M.A., Sparke, L.S., Gallagher, J.S., Wilcots, E.M.,
 van Driel, W., \& Monnier-Ragaigne, D. 2006, ApJS, 166, 505

\noindent Angiras, R.A., Jog, C.J., Omar, A., \& Dwarakanath, K.S.
 2006, MNRAS, 369, 1849

\noindent Angiras, R.A., Jog, C.J., Dwarakanath, K. S., \&
 Verheijen, M.A.W. 2007, MNRAS, 378, 276

\noindent Athanassoula, L. 2002, ApJ, 569, L83

\noindent Bacon, R., Emsellem, E., Monnet, G., \& Nieto, J.~L., 1994, A\&A, 281, 691

\noindent Bacon, R., Emsellem, E., Combes, F., Copin, Y.,
   Monnet, G., \& Martin, P. 2001, A \& A, 371, 409

\noindent Bailin, J., Simon, J. D., Bolatto, A.D., Gibson, B.K., \& Power, C.
  2007, ApJ, 667, 191

\noindent Baldwin, J.E., Lynden-Bell, D., \& Sancisi, R. 1980,
 MNRAS, 193, 313

\noindent Bally, J., Stark, A. A., Wilson, R. W., \& Henkel, C. 1988, ApJ, 324, 223

\noindent Banerjee, A., \& Jog, C.J. 2008, ApJ, 685, 254

\noindent Barnes, J.E. 1992, ApJ, 393, 484

\noindent Batcheldor D., Axon D., Merritt D. et al.: 2005, ApJS 160, 76

\noindent Beale, J.S., \& Davies, R.D. 1969, Nature, 221, 531

\noindent Becquaert, J.-F., \& Combes, F. 1997, A \& A, 325, 41

\noindent Begeman, K.G. 1987, Ph.D. thesis, University of Groningen 

\noindent Bekki, K. 2000a, ApJ, 540, L79 

\noindent Bekki, K. 2000b, ApJ, 545, 753 

\noindent Binney, J., \& Tremaine, S. 1997, Galactic Dynamics (Princeton: Princeton Univ. Press)

\noindent Block, D.L., Bertin, G., Stockton, A., Grosbol, P.,
 Moorwood, A.F.M., \& Peletier, R.F. 1994, A \& A, 288, 365

\noindent Block, D.L., Bournaud, F., Combes, F., Puerari, I., \&
Buta, R. 2002, A \& A, 394, L35

\noindent Block, D.L., Bournaud, F., Combes, F., Groess, R., Barmby, P., Ashby, M. L. N., 
Fazio, G. G., Pahre, M. A., \& Willner, S. P. 2006, Nature, 443, 832

\noindent Bontekoe, Tj. R., \& van Albada, T. S. 1987, MNRAS, 224, 349

\noindent Bosma, A. 1981, AJ, 86, 1791

\noindent Bournaud, F., Combes, F., \& Jog, C.J. 2004, A \& A, 418, L27

\noindent Bournaud, F., Jog, C.J., \& Combes, F. 2005 a, A \& A, 437, 69

\noindent Bournaud, F., Combes, F., Jog, C.J., \& Puerari, I.
 2005 b, A \& A , 438, 507

\noindent Braun, R. 1995, A \& A S, 114, 409

\noindent Briggs, F. 1990, ApJ, 352, 15

\noindent Buta, R., Vasylyev, S., Salo, H., \& Laurikainen, E. 2005, AJ,
130, 506

\noindent Chan, R. \& Junqueira, S. 2003, ApJ, 586, 780

\noindent Charlton J.C., \& Laguna P. 1995, ApJ, 444, 193

\noindent Chatterjee, P., Hernquist, L., \& Loeb, A. 2002, ApJ, 572, 371

\noindent Chemin, L., Cayattte, V., Carignan, C., Amram, P., Garrido, O., 
Hernandez, O., Maercelin, M., Adami, C., Boselli, A., \& Boulesterix, J.
  2006, MNRAS, 366, 812

\noindent Chitre, A., \& Jog, C.J. 2002, A \& A, 388, 407

\noindent Colin J., \& Athanassoula E. 1981, A\&A, 97, 63

\noindent Combes, F. 2000, in GH Advanced Lectures on the Starburst-AGN
Connection, INAOE, June 2000, ed. D. Kunth, I. Aretxaga (astro-ph/0010570)

\noindent Combes, F., Boisse, P., Mazure, A., \& Blanchard, A. 2004,
"Galaxies \& Cosmology" , second edition (Berlin: Springer-Verlag)

\noindent Combes F., Leon S., \& Meylan G. 1999, A\&A, 352, 149

\noindent Combes, F. 2008, in "Formation and Evolution of Galaxy Disks", ed. J. Funes \& E. Corsini
(arXiv:0801.0343) 

\noindent Comins, N.F., Lovelace, R. V. E., Zeltwanger, T., \& Shorey, P. 1997, ApJ, 484, L33

\noindent Conselice, C.J., Bershady, M.A., \& Jangren, A. 2000, ApJ,
529, 886

\noindent Davidge, T. J., Rigaut, F., Doyon, R., \& Crampton, D. 1997, AJ, 113, 2094

\noindent Davis, D.S., Keel, W.C., Mulchaey, J.S., \& Henning, P.A. 1997, AJ, 114, 613

\noindent Debattista, V., \& Sambhus, N. 2008, preprint

\noindent de Oliveira, M., \& Combes, F. 2008, in preparation

\noindent de Vaucouleurs G., \& Freeman K.C. 1970, IAU 38, 356

\noindent Downes, D., \& Solomon, P.M. 1998, ApJ, 507, 615

\noindent Dressler, A., \& Richstone, D.~O., 1988, ApJ 324, 701

\noindent Dury, V. De Rijcke, S., Debattista, V.P., \& Dejonghe H. 2008, MNRAS, 387, 2

\noindent Earn D., \& Lynden-Bell, D. 1996, MNRAS, 278, 395

\noindent Elmegreen, D.M. \& Elmegreen, B. G. 1982, MNRAS, 201, 1021

\noindent Elmegreen, D.M., Elmegreen, B. G., \& Hirst, A. C. 2004, ApJ, 604, L21

\noindent Emsellem, E., \& Combes, F., 1997, A\&A, 323, 674

\noindent Emsellem, E. , Cappellari M., Peletier R., McDermid, R.M., Bacon, R., Bureau, M., Copin, Y., Davies, Roger L., Krajnovic, D., Kuntschner, H. and 2 coauthors 2004, MNRAS, 352, 721

\noindent Fuchs, B. 2004, A \& A, 419, 941

\noindent Gao, L., \& White, S.D.M. 2006, MNRAS, 373, 65

\noindent Garcia, M. R., Murray, S. S., \& Primini, F. A. et al. 2000, ApJ, 537, 23

\noindent Garcia-Burillo, S., Sempere, M. J., Combes, F., Hunt, L. K., \& Neri, R.
 2000, A\&A, 363, 869

\noindent Garcia-Burillo, S., Combes, F., Hunt, L. K. et al.:   2003, A\&A, 407, 485

\noindent Gebhardt, K., Bender, R., Bower, G. et al. 2000, ApJ, 539, L13

\noindent Gentile, G., Burkert, A., Salucci, P., Klein, U., \& Walter, F. 2005, ApJ, 
  634, L145

\noindent Gerin M., Combes F., \& Athanassoula E. 1990, A\&A, 230, 37

\noindent Giovanelli, R., \& Haynes, M.P. in Galactic and Extragalactic Radio 
Astronomy, eds. G.L. Verschuur \& K.I. Kellermann, Second edition, New York- Springer,
522

\noindent Goldreich, P., \& Tremaine, S., 1979, AJ, 84, 1638

\noindent Gonzalez-Serrano J.I., \& Carballo R. 2000, A\&AS, 142, 353

\noindent Haynes, M.P., Giovanelli, R., \& Kent, B.R., 2007, ApJ, 665, L19

\noindent Haynes, M.P., Hogg, D.E., Maddalena, R.J., Roberts, M.S.,
\& van Zee, L. 1998, AJ, 115, 62

\noindent Heemskerk, M. H. M.. Papaloizou, J. C., \& Savonije, G. J., 1992, A\&A 260, 161

\noindent Heller, A.B., Brosch, N., Almoznino, E., van Zee, L., \&
Salzer, J. 2000, MNRAS, 316, 569

\noindent Hozumi, S. \& Fujiwara, T. 1989, PASJ, 41, 841

\noindent Ibata, R., Irwin, M., Lewis, G., Ferguson, A. M. N., \& Tanvir, N. 2001, Nature, 412, 49

\noindent Ideta, M. 2002, ApJ, 568, 190

\noindent Ideta, M., Hozumi, S., Tsuchiya, T., \& Takizawa, M. 2000, MNRAS, 311, 733

\noindent Irwin, M., Ferguson, A. M. N., Ibata, R., Lewis, G., \& Tanvir, N. 2005, ApJ, 628, L105

\noindent Jacobs, V., \& Sellwood, J.~A., 2001, ApJ, 555, L25

\noindent Jarrett, T.H., Chester, T., Cutri, R., Schneider, S.E., \&
Huchra, J.P. 2003, AJ, 125, 525

\noindent Jog, C.J. 1992, ApJ, 390, 378

\noindent Jog, C.J. 1997, ApJ, 488, 642

\noindent Jog, C.J. 1999, ApJ, 522, 661

\noindent Jog, C.J. 2000, ApJ, 542, 216

\noindent Jog, C.J. 2002, A \& A, 391, 471

\noindent Jog, C. J., \& Maybhate, A. 2006, MNRAS, 370, 891

\noindent Jog, C.J., \& Solomon, P.M. 1992, ApJ, 387, 152 

\noindent Johnston K.V., Sigurdsson S., \& Henquist L. 1999, MNRAS, 302, 771

\noindent Junqueira, S. \& Combes, F. 1996, A\&A, 312, 703

\noindent Kamphuis, P. 1993, Ph.D. thesis, University of Groningen.

\noindent Kannappan, S.J., \& Fabricant, D.G. 2001, AJ, 121, 140

\noindent Kato, S. 1983, PASJ, 35, 248

\noindent King, I.~R., Stanford, S.~A., \& Crane, P., 1995, AJ, 109, 164

\noindent Kormendy, J., 1988, ApJ, 325, 128

\noindent Kormendy, J., \& Bender, R., 1999, ApJ, 522, 772 

\noindent Kornreich, D.A., Haynes, M.P., \& Lovelace, R.V.E.
1998, AJ, 116, 2154

\noindent Kornreich, D.A., Lovelace, R.V.E., \& Haynes, M.P. 2002,
 ApJ, 580, 705

\noindent Kuijken, K. 1993, ApJ, 409, 68

\noindent Kuijken, K., Fisher, D., \& Merrifield, M.R. 1996, MNRAS,
283, 543

\noindent Laine, S., \& Heller, C.H. 1999, MNRAS, 308, 557 
7479

\noindent Lake, G., Katz, N. \& Moore, B. 1998, ApJ, 495, 152

\noindent Lallemand A., Duchesne M., \& Walker M.F., 1960, PASP, 72, 76

\noindent Lauer, T.~R., Faber, S.~M., Gebhardt K. et al 2005, AJ, 129, 2138

\noindent Lauer, T.~R., Gebhardt K., Richstone D. et al. 2002, AJ, 124, 1975

\noindent Lauer, T.~R., Faber, S.~M., Ajhar, E.~A., Grillmair, C.~J., \& Scowen, P.~A., 1998, AJ, 116, 2263

\noindent Lauer, T.~R., Tremaine, S., Ajhar, E. A. et al. 1996, ApJ, 471, L79 

\noindent Lauer, T.~R., Faber, S.~M., Groth, E.~J., et al. 1993, AJ, 106, 1436

\noindent Laurikainen, E., Salo, H., \& Rautiainen, P. 2002, MNRAS,
331, 880

\noindent Lee E., \& Goodman J. 1999, MNRAS, 308, 984

\noindent Levine, E.S., Blitz, L., \& Heiles, C. 2006, ApJ, 643, 881

\noindent Levine, S.E., \& Sparke, L.S. 1998, ApJ, 496, L13

\noindent Light, E.~S., Danielson, R.~E., \& Schwarzschild, M., 1974, ApJ, 194, 257

\noindent Lin C.C., \& Shu F.H. 1964, ApJ, 140, 646

\noindent Lindblad, P. 1959, Galactic Dynamics in Handbuch der Physics, 53, 21

\noindent Liszt, H. 2006, A\&A, 447, 533

\noindent Lord, S.D., \& Young, J.S. 1990, ApJ, 356, 135

\noindent Lovelace, R.V.E., Zhang, L., Kornreich, D.A., \& Haynes, M.P., 1999, ApJ, 524, 634

\noindent Lynden-Bell, D., \& Kalnajs, A. 1972, MNRAS, 157, 1

\noindent Makino, J., \& Sugimoto, D. 1987,  PASJ, 39, 589

\noindent Manthey, E. et al. 2008, in preparation

\noindent Mapelli, M., Moore, B., \& Bland-Hawthorn, J. 2008, MNRAS, 388, 697

\noindent Marshall, D.J., Fux R., Robin, A.C., \& Reyl\'e, C. 2008, A\&A 477, L21

\noindent Masset, F., \& Tagger, M. 1997, A \& A, 332, 442

\noindent Matthews, L.D., van Driel, W., \& Gallagher, J.S. 1998, AJ, 116, 2196

\noindent McDermid, R. M., Emsellem, E., Shapiro, K. L. et al. 2006, MNRAS, 373, 906

\noindent Melchior, A-L., Viallefond, F., Guelin, M., \& Neininger, N., 2000, MNRAS 312, L29

\noindent Merritt, D., \& Cruz, F. 2001, ApJ, 551, L41 

\noindent Miller, R.H., \& Smith, B.F., 1992, ApJ, 393, 508

\noindent Miller, R.H., \& Smith, B.F., 1994, CeMDA, 59, 161

\noindent Mould, J., Graham, J., Matthews, K., Soifer, B. T., \& Phinney, E. S. 1989, ApJ, 339, L21

\noindent Narayan, C.A., Saha, K., \& Jog, C.J. 2005, A \& A, 440, 523

\noindent Natarajan, P. 2002, The shape of the galaxies and their dark halos, Yale cosmology workshop 
(Singapore: World Scientific)

\noindent Nelson, R.W., \& Tremaine, S.D. 1995, MNRAS,  275, 897

\noindent Nieto, J.-L., Macchetto, F. D., Perryman, M. A. C., di Serego Alighieri, S., \& Lelievre, G. 1986, A\&A, 165, 189

\noindent Nishiura, S., Shimada, M., Ohyama, Y., Murayama, T., \&
Taniguchi, Y. 2000, AJ, 120, 1691

\noindent Noordermeer, E., Sparke, L., \& Levine, S.E. 2001, MNRAS,
 328, 1064

\noindent Noordermeer, E., van der Hulst, J. M., Sancisi, R., Swaters, R. A., \& van
Albada, T. S. 2005, A\&A 442, 137

\noindent Olling, R.P. 1996, AJ, 112, 481

\noindent Pasha, I.I. 1985, Soviet Astr. Letters, 11, 1 

\noindent Peiris H., \& Tremaine S. 2003, ApJ 599, 237

\noindent Phookun B., \& Mundy L.G. 1995, ApJ, 453, 154

\noindent Phookun, B., Vogel, S.N., \&
 Mundy, L.G. 1993, ApJ, 418, 113

\noindent Pranav, P., \& Jog, C.J. 2008, in preparation.

\noindent Quillen A. C., \& Hubbard A. 2003, AJ, 125, 2998

\noindent Rauch K.P., \& Tremaine S. 1996, New A., 1, 149

\noindent Reichard, T.A., Heckman, T.M., Rudnick, G., Brinchmann, J., \& Kauffmann G. 2008, 
ApJ, 677, 186

\noindent Reshetnikov, V., \& Combes, F. 1998, A \& A, 337, 8

\noindent Richmond, M.W., \& Knapp, G.R. 1986, AJ, 91, 517

\noindent Richter O.-G., \& Sancisi, R. 1994, A \& A, 290, L9

\noindent Rix, H.-W., \& Rieke, M.J. 1993, ApJ, 418, 123

\noindent Rix, H.-W., \& Zaritsky, D. 1995, ApJ, 447, 82

\noindent Rodriguez-Fernandez N., \& Combes F. 2008, A\&A, 489, 115

\noindent Rohlfs, K. 1977, Lecture notes on density wave theory,
(Berlin: Springer-Verlag)

\noindent Rubin, V.C., Hunter, D.A., \& Ford, W.K. Jr. 1991, ApJS,
76, 153

\noindent Rubin, V.C., Waterman, A.H., \& Kenney, J. D. 1999, AJ,
118, 236

\noindent Rudnick, G. \& Rix, H.-W. 1998, AJ, 116, 1163

\noindent Sackett, P.D., \& Sparke, L.S. 1990, ApJ, 361, 408

\noindent Sage, L.J., \& Solomon, P.M. 1991, ApJ, 380, 392

\noindent Saha, K., Levine, E., Jog, C.J., \& Blitz, L. 2008, ApJ, submitted.

\noindent Saha, K. 2008, MNRAS, 386, L101

\noindent Saha, K., Combes, F., \& Jog, C.J. 2007, MNRAS, 382, 419

\noindent Saha, K., \& Jog, C.J. 2006, MNRAS, 367, 1297

\noindent Sakamoto K., Matsushita S., Peck A. et al. 2004, ApJ, 616, L59

\noindent Salow, R.M., \& Statler, T.S. 2004, ApJ, 611, 245 

\noindent Salow, R.M., \& Statler, T.S. 2001, ApJ, 551, L49

\noindent Sambhus, N., \& Sridhar, S. 2002, A\&A 388, 766 R

\noindent Sancisi, R., Fraternali, F., Oosterloo, T., \& van der Hulst J.M. 2008,
Astronomy \& Astrophysics Reviews (arXiv0803.0109)

\noindent Sandage, A. 1961, The Hubble Atlas Of Galaxies
 (Washington: Carnegie Institution Of Washington)

\noindent Schoenmakers, R.H.M. 1999, Ph.D. thesis, University of
Groningen.

\noindent Schoenmakers, R.H.M. 2000, in Small Galaxy Groups, eds.
Valtonen,M. \& Flynn,C., ASP conf. series vol. 209, pg. 54

\noindent Schoenmakers, R.H.M., Franx, M., \& de Zeeuw, P.T.
  1997, MNRAS, 292, 349

\noindent Schweizer, F. 1996, AJ, 111, 109

\noindent Sellwood, J.A., \& Valluri, M. 1997, MNRAS, 287, 124

\noindent Semelin, B. \& Combes, F. 2005, A \& A, 441, 55

\noindent Shlosman, I. 2005, in The Evolution of Starbursts, Eds. S. Huettemeister and E. Manthey (Melville: AIP) (astro-ph/0412163) 

\noindent Shu F.H., Tremaine S., Adams F.C., \& Ruden S.P., 1990, ApJ, 358, 495

\noindent Simard, L., Willmer, C. N. A., Vogt, N. P. et al. 2002, ApJS, 142, 1

\noindent Sofue, Y. \& Rubin, V. 2001, ARAA, 39, 137

\noindent Statler, T.~S., King, I.~R., Crane, P., and Jedrzejewski, R.~I. 1999, AJ, 117, 894 

\noindent Statler, T. 2001, AJ, 122, 2257

\noindent Swaters, R.A., Schoenmakers, R.H.M., Sancisi, R., \& van
 Albada, T.S. 1999, MNRAS, 304, 330

\noindent Swaters, R.A., van Albada, T.S., van der Hulst, J.M., \&
Sancisi, R. 2002, A \& A, 390, 829

\noindent Swaters, R.A. 1999, Dark matter in late-type dwarf
galaxies, Ph.D. thesis, University of Groningen.

\noindent Sweatman W.L. 1993, MNRAS 261, 497

\noindent Syer, D., \& Tremaine, S., 1996, MNRAS, 281, 925

\noindent Taga, M., \& Iye, M. 1998a, MNRAS, 299, 111

\noindent Taga, M., \& Iye, M. 1998b, MNRAS, 299, 1132

\noindent Thatte N., Tecza M., \& Genzel R. 2000, A\&A,  364, L47

\noindent Thomasson, M., Donner, K.J., Sundelius, B., Byrd, G.G., Huang, T.-Y., \& Valtonen, M.J.
1989, A \& A, 211, 25

\noindent Toomre, A. 1981, in Structure \& Evolution of normal
galaxies, eds. S.M. Fall \& D. Lynden-Bell (Cambridge: Cambridge
Univ. Press)

\noindent Toth, G., \& Ostriker, J.P. 1992, ApJ, 389, 5

\noindent Touma J. 2002, MNRAS, 333, 583

\noindent Tremaine, S.D., Ostriker J.P., \& Spitzer L. 1975, ApJ, 196, 407

\noindent Tremaine, S. 1995, AJ, 110, 628

\noindent Tremaine S. 2001, AJ, 121, 1776

\noindent Tremaine S. 2005, ApJ, 625, 143

\noindent Verheijen, M.A.W. 1997, Ph.D. thesis, University of
Groningen

\noindent Vesperini, E., \& Weinberg, M.D. 2000, ApJ, 534, 598

\noindent Vollmer, B., Braine, J., Balkowski, C., Cayatte, V., \& Duschl, W. J. 2001, A \& A, 
 374, 824

\noindent Walker, I. R., Mihos, J.C., \& Hernquist, L. 1996, ApJ,
 460, 121

\noindent Weinberg, M.D. 1994, ApJ, 421, 481

\noindent Weinberg, M.D. 1995, ApJ, 455, L31

\noindent Weinberg, M.D. 1998, MNRAS 299, 499

\noindent Wilcots, E. M., \& Prescott, M.K.M. 2004, AJ, 127, 1900

\noindent Wilson, A.S., \& Baldwin, J.A. 1985, ApJ, 289, 124 

\noindent Woodward, J.W., Tohline, J.E., \& Hachisu, J. 1994, ApJ, 420, 247 

\noindent Young, J.S. 1990, in The Interstellar Medium in Galaxies, eds. H.A. Thronson and M.J. Shull, Kluwer: Dordrecht, pg. 67

\noindent Young, L.M., Rosolowski, E., van Gorkom, J.H., \& Lamb,
S.A. 2006, ApJ, 650, 166

\noindent Zaritsky, D., \& Rix, H.-W. 1997, ApJ, 477, 118

\end{document}